\documentclass[11pt, twoside]{article}

\usepackage{latexsym}
\usepackage{amsmath,amssymb,amscd}
\usepackage{amsthm}
\usepackage[pdftex]{graphicx,color}
\usepackage{epsfig}
\usepackage{stmaryrd}
\usepackage{multicol}
\usepackage{pdfsync}
\usepackage{dsfont}
\usepackage{natbib}
\usepackage[colorlinks=true,breaklinks=true,bookmarks=true,urlcolor=blue,
     citecolor=blue,linkcolor=blue,bookmarksopen=false,draft=false]{hyperref}
\usepackage{subfig}
\usepackage[nohead, margin=0.8in, truedimen]{geometry}
\usepackage{mathrsfs, mathabx}   % nice rounded characters
\usepackage{array}

\newif\ifdraft
\drafttrue
\newtheorem{problem}{Problem}
\newtheorem{theorem}{Theorem}[section]
\newtheorem{lemma}{Lemma}[section]
\newtheorem{definition}{Definition}[section]
\newtheorem{corollary}{Corollary}[section]

\newtheorem{example}{Example}[section]
\newtheorem{assumption}{Assumption}[section]
\newtheorem{proposition}{Proposition}[section]
\newtheorem{remark}{Remark}[section]

\newcommand{\reals}             {\mathbb{R}}

\renewcommand{\Pr}              {\mathbb{P}}

\newcommand{\E}                 {\mathbb{E}}

\def\argmax{\operatornamewithlimits{arg\,max}}
\def\argmin{\operatornamewithlimits{arg\,min}}

\def\ext{\operatornamewithlimits{ext}}

\def\bydef{\ensuremath{\stackrel{\textup{\tiny def}}{=}}}
\newcommand{\tr}{^T}
\newcommand{\scprod}[2]{\ensuremath{\mb{#1}\tr\mb{#2}}}
\newcommand{\mb}[1]{\ensuremath{\boldsymbol{#1}}}

\newcommand{\oneN}[1]{\ensuremath{\mb{1}_{#1}}}
\newcommand{\opt}{\star}

\newcommand{\refmeas}{\ensuremath{\mathbb{P}}}
\newcommand{\var}{\ensuremath{\textup{VaR}}}
\newcommand{\cvar}{\ensuremath{\textup{AVaR}}}
\newcommand{\lev}{\varepsilon}
\newcommand{\cons}{\ensuremath{\mu_C}}
\newcommand{\incons}{\ensuremath{\mu_I}}
\newcommand{\qincons}{\ensuremath{\mathcal{Q}_I}}
\newcommand{\qcons}{\ensuremath{\mathcal{Q}_C}}
\newcommand{\qcondit}[1]{\ensuremath{\mathcal{Q}_{#1}}}
\newcommand{\nodes}[1]{\ensuremath{\Omega_{#1}}}
\newcommand{\child}[1]{\ensuremath{\mathscr{C}_{#1}}}
\newcommand{\desc}[1]{\ensuremath{\mathscr{D}_{#1}}}
\newcommand{\spacenodes}[1]{\ensuremath{\reals^{|\nodes{#1}|}}}
\newcommand{\spacechild}[1]{\ensuremath{\reals^{|\child{#1}|}}}
\newcommand{\filt}{\mathcal{F}}
\newcommand{\rvspace}{\mathcal{X}}
\newcommand{\expectmeas}[2]{\E_{#1}\left[#2\right]}
\newcommand{\alphagen}[2]{\alpha_{#1,#2}}
\newcommand{\alphaopt}[2]{\alpha^{\opt}_{#1,#2}}

\newcommand{\qcal}{\mathcal{Q}}

\newcommand{\dc}{\textup{DC}}

\newcommand{\suchthat}{\ensuremath{\,:\,}}
\newcommand{\mysetminus}{\ensuremath{\,\setminus\,}}

\newcommand{\sub}[1]{\ensuremath{\textup{sub}(#1)}}
\newcommand{\basepoly}[1]{\ensuremath{\mathcal{B}_{#1}}}
\newcommand{\polymat}[1]{\ensuremath{\mathcal{P}_{#1}}}
\newcommand{\extpolymat}[1]{\ensuremath{\mathcal{EP}_{#1}}}

\newcommand{\face}{\operatorname{face}}

\begin{document}
%%%%%%%%%%%%%%%%

\title{\LARGE \bf Tight Approximations of Dynamic Risk Measures}
\author{Dan A. Iancu\thanks{Graduate School of Business, Stanford University, 655 Knight Way, Stanford, CA 94305. Email: daniancu@stanford.edu. This research was conducted while the author was with the Department of Mathematical Sciences of the IBM T.J. Watson Research Center, whose financial support is gratefully acknowledged.} \and Marek Petrik\thanks{IBM T.J. Watson Research Center, Yorktown Heights, NY 10598. Email: \{mpetrik, dharmash\}@us.ibm.com.} \and Dharmashankar Subramanian\footnotemark[2]}
\date{November 9, 2012}
\maketitle

\begin{abstract}
This paper compares two different frameworks recently introduced in the literature for measuring risk in a multi-period setting. The first corresponds to applying a single coherent risk measure to the cumulative future costs, while the second involves applying a composition of one-step coherent risk mappings. We summarize the relative strengths of the two methods, characterize several necessary and sufficient conditions under which one of the measurements always dominates the other, and introduce a metric to quantify how close the two risk measures are. 

Using this notion, we address the question of how tightly a given coherent measure can be approximated by lower or upper-bounding compositional measures. We exhibit an interesting asymmetry between the two cases: the tightest possible upper-bound can be exactly characterized, and corresponds to a popular construction in the literature, while the tightest-possible lower bound is not readily available. 
We show that testing domination and computing the approximation factors is generally NP-hard, even when the risk measures in question are comonotonic and law-invariant. However, we characterize conditions and discuss several examples where polynomial-time algorithms are possible. One such case is the well-known Conditional Value-at-Risk measure, which is further explored in our companion paper \citet{Huang_I_Petrik_Subraman_cvar_2012}. Our theoretical and algorithmic constructions exploit interesting connections between the study of risk measures and the theory of submodularity and combinatorial optimization, which may be of independent interest.
\end{abstract}

\section{Introduction.}
\label{sec:introduction}
Measuring the intrinsic risk in a particular unknown outcome and comparing multiple risky alternatives has been a topic of central concern in a wide range of academic disciplines, resulting in the development of numerous frameworks, such as expected utility, stochastic ordering, and, in recent years, convex and coherent risk measures. 

The latter class has emerged as an axiomatically justified and computationally tractable alternative to several classical approaches, and has provided a strong bridge across a variety of parallel streams of research, including ambiguous representations of preferences in economics (e.g., \citet{Gilboa_Schmeidler_89}, \citet{Schmeidler_1989}, \citet{Epstein_Schneid_03}, \citet{Mache_Marin_Rustic_06}), axiomatic treatments of market risk in financial mathematics (\citet{Artzner_Delbaen_1999,follmer_schied_2001}), actuarial science (\citet{LynnWirch1999337,Wang_2000,acerbi_spectral_rm_2002,Kusuoka_2001,Tsanakas2004223}), operations research (\citet{Bental_Teboulle_2007}) and statistics (\citet{Huber_book}). As such, our goal in the present paper is not to \emph{motivate} the use of risk measures -- rather, we take the framework as given, and investigate two distinct ways of using it to ascribe risk in dynamic decision settings.

A first approach, prevalent among practitioners, entails applying a static risk measure to the total future costs accumulated over the remaining problem horizon, and conditioned on the available information. More formally, a decision maker faced with a future sequence of random costs $X_t,\dots,X_T$, respectively dispensed over a finite horizon $t,t+1,\dots,T$, would measure the risk at time $t$ by $\mu_t( X_t + \dots + X_T | \mathcal{F}_t)$, where $\mathcal{F}_t$ denotes the filtration containing all information at time $t$, and $\mu_t$ is a static risk measure. In practice, the same $\mu_t = \mu$ is often used at every time $t$, resulting in a risk preference that is easy to specify and calibrate. Apart from simplicity, the approach also has one other key advantage: when the risk measure used is convex, \emph{static} decisions can be efficiently computed by combining simulation procedures with convex optimization (e.g., \citet{Rock_Uryas_2000}, \citet{Ruszczynski_shapiro_opt_cvx_06}). This has lead to a wide adoption of the methodology in practice, as well as in several academic papers (see, e.g., \citet{Basak_Shapiro}, \citet{CuocoHeIssaenko_2008} and references therein).

The paradigm above, however, is known to suffer from several serious shortcomings. It can result in inconsistent preferences over risk profiles in time, whereby a decision maker faced with two alternative cumulative costs $Y$ and $Z$ can deem $Y$ riskier than $Z$ in every state of the world at some time $t+1$, but nonetheless deem $Z$ riskier than $Y$ at time $t$. This \emph{dynamic} or \emph{time} \emph{ inconsistency} has been criticized from an axiomatic perspective, as it is a staple of irrational behavior \citep{Epstein_Schneid_03, Roorda_Schum_Engw_2005_coherent, delbaen_multiperiod_2007}. Furthermore, time inconsistent objectives couple risk preferences over time, which is very undesirable from a dynamic optimization viewpoint, since it prevents applying the principles of Dynamic Programming to decompose the problem in stages (\citet{Epstein_Schneid_03,Ruszczynski_Shapiro_06,Nilim_ElGhaoui_2005,Iyengar_2005_robust_DP}).

In order to correct such undesirable effects, additional conditions must be imposed on the risk measurement process at distinct time periods. Such requirements have been discussed extensively in the literature, and it has been shown that any risk measure that is \emph{time consistent} is obtained by composing one-step conditional risk mappings. More formally, a time consistent decision maker faced with costs $X_1,\dots,X_T$ would assess the risk at time $t$ by $\mu_t\bigl(\mu_{t+1} (\dots \mu_T( X_t + \dots + X_T | \mathcal{F}_t ) \dots) \bigr)$, for a set of suitable mappings $\{\mu_\tau\}_{\tau \in \{t,\dots,T\}}$ (see, e.g., \citet{Epstein_Schneid_03}, \citet{Riedel2004185}, \citet{Cheridito_Delbaen_Kupper_2006}, \citet{delbaen_multiperiod_2007}, \citet{Roorda_Schum_Engw_2005_coherent}, \citet{Follmer_Penner_2006_convex_rm_dynamic}, \citet{Rusz_2010_risk_averse_DP}). Apart from yielding consistent preferences, this compositional form also allows a recursive estimation of the risk, and an application of the Bellman principle in optimization problems involving dynamic risk measures \citep{Nilim_ElGhaoui_2005, Iyengar_2005_robust_DP,Ruszczynski_Shapiro_06}.

From a pragmatic perspective, however, the compositional form entails a significantly more complicated risk assessment than the na\"{i}ve inconsistent approach. A risk manager would need to specify single-period conditional risk mappings for every future time-point; furthermore, even if these corresponded to the same risk measure $\mu$, the exact result of the composition would no longer be easily interpretable, and would bear no immediate relation to the original $\mu$. Our conversations with managers also revealed a certain feeling that such a measurement could result in ``overly conservative'' assessments, since risks are compounded in time -- for instance, by composing $\var$, one would obtain extreme quantiles of quantities that are already extreme quantiles. This has been recognized informally in the literature by \citet{Roorda_Schumach_2007_tail_var,Roorda_Schumacher_2008}, who proposed new notions of time consistency that avoided the issue, but without establishing formally if or to what degree the conservatism is actually true. Furthermore, it is not obvious how ``close'' a particular compositional measure is to a given inconsistent one, and how one could go about constructing the latter in a way that tightly approximates the former. This issue should be very relevant when considering dynamic decision problems under risk, but it seems to have been largely ignored by the literature (most papers examining operational problems under dynamic risk typically start with a set of given dynamic risk measures, e.g., \citet{Ahmed_Cakmak_Shapiro_2007}, \citet{Shapiro_2011},  \citet{Choi_Ruszcz_2011}).

With this motivation in mind, the goal of the present paper is to better understand the relation and exact tradeoffs between the two measurement processes outlined above, and to provide guidelines for constructing and/or estimating safe counterparts of one from the other. Our contributions are as follows.
\begin{itemize}
\item We provide several equivalent necessary and sufficient conditions guaranteeing when a time consistent risk measure $\cons$ always over (or under) estimates risk as compared with an inconsistent measure $\incons$. We argue that iterating the same $\incons$ does not necessarily over (or under) estimate risk as compared to a single static application of $\incons$, and this is true even in the case considered by \citet{Roorda_Schumach_2007_tail_var,Roorda_Schumacher_2008}. We show that composition with conditional expectation operators at \emph{any} stage of the measurement process results in valid, time consistent lower bounds. By contrast, upper bounds are obtained only when composing with worst-case operators in the \emph{last} stage of the measurement process.

\item We formalize the problem of characterizing and computing the smallest $\alphagen{\cons}{\incons}$ and $\alphagen{\incons}{\cons}$ such that $\cons \leq \incons \leq \alphagen{\cons}{\incons} \cdot \cons$ and $\incons \leq \cons \leq \alphagen{\incons}{\cons}$, respectively. The smallest such factors, $\alphaopt{\cons}{\incons}$ and $\alphaopt{\incons}{\cons}$, provide a compact notion of how closely a given $\incons$ can be multiplicatively approximated through lower (respectively, upper) bounding consistent measures $\cons$, respectively. Since, in practice, $\incons$ may be far easier to elicit from observed preferences or to estimate from empirical data, characterizing and computing $\alphaopt{\cons}{\incons}$ and $\alphaopt{\incons}{\cons}$ can be seen as the first step towards \emph{constructing} the time-consistent risk measure $\cons$ that is ``closest'' to a given $\incons$. 

\item Using results from the theory of submodularity and matroids, we particularize our results to the case when $\incons$ and $\cons$ are both comonotonic risk measures. We show that computing $\alphaopt{\cons}{\incons}$ and $\alphaopt{\incons}{\cons}$ is generally NP-hard, even when the risk measures in question are law-invariant. However, we provide several conditions under which the computation becomes simpler. Using these results, we compare the strength of approximating a given $\incons$ by time-consistent measures obtained through composition with conditional expectation or worst-case operators.

\item We characterize the tightest possible time-consistent and coherent upper bound for a given $\incons$, and show that it corresponds to a construction suggested in several papers in the literature \citep{Epstein_Schneid_03,Roorda_Schum_Engw_2005_coherent,delbaen_multiperiod_2007,Shapiro_2011}, which involves ``rectangularizing'' the set of probability measures corresponding to $\incons$. This yields not only the smallest possible $\alphaopt{\incons}{\cons}$, but also the uniformly tightest upper bound among all coherent upper bounds.

\item We summarize results from our companion paper \citep{Huang_I_Petrik_Subraman_cvar_2012}, which applies the ideas derived here to the specific case when both $\incons$ and $\cons$ are given by Average Value at Risk, a popular measure in financial mathematics. In this case, the results take a considerably simpler form: analytical expressions are available for two-period problems, and polynomial-time algorithms are available for some multi-period problems. We give an exact analytical characterization for the tightest uniform upper bound to $\incons$, and show that it corresponds to a compositional $\cvar$ risk measure that is increasingly conservative in time. For the case of lower bounds, we give an analytical characterization for two-period problems. Interestingly, we find that the best lower-bounds always provide tighter approximations than the best upper bounds in two-period models, but are also considerably harder to compute than the latter in multi-period models.
\end{itemize}

%We view the paper as a first step towards a systematic way of constructing risk-adjusted objective functions that are compatible with the modern theory of risk measures, but are also computationally tractable and ``easy'' to calibrate and explain to managers.

The rest of the paper is organized as follows. Section~\ref{sec:problem_statement} provides the necessary background in static and dynamic risk measures, and introduces the precise mathematical formulation for the questions addressed in the paper. Section~\ref{sec:find-optim-alpha} discusses the case of determining upper or lower bounding relations between two arbitrary consistent and inconsistent risk measures, and characterizes the resulting factors $\alphaopt{\cons}{\incons}$ and $\alphaopt{\incons}{\cons}$. Section~\ref{sec:discussion} discussed our results in detail, touching on the computational complexity, and introducing several examples of how the methodology can be used in practice. Section~\ref{sec:conclusions} concludes the paper and suggests directions for future research.

\subsection{Notation.}
\label{sec:notation}
With $i < j$, we use $[i,j]$ to denote the index set $\{i,\dots,j\}$. For a vector $\mb{x} \in \reals^{n}$ and $i \in \{1,\dots, n\}$, we use $x_i$ to denote the $i$-th component of $\mb{x}$. For a set $S \subseteq \{1,\dots,n\}$, we let $\mb{x}(S) \bydef \sum_{i \in S} x_i$. Also, we use $\mb{x}_S \in \reals^{n}$ to denote the vector with components $x_i$ for $i \in S$ and $0$ otherwise (e.g., $\oneN{S}$ is the characteristic vector of the set $S$), and $\mb{x}\vert_S \in \reals^{|S|}$ to denote the projection of the vector $\mb{x}$ on the coordinates $i \in S$. When no confusion can arise, we denote by $\oneN{}$ the vector with all components equal to 1. We use $\mb{x}\tr$ for the transpose of $\mb{x}$, and $\scprod{\mb{x}}{\mb{y}} \bydef \sum_{i=1}^n x_i\, y_i$ for the scalar product in $\reals^n$.

For a set or an array $S$, we denote by $\Pi(S)$ the set of all permutations on the elements of $S$. $\pi(S)$ or $\sigma(S)$ designate one particular such permutation, with $\pi(i)$ denoting the element of $S$ appearing in the $i$-th position under permutation $\pi$.

We use $\Delta^{n}$ to denote the probability simplex in $\reals^n$, i.e., $\Delta^n \bydef \{ \, \mb{p} \in \reals^{n}_+ \,:\, \scprod{1}{p}=1 \,\}$. For a set $P \subseteq \reals^n$, we use $\ext(P)$ to denote the set of its extreme points.

Throughout the exposition, we adopt the convention that $\frac{0}{0}=0$.

\section{Consistent and Inconsistent Risk Measures.}
\label{sec:problem_statement}
As discussed in the introduction, the goal of the present paper is to analyze two paradigms for assessing risk in a dynamic setting: a ``na\"{i}ve'' one, obtained by applying a static risk measure to the conditional cumulative future costs, and a ``sophisticated'', time-consistent method, obtained by composing one-period risk mappings.

In the present section, we briefly review the relevant background material in risk theory, describe the two approaches formally, and then introduce the main questions addressed in the paper. %To keep the paper self-contained, we also provide a short overview of static risk measures, and direct the interested reader to \citep{Follmer_Schied} for more details.

\subsection{Probabilistic Model.}
\label{sec:prob-model-backgr}
We begin by describing the probabilistic model. Our notation and framework are closely in line with that of \citep{Shapiro_Ruszczynski_Dentcheva_2009_Stochastic_Prog}, to which we direct the reader for more details.

For simplicity, we consider a scenario tree representation of the uncertainty space, where $t \in [0,T]$ denotes the time, $\nodes{t}$ is the set of nodes at stage $t\in [0,T]$, and $\child{i}$ is the set of children\footnote{In other words, $\{\child{i}, \, i\in \nodes{t}\}$ is a partition of the nodes in $\Omega_{t+1}, \, \forall \, t \in \{0,\dots,T-1\}$.} of node $i \in \nodes{t}$. We also use $\desc{i}$ to denote the set of all leaves descending from node $i$, i.e., with $\desc{i} = \{i\}, \, \forall \, i \in \nodes{T}$, we recursively define $\desc{i} \bydef \cup_{j \in \child{i}} \desc{j}, \, \forall \, i \in \cup_{t=0}^{T-1} \nodes{t}$. Similarly, we define $\desc{U} \bydef \cup_{i \in U} \desc{i}$ for any set $U \subseteq \nodes{t}$.

With the set $\nodes{T}$ of elementary outcomes, we associate the $\sigma$-algebra $\filt_T = 2^{\nodes{T}}$ of all its subsets, and we consider the filtration $\filt_0 \subseteq \filt_1 \subseteq \dots \subseteq \filt_T$, where $\filt_t$ is the sub-algebra of $\filt_{t+1}$ that is generated by the sets $\{\child{i}\}_{i \in \Omega_t}$, for any $t \in [0,T-1]$. %We also take $\filt_0 \equiv \{\emptyset, \nodes{T}\}$, the trivial $\sigma$-algebra.

We construct a probability space $(\nodes{T}, \filt_T, \refmeas)$ by introducing a \emph{reference measure} $\refmeas \in \Delta^{|\nodes{T}|}$, assumed to satisfy\footnote{This is without loss of generality - otherwise, all arguments can be repeated on a tree where leaves with zero probability are removed.} $\refmeas > 0$. On the space $(\nodes{T}, \filt_T, \refmeas)$, we use $\rvspace_T$ to denote the space of all functions $X_T \,:\, \nodes{T} \rightarrow \reals$ that are $\filt_T$-measurable. Since $\rvspace_T$ is isomorphic with $\spacenodes{T}$, we denote by $X_T$ the random variable, and by $\mb{X}_T$ the vector in $\spacenodes{T}$ of induced scenario-values, and we identify the expectation of $X_T$ with respect to a measure $\mb{q} \in \Delta^{|\nodes{T}|}$ as the scalar product $\scprod{q}{X}_T$. In a similar fashion, we introduce the sequence $\rvspace_t, \, t \in [0,T-1]$, where $\rvspace_t$ is the sub-space of $\rvspace_T$ containing functions which are $\filt_t$-measurable. Note that any function $X_t \in \rvspace_t$ is constant on every set $\child{i}, \, i \in \nodes{t}$, so that $X_t$ can also be identified with the vector $\mb{X}_t \in \spacenodes{t}$. %We also introduce $\rvspace_{[t,\tau]} \equiv \rvspace_t \times \dots \times \rvspace_{\tau}$, for any $\tau \geq t$, and, similarly, $X_{[t,\tau]} \equiv (X_t,\dots,X_\tau)$.
To this end, in order to simplify the notation, we identify any function $f : \rvspace_{t+1} \rightarrow \rvspace_t$ with a set of $|\nodes{t}|$ functions, and we write $f \equiv (f_i)_{i \in \nodes{t}}$, where $f_i : \reals^{|\nodes{t+1}|} \rightarrow \reals$. Furthermore, since all the functions of this form that we consider correspond to conditional evaluations on the nodes of the tree, we slightly abuse the notation and write $f \equiv (f_i)_{i \in \nodes{t}}$, where $f_i : \reals^{|\child{i}|} \rightarrow \reals$.

\subsection{Static Risk Measures.}
\label{sec:background_risk_theory}
Consider a discrete probability space $(\Omega,\mathcal{F},\refmeas)$, and let $\rvspace$ be a linear space of random variables on $\Omega$. %, typically restricted to be a subspace of $L^{p}(\Omega,\filt,\refmeas)$, for some $p>1$.
%In the context of the scenario tree outlined in Section~\ref{sec:prob-model-backgr}, examples of such a space could be $(\Omega_T, \filt_T, \mb{p})$, but also $(\child{i}, 2^{\child{i}}, \mb{p}_i)$, for some $i \in \nodes{t}, \, t \in [0,T-1]$. 
In this setup, we are interested in appropriate ways of assessing the riskiness of a random \emph{cost} (or loss) $X \in \rvspace$. The standard approach in the literature \citep{Artzner_Delbaen_1999, Follmer_Schied} is to use a functional $\mu : \rvspace \rightarrow \reals$ such that $\mu(X)$ represents the minimal reduction making a cost $X$ acceptable to the risk manager. The following axiomatic requirements are typically imposed.
\begin{itemize}
\item[{\bf [P1]}] \emph{Monotonicity}. For any $X, Y \in \rvspace$ such that $X \geq Y$, $\mu(X) \geq \mu(Y)$.
\item[{\bf [P2]}] \emph{Translation invariance}. For any $X \in \rvspace$ and any $m \in \reals$, $\mu(X+m) = \mu(X) + m$.
\item[{\bf [P3]}] \emph{Convexity}. For any $X, Y \in \rvspace$, and any $\lambda \in [0,1]$, $\mu \bigl(\lambda \, X + (1-\lambda) \, Y \bigr) \leq \lambda \, \mu(X) + (1-\lambda) \, \mu(Y)$.
\item[{\bf [P4]}] \emph{Positive homogeneity}. For any $X \in \rvspace$, and any $\lambda \geq 0$, $\mu(\lambda X) = \lambda \, \mu(X)$.
\item[{\bf [P5]}] \emph{Comonotonicity}. $\mu(X+Y) = \mu(X) + \mu(Y)$ for any $X, Y \in \rvspace$ that are comonotone, i.e., $\bigl[ X(\omega) - X(\omega') \bigr]\, \bigl[ Y(\omega) - Y(\omega') \bigr] \geq 0$, for any $\omega, \omega' \in \Omega$.
\item[{\bf [P6]}] \emph{Law-invariance}. $\mu(X) = \mu(Y)$ for any $X, Y \in \rvspace$ such that $F_X(\cdot) = F_Y(\cdot)$.
\end{itemize}

\emph{Monotonicity} requires that a larger cost should always be deemed riskier. \emph{Translation} (or \emph{cash}) \emph{invariance} gives $\mu$ an interpretation as capital requirement: typically, a cost $X$ is deemed acceptable if $\mu(X) \leq 0$, so cash invariance implies that $\mu\bigl( X - \mu(X) \bigr) = 0$, i.e., $\mu(X)$ is the smallest amount of cost reduction making $X$ acceptable. \emph{Convexity} suggests that diversification of costs should never increase the risk (or, conversely, that a convex combination of two acceptable costs $X$ and $Y$ should also be acceptable), while \emph{positive homogeneity} implies that risk should scale linearly with the size of the cost. \emph{Comonotonicity} implies that the risk in costs that move together (i.e., are comonotone) cannot be diversified by mixtures, while \emph{law-invariance} requires the risk measures to only depend on the probability distribution of the random costs. For an in-depth discussion and critique of these axioms, we direct the reader to \citep{Artzner_Delbaen_1999,Follmer_Schied} and references therein.

Following the common terminology in the literature, we call any functional satisfying {\bf [P1-2]} a \emph{risk measure}. Any \emph{risk measure} satisfying {\bf [P3]} is said to be \emph{convex}, and any \emph{convex risk measure} that satisfies {\bf [P4]} is said to be \emph{coherent}. The main focus of the present paper are functionals that satisfy\footnote{It is known that comonotonicity actually implies positive homogeneity \citep{Follmer_Schied}, so the we can define comonotonic risk measures as those satisfying {\bf [P1-3]} and {\bf [P5]} (the reverse is not true, i.e., not all coherent risk measures are comonotonic \citep{acerbi_book_chapter}).} {\bf [P1-5]}, which are called \emph{comonotonic risk measures}. Some of our results take a simpler form when further restricting attention to the class of \emph{distortion risk measures}, which are all comonotonic risk measures additionally satisfying {\bf [P6]}. Such measures have been examined in economics, actuarial science, and financial mathematics, and form a well-established class of risk metrics (see, e.g., \citep{Schmeidler_1986_choquet,Wang_2000,Tsanakas2004223,Cotter20063469,Kusuoka_2001,acerbi_spectral_rm_2002,acerbi_book_chapter,Follmer_Schied} for more references and details).

One of the main results in the literature is a universal representation theorem for any coherent risk measure, which takes a specialized form in the comonotonic case \citep{Schmeidler_1986_choquet, Follmer_Schied}.
\begin{theorem}%[Representation Theorem for Convex, Comonotonic Risk Measures]
  \label{thm:representation_comonotonic_rm}
  A risk measure $\mu$ is coherent if and only if it can be represented as 
  \begin{align}
    \label{eq:submodular_representation_comonotonic}
    \mu(X) &= \max_{\mathbb{Q} \in \mathcal{Q}} \expectmeas{\mathbb{Q}}{X},
  \end{align}
  for some $\mathcal{Q} \subseteq \Delta^{|\Omega|}$. Furthermore, if $\mu$ is comonotonic, then $\mathcal{Q} = \bigl\{ \, \mathbb{Q} \in \Delta^{|\Omega|} \suchthat \mathbb{Q}(S) \leq c(S), \, \forall S \in \mathcal{F} \,\bigr\}$, where $c$ is a Choquet capacity.
\end{theorem}

The result essentially states that any coherent risk measure is an expectation with respect to a worst-case probability measure, chosen adversarially from a suitable set of test measures (or generalized scenarios) $\qcal$. For comonotonic risk measures, this set is uniquely determined by a particular function $c$, known as a \emph{Choquet capacity}.
\begin{definition}
  \label{def:choquet_capacities}
  A set function $c : 2^\Omega \rightarrow [0,1]$ is said to be a \emph{Choquet capacity} if it satisfies the following properties:
  \begin{itemize}
  \item nondecreasing: $c(A) \leq c(B), \, \forall \, A \subseteq B \subseteq \Omega$
  \item normalized: $c(\emptyset) = 0$ and $c(\Omega) = 1$
  \item submodular: $c(A \cap B) + c(A \cup B) \leq c(A) + c(B), \, \forall \, A, \, B \subseteq \Omega$.
  \end{itemize}
\end{definition}

When a comonotonic risk measure is additionally law-invariant (i.e., it is a \emph{distortion} measure), the Choquet capacities are uniquely determining by a concave distortion function, i.e., 
\begin{align}
  \label{eq:distortion_measure_repres}
  c(S) = \Psi ( \refmeas(S) ) , \, \forall \, S \in \mathcal{F},
\end{align}
where $\Psi : [0,1] \rightarrow [0,1]$ is a concave, nondecreasing function satisfying $\Psi(0) = 0$ and $\Psi(1) = 1$. 

A popular example of comonotonic (in fact, distortion) risk measure, studied extensively in the literature, is Average Value-at-Risk at level $\lev \in [0,1]$ ($\cvar_\lev$), also known as \emph{Conditional Value-at-Risk}, \emph{Tail Value-at-Risk} or \emph{Expected Shortfall}. It is defined as
\begin{subequations}
  \begin{align}
    % \var_\lev(X) &\bydef \inf\{ \,m \in \reals \suchthat \refmeas[X - m > 0] \leq \lev \,\} \label{eq:var_definition} \\
    \cvar_\lev(X) &\bydef \frac{1}{\lev} \int_{1-\lev}^1 \var_{1-t}(X) \, dt.
    \label{eq:cvar_definition}
  \end{align}
\end{subequations}
where $\var_\lev(X) \bydef \inf\{ \,m \in \reals \suchthat \refmeas[X - m > 0] \leq \lev \,\}$ is the Value at Risk at level $\lev$. As the name suggests, $\cvar_\lev$ represents an average of all $\var$ measures with level at most $\lev$. When the underlying reference measure $\refmeas$ is non-atomic, it can be shown \citep{Follmer_Schied} that $\cvar_\lev(X) = \expectmeas{\refmeas}{ X \,|\, X \geq \var_\lev(X) }$, which motivates the second and third names that the latter measure bears. While $\cvar$ is a distortion measure, $\var$ is not even convex, since it fails requirement {\bf [P3]}. 

\subsection{Dynamic Risk Measures.}
\label{sec:dynam-risk-meas}
As stated in the introduction, the main focus of the present paper are dynamic risk measures, i.e., risk measures defined for cash streams that are received or dispensed across several time-periods. More precisely, a dynamic risk measure entails the specification of an entire sequence of risk measures $\{\mu_{[t,T]}\}_{t=0}^{T-1}$, such that $\mu_{[t,T]}$ maps a future stream of random costs $X_{[t,T]} \bydef (X_t,\dots,X_{T})$ into risk assessments at $t$. 

Following a large body of literature \citep{Riedel2004185,delbaen_multiperiod_2007,Detlefs_Scandolo_05,Roorda_Schum_Engw_2005_coherent,Cheridito_Delbaen_Kupper_2006,Follmer_Penner_2006_convex_rm_dynamic,Ruszczynski_Shapiro_06,Rusz_2010_risk_averse_DP,Cheridito_Kupper_2011}, we furthermore restrict the risk measurements at time $t$ to only depend on the cumulative costs in the future, i.e., we take $\mu_{[t,T]} : \rvspace_{T} \rightarrow \rvspace_t$, and the risk of $X_{[t,T]}$ is $\mu_{[t,T]}(X_t + \dots + X_T)$. While such measures have been criticized for ignoring the timing when future cashflows are received, they are consistent with the assumptions in many academic papers focusing on portfolio management under risk \citep{Chabakauri_Basak_2010,CuocoHeIssaenko_2008}, as well as with current risk management practice \citep{Jorion_VaR_2006}, and provide a natural, simpler first step in our analysis.

In this framework, we introduce the first way of measuring dynamic risk, whereby $\mu_{[t,T]}$ is obtained by applying a static risk measure, conditioned on information available at time $t$. In the context of the probabilistic space of Section~\ref{sec:prob-model-backgr}, this can be formalized as follows.
\begin{definition}
  \label{def:dynamically_inconsistent_risk_measure}
  A \emph{time inconsistent (dynamic) risk measure} is any set of mappings $\{\mu_{[t,T]}\}_{t=1}^T$ of the form $\mu_{[t,T]} = (\mu^i)_{i \in \nodes{t}}, \forall \, t \in [0,T]$, where $ \mu^i : \rvspace_{T} \rightarrow \reals$ is a risk measure, for any node $i \in \nodes{t}$.
\end{definition}

In other words, conditional on reaching node $i \in \nodes{t}$ at time $t$, the risk of a future cashflow $X_{[t,T]}$ is given by $\mu^i(\sum_{\tau=t}^T X_\tau)$, where every $\mu^i$ is a static risk measure, which can be furthermore required to satisfy additional axiomatic properties, as per Section~\ref{sec:background_risk_theory}. 
% (for the remainder of the paper, we mainly focus on $\mu^i$ that are comonotonic).

The choice above is eminently sensible - the specification of risk can be done in a unified fashion, by means of a single risk measure at every node and time. This makes for a compact representation of risk preferences, which can be more easily calibrated from empirical data, more readily comprehended and adopted by practitioners, and more uniformly applied across a variety of businesses and products. For instance, it is by far the most common paradigm in financial risk management, where a 10-day $\var$ is typically calculated at level $\lev = 1\%$, assuming the trading portfolio remains fixed during the assessment period \citep{Jorion_VaR_2006}.

\begin{figure}[h]
  \centering
  \includegraphics[width=0.3\textwidth]{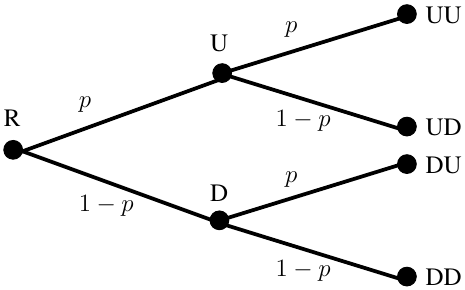}
  \caption{Example showing time inconsistency of a static risk measure. The random cost $X$ with $X(UU) = X(DD) = 0$, $X(UD) = X(DU) = 100$ is deemed strictly riskier in all states of nature at time $t=1$ than a deterministic cost $Y = 50$, but nonetheless $Y$ is deemed riskier than $X$ at time $t=0$.}
  \label{fig:inconsist_measure}
\end{figure}

However, as the name suggests, such risk measures readily result in time inconsistent behavior. To see this, consider the following example, adapted from \citet{Roorda_Schum_Engw_2005_coherent}.
\begin{example}
  \label{example:naive_inconsistencies}
  Consider the tree in Figure~\ref{fig:inconsist_measure}, with the elementary events $\Omega = \{UU, UD, DU, DD\}$. Consider the risk measure given by
  \[ 
  \mu^i( X ) = \max_{\Pr \in \mathcal{P}} \E_{\Pr} [ X | i ], ~ \forall \, i \in \{R, U, D\},
  \] 
  where $\mathcal{P}$ contains two probability measures, one corresponding to $p = 0.4$, and the other to $p = 0.6$. Clearly, all $\{\mu^i\}_{i \in \{R, U, D\}}$ correspond to coherent risk measures. For the random cost $X$ such that $X(UU) = X(DD) = 0$, and $X(UD) = X(DU) = 100$, we have $\mu^U( X ) = \mu^D( X ) = 60$, and $\mu^R( X ) = 48$. Therefore, $X$ is deemed strictly riskier than a deterministic cost $Y = 50$ in all states of nature at time $t=1$, but nonetheless $Y$ is deemed riskier than $X$ at time $t=0$.
\end{example}

We note that there is nothing peculiar in the choices above, in that similar counterexamples can be constructed for any risk measures $\mu^i$, even when the latter are comonotonic. Rather, the issue at play is the key feature distinguishing dynamic from static risk assessment, namely the consistency in the risk preference profile over time. This is summarized in the axiom\footnote{We note that there are several notions of time consistency in the literature (see \citet{Penner_thesis,acciaio_penner_2011_dc,Roorda_Schumach_2007_tail_var} for an in-depth discussion and comparison). The one we adopt here is closest in spirit to strong dynamic consistency, and seems to be the most widely accepted notion in the literature.} of \emph{time} (or \emph{dynamic}) \emph{consistency}, which asks that 
%\begin{definition}
% \label{def:dynamically_consistent_measures}
  a dynamic risk measure $\bigl\{ \mu_{t,T} \bigr\}_{t=0}^{T-1}$ should satisfy, for all $t \in [0,T-1]$ and all $X, Y \in \rvspace_{T}$,
  \begin{align*}
    \mu_{t+1,T}(X) \geq \mu_{t+1,T}(Y) ~\textup{implies}~ \mu_{t,T}(X) \geq \mu_{t,T}(Y).
  \end{align*}
%\end{definition}
%In other words, if a cost $X$ is considered riskier than $Y$ in all states of the world at time $t+1$, it should be considered riskier at time $t$. 

This is a requirement on the particular functional forms that can be considered for $\mu_{[t,T]}$, which is typically violated by the na\"{i}ve dynamic measures of Definition~\ref{def:dynamically_inconsistent_risk_measure}. A central result in the literature \citep{Riedel2004185,delbaen_multiperiod_2007,Detlefs_Scandolo_05,Roorda_Schum_Engw_2005_coherent,Cheridito_Delbaen_Kupper_2006,Roorda_Schumach_2007_tail_var,Penner_thesis, Follmer_Penner_2006_convex_rm_dynamic,Rusz_2010_risk_averse_DP} is the following theorem, stating that any consistent measure has a compositional representation in terms of one-period risk mappings.
\begin{theorem}
  \label{thm:compositional_form_representation_dynamic_rm}
  Any dynamic risk measure $\bigl\{ \mu_{t,T} \bigr\}_{t=0}^{T-1}$ that is time consistent can be written as
\begin{align}
  \mu_{t,T}(X_t + \dots + X_T) = \mu_{t+1} \Bigl( \mu_{t+2} \bigl( \dots (\mu_T(X_t + \dots + X_T)) \dots \bigr) \Bigr).
  \label{eq:compositional_form_consistent_measure}
\end{align}
where $\mu_{t} : \mathcal{X}_{t} \rightarrow \mathcal{X}_{t-1}, \, t \in [1,T]$  are a set of single-period conditional risk mappings.
\end{theorem}

This leads us to define the second way of measuring dynamic risk on the scenario tree of Section~\ref{sec:prob-model-backgr}, by means of composing risk measures.
% every such conditional risk mapping is exactly given by
%  $ \mu_{t} = (\mu^i)_{i \in \nodes{t}} , \forall \, t \in [1,T]$,
% where
% $  \mu^i : \spacechild{i} \rightarrow \reals, \, \forall \, i \in \nodes{t}$
% represents a (conditional) risk measure associated with node $i \in \nodes{t}$ (see \citet{Shapiro_Ruszczynski_Dentcheva_2009_Stochastic_Prog}). We note that, here, we could also require 
% Since every $\mu^i$ is now a static (i.e., single period measure), it can be required to satisfy additional axiomatic properties, such as the ones introduced in Section~\ref{sec:background_risk_theory}. %, to define the class of \emph{time consistent distortion risk measures}.
\begin{definition}
  \label{def:dynamic_consist_distortion_risk_measures}
  A set of mappings $\{\mu_{[t,T]}\}_{t=0}^{T-1}$ is said to be a \emph{time consistent (dynamic) risk measure} if $\mu_{[t,T]} = \mu_{t+1} \circ \mu_{t+2} \circ \dots \circ \mu_T$ for any $t \in [0,T-1]$, where $\mu_{t+1} \equiv (\mu^i)_{i \in \nodes{t}}$, and $\mu^i : \spacechild{i} \rightarrow \reals$ are risk measures, for any $i \in \nodes{t}$.
\end{definition}

We say that $\{\mu_{[t,T]}\}_{t=0}^{T-1}$ is a time-consistent, coherent (comonotonic) risk measure if every $\mu^i$ is coherent (respectively, comonotonic), for any $i \in \nodes{t}$ and $t \in [0,T-1]$.

Apart from being axiomatically justified, this compositional form has the advantage of allowing a recursive estimation of the risk, and an application of the Bellman optimality principle in optimization problems involving dynamic risk measures \citep{Nilim_ElGhaoui_2005, Iyengar_2005_robust_DP,Ruszczynski_Shapiro_06,Rusz_2010_risk_averse_DP}. This has lead to its adoption in actuarial science \citep{Hardy_Wirch_2004,Brazauskas_2008}, as well as in several recent papers that re-examine operational problems under coherent measures of risk \citep{Ahmed_Cakmak_Shapiro_2007,Choi_Ruszcz_2011}.

The main downside of the compositional form is that it requires a specification of all the mappings $\mu^i$, which furthermore no longer lends itself to an easy interpretation, particularly as seen from the perspective of time $t=0$. In particular, even if $\mu^i$ corresponded to the same primitive risk measure $\mu$, the overall compositional measure\footnote{Here and throughout the paper, we use the shorthand notation $\mu \circ \mu$ with the understanding that the elementary risk measure $\mu$ is applied in stages $t \geq 1$ in a conditional fashion.} $\mu_{[0,T]} = \mu \circ \mu \circ \dots \circ \mu$ would bear no immediate relation to $\mu$. As an example, when $\mu^i = \cvar_\lev$, the overall $\mu_{[0,T]}$ corresponds to the so-called ``iterated CTE'' \citep{Hardy_Wirch_2004,Brazauskas_2008,Roorda_Schumach_2007_tail_var}, which does not lend itself to the same simple interpretation as a single $\cvar$. Furthermore, practitioners often feel that the overall risk measure $\mu_{[0,T]}$ is \emph{overly conservative}, since it composes what are already potentially conservative risk evaluations backwards in time -- for instance, for the iterated TCE, one is taking tail conditional expectations of tail conditional expectations. This has been recognized informally in the literature by \citet{Roorda_Schumach_2007_tail_var,Roorda_Schumacher_2008}, who propose new notions of time consistency that avoid the issue, but without establishing precisely whether or to what extent the conservatism is actually true. From a different perspective, it is not obvious how ``close'' a particular compositional measure is to a given inconsistent one, and how one could go about constructing the latter in a way that tightly approximates the former. 

\subsection{Main Problem Statement.}
\label{sec:cons-meas-scen}
In this context, the goal of the present paper is to take the first step towards better understanding the tradeoffs between the two ways of measuring risk. More precisely, we consider dynamic risk as viewed from the perspective of time $t=0$, and examine two potential metrics: a time-inconsistent (comonotonic) risk measure $\incons : \rvspace_T \rightarrow \reals$, and a time-consistent (comonotonic) risk measure $\cons : \rvspace_T \rightarrow \reals$. For two such metrics, we seek to address the following related problems.

% \begin{problem}
%   \label{prob:problem_statement_test_inequalities}
%   Given $\incons$ and $\cons$, test whether $\cons(Y) \leq \incons(Y), \, \forall \, Y \in \rvspace_T$ or whether $\incons(Y) \leq \cons(Y), \, \forall \, Y \in \rvspace_T$.
% \end{problem}

% \begin{problem}
%   \label{prob:problem_statement_alpha_beta}
%   For $\incons$, $\cons$ such that $\cons(Y) \leq \incons(Y), \, \forall \, Y \in \rvspace_T$, find the smallest $\alphagen{\cons}{\incons} > 0$ such that
%   \begin{align}
%     \cons(Y) &\leq \incons(Y) \leq \alphagen{\cons}{\incons} \cdot \cons(Y), \, \forall \, Y \in \rvspace_T.
%     \label{eq:tight_approximation_consistent_inconsistent}
%     % \cons(Y) &\leq \beta \cdot \incons(Y), \, \forall \, Y \in \rvspace_T, \, Y \geq 0.
%   \end{align}
%   Similarly, if $\incons(Y) \leq \cons(Y), \, \forall \, Y \in \rvspace_T$, find the smallest $\alphagen{\incons}{\cons}$ such that $\incons(Y) \leq \cons(Y) \leq \alphagen{\incons}{\cons} \cdot \incons(Y), \, \forall \, Y \in \rvspace_T$.
% \end{problem}

\begin{problem}
  \label{prob:problem_statement_test_inequalities}
  Given $\incons$ and $\cons$, test whether 
  \[ \cons(Y) \leq \incons(Y), \, \forall \, Y \in \rvspace_T \quad \textup{or} \quad \incons(Y) \leq \cons(Y), \, \forall \, Y \in \rvspace_T. \]
\end{problem}

\begin{problem}
  \label{prob:problem_statement_alpha_beta}
  Given $\incons$, $\cons$, find the smallest $\alphagen{\cons}{\incons} > 0$ and $\alphagen{\incons}{\cons} > 0$ such that
  \begin{subequations}
    \begin{align}
      \textup{if}~ \cons(Y) \leq \incons(Y), \, \forall \, Y \qquad & \incons(Y) \leq \alphagen{\cons}{\incons} \cdot \cons(Y), \, \forall \, Y \in \rvspace_T, \, Y \geq 0 
      \label{eq:tight_approximation_consistent_inconsistent}\\
      \textup{if}~ \incons(Y) \leq \cons(Y), \, \forall \, Y \qquad &  \cons(Y) \leq \alphagen{\incons}{\cons} \cdot \incons(Y), \, \forall \, Y \in \rvspace_T, \, Y \geq 0. 
      \label{eq:tight_approximation_inconsistent_consistent}
    \end{align}
  \end{subequations}
\end{problem}

A satisfactory answer to Problem~\ref{prob:problem_statement_test_inequalities} would provide a test for whether one of the formulations is always over or under estimating risk as compared to the other. As we show, consistent measures obtained by iterating the same primitive measure $\mu$ do not necessarily over (or under) estimate risk as compared to $\mu$, and this is true even when $\mu = \cvar_\lev$, the case considered in \citep{Roorda_Schumach_2007_tail_var,Roorda_Schumacher_2008}. However, by composing $\mu$ with conditional expectation operators, one always obtains lower bounds to the static risk measurement under $\mu$. For instance, $\mu \circ \E$ and $\E \circ \mu$ are both lower bounds to a given static evaluation by $\mu$. By contrast, upper bounds are obtained only when composing with worst-case operators in the \emph{final} periods of the horizon: e.g., $\mu \circ \max$ is necessarily an upper bound for $\mu$, but $\max \circ \mu$ is not.

To understand the relevance of Problem~\ref{prob:problem_statement_alpha_beta}, note that the minimal factors $\alphaopt{\cons}{\incons}$ and $\alphaopt{\incons}{\cons}$ satisfying \eqref{eq:tight_approximation_consistent_inconsistent} and~\eqref{eq:tight_approximation_inconsistent_consistent}, respectively, provide a compact notion of how closely $\incons$ can be approximated through lower or upper bounding consistent measures $\cons$, respectively. Since, in practice, it may be far easier to elicit or estimate a single static risk measure $\incons$, characterizing and computing $\alphaopt{\cons}{\incons}$ and $\alphaopt{\incons}{\cons}$ constitutes the first step towards constructing the time-consistent risk measure $\cons$ that is ``closest'' to a given $\incons$. We note that a similar concept of inner and outer approximations by means of distortion risk measures appears in \citep{Bertsimas_Brown_2009}. However, the goal and analysis there are quite different, since the question is to approximate a static risk measure by means of another static distortion risk measure.

In a different sense, the smallest $\alphaopt{\incons}{\cons}$ could be used to scale up risk measurements according to $\incons$ in order to turn them into ``safe'' (i.e., conservative) estimates of measurements according to $\cons$. Scaling risk assessments by particular factors is actually common practice in financial risk management: according to the rules set forth by the Basel Committee for banking regulation, banks are required to report the 10-day $\var$ calculated at $1\%$ level, which is then \emph{multiplied by a factor of 3} to provide the minimum capital requirement for regulatory purposes; the factor of 3 is meant to account for losses occurring beyond $\var$, and also for potential model misspecification (see Chapter~5 of~\citep{Jorion_VaR_2006}). Therefore, this usage of $\alphaopt{\incons}{\cons}$ could integrate well with practice. 

We conclude the section by a remark pertinent to the two problems and our analysis henceforth.
\begin{remark}
  \label{rem:nonnegative_Y}
  On first sight, the requirement of non-negative $Y$ in the text of Problem~\ref{prob:problem_statement_alpha_beta} might seem overly restrictive. However, note that, if we insisted on $\cons(Y) \leq \incons(Y)$ holding for \emph{any} cost $Y$, and if $\cons, \, \incons$ were allowed to take both positive and negative values, then the questions in Problem~\ref{prob:problem_statement_alpha_beta} would be meaningless, in that no feasible $\alphagen{\cdot}{\cdot}$ values would exist satisfying~\eqref{eq:tight_approximation_consistent_inconsistent} or~\eqref{eq:tight_approximation_inconsistent_consistent}. To this end, we are occasionally forced to make the assumption that the stochastic losses $Y$ are non-negative.
  % introduce the following standing assumption throughout the remainder of the analysis.
  % \begin{assumption}[Non-negative Losses]
  %   \label{assum:nonnegative_losses}
  %   The stochastic losses $Y$ are non-negative.
  % \end{assumption}
  This is not too restrictive whenever a lower bound $Y_L$ is available for $Y$. By using the cash-invariance property ({\bf [P2]}) of the risk measures involved, one could reformulate the original question with regards to the random loss $Y-Y_L$, which would be nonnegative. Furthermore, in specific applications (e.g., multi-period inventory management \citep{Ahmed_Cakmak_Shapiro_2007}), $Y$ is the sum of intra-period costs $X_t$ that are always non-negative, so requiring $Y$ to be nonnegative is quite sensible.
\end{remark}

\section{Bounds for Coherent and Comonotonic Risk Measures.}
\label{sec:find-optim-alpha}
In this section, we seek appropriate answers to Problem~\ref{prob:problem_statement_test_inequalities} and Problem~\ref{prob:problem_statement_alpha_beta}, with the end-goal of characterizing the % smallest factors $\alpha$ and $\beta$ satisfying
% \begin{align*}
%   \cons(Y) &\leq \incons(Y) \leq \alpha \cdot \cons(Y), \, \forall \, Y \in \rvspace_{T}, \, Y \geq 0 \quad \textup{and} \\
%   \incons(Y) &\leq \cons(Y) \leq \beta \cdot \incons(Y), \, \forall \, Y \in \rvspace_{T},
% \end{align*}
% for any $Y \in \rvspace_{T}, \, Y \geq 0$.
tightest multiplicative approximation of a given inconsistent risk measure by means of lower (or upper) bounding consistent risk measures.

To keep the discussion compact and avoid repetitive arguments, we first treat an abstract setting of comparing two coherent measures on the same space of outcomes. The conditions that we derive will be quite general, since no further structure will be imposed on the two measures. Section~\ref{sec:charact_qc_qi_subqc_subqi} will then discuss in detail the comparison between a \emph{time-inconsistent, comonotonic} risk measure $\incons$ and a \emph{consistent, comonotonic} risk measure $\cons$, deriving particular forms for the results and conditions. In Section~\ref{sec:basic bounds}, we derive the main technical result needed to obtain multiplicative bounds on inconsistent, comonotonic risk measures, which we then use in Section~\ref{sec:charact_qc_qi_subqc_subqi} to address the main problems of interest.

Consider a discrete probability space $(\Omega,\mathcal{F},\refmeas)$, and let $\rvspace$ be the space of all random variables on $\Omega$ (isomorphic with $\reals^{|\Omega|}$). On this space, we are interested in comparing two \emph{coherent} risk measures $\mu_{1,2} : \rvspace \rightarrow \reals$ given by \emph{polyhedral} sets of measures, i.e.,
\begin{align*}
  \mu_i(Y) = \max_{\mb{q} \in \qcal_i} \, \scprod{q}{Y}, \, \forall \, Y \in \rvspace, ~ \forall \, i \in \{1,2\},
\end{align*}
where $\qcal_{1,2} \subseteq \Delta^{|\Omega|}$ are (bounded) polyhedra\footnote{Several of the results discussed here readily extend to arbitrary closed, convex sets of representing measures. We restrict attention to the polyhedral case since it captures the entire class of comonotonic risk measures, it is simpler to describe, and computationally advantageous, since evaluating the risk measure entails solving a linear program.}. Our main focus is on (1) characterizing conditions such that $\mu_1(Y) \leq \mu_2(Y), \, \forall \, Y \in \rvspace$, and (2) finding the smallest factor $\alpha$ such that
\begin{align*}
  \mu_1(Y) \leq \mu_2(Y) \leq \alpha \, \mu_1(Y), \, \forall \, Y \in \rvspace ~ (Y \geq 0).
\end{align*}

In this context, the risk measure $\mu_i(Y)$ can be identified as the support function of the convex set $\qcal_i$, so that the following standard result in convex analysis (see, e.g., \citep[Corollary 13.1.1]{Rock70}) can be invoked to test whether one risk measure dominates the other.
\begin{proposition}
  \label{prop:inequality_risk_measures}
  The inequality $\mu_1(Y) \leq \mu_2(Y), \, \forall \, Y \in \rvspace$ holds if and only if $\qcal_1 \subseteq \qcal_2$.
\end{proposition}
The usefulness of the latter condition critically depends on the representation of the sets of measures $\qcal_i$. For instance, if $\qcal_i$ are polytopes, the containment problem $\qcal_1 \subseteq \qcal_2$ is co-NP-complete when $\qcal_1$ is given by linear inequalities and $\qcal_2$ is given by its extreme points, but it can be solved in polynomial time, by linear programming (LP), for all the other three possible cases \citep{Freund_Orlin_85}. 

Proposition~\ref{prop:inequality_risk_measures} also sheds light on the second question of interest, through the following corollary.
\begin{corollary}
  \label{corol:impossible_alpha_forall_Y}
  There does not exist any $\alpha \neq 1$ such that $\mu_2(Y) \leq \alpha \, \mu_1(Y), \, \forall \, Y \in \rvspace$.
\end{corollary}
\begin{proof}{Proof.}
  For any $\alpha > 0$, the condition $\{\mu_2(Y) \leq \alpha \, \mu_1(Y), \, \forall \, Y \in \rvspace\}$ is equivalent to $\qcal_2 \subseteq \alpha \qcal_1$. Since $\qcal_{1,2} \subseteq \Delta^{|\Omega|}$, the containment cannot hold for any $\alpha \neq 1$.
  \qed
\end{proof}

This result prompts the need to restrict the space of random losses considered. As suggested in Section~\ref{sec:cons-meas-scen}, an eminently sensible choice is to take $Y \geq 0$, which is always reasonable when a lower bound on the losses is available. This allows us to characterize the desired conditions by examining inclusions of \emph{down-monotone closures} of the sets $\qcal_i$. To this end, we introduce the following two definitions (see Section~\ref{sec:down_monot_closures} of the Appendix for more details and references).

\begin{definition}
  \label{def:down_monotone_polytope}
  A non-empty set $Q \subseteq \reals^n_{+}$ is said to be \emph{down-monotone} if for any $\mb{x} \in Q$ and any $\mb{y}$ such that $0 \leq \mb{y} \leq \mb{x}$, we also have $ \mb{y} \in Q$.
  %\begin{align*}
%  $  0 \leq \mb{y} \leq \mb{x}  ~\Rightarrow~ \mb{y} \in Q$.
 % \end{align*}
\end{definition}

\begin{definition}
  \label{def:down_monotone_closure}
  The \emph{down-monotone closure} of a set $Q \subseteq \reals^n_{+}$, denoted by $\sub{Q}$, is the smallest down-monotone set containing $Q$, i.e.,
  \begin{align*}
    \sub{Q} \bydef \bigl\{\, \mb{x} \in \reals^n_{+} \suchthat \exists \, \mb{q} \in Q, \,\mb{x} \leq \mb{q} \,\bigr\}.
  \end{align*}
\end{definition}

When restricting attention to nonnegative losses, one can readily show that a coherent risk measure can be obtained by evaluating the worst case over an \emph{extended} set of generalized scenarios, given by the down-monotone closure of the original set. This is summarized in the following extension of representation Theorem~\ref{thm:representation_comonotonic_rm}. %can be rewritten as follows for the case of nonnegative losses. %in terms of down-monotone sets of measures. In particular, replacing $\qcal$ by its down-monotone closure $\sub{\qcal}$ does not affect the risk evaluation, which is summarized in following proposition.
\begin{proposition}
  \label{prop:can_close}
  Let $\mu(Y) = \max_{\mb{q} \in \qcal} \, \scprod{q}{Y}$ be a coherent risk measure. Then,
  \begin{equation}
    \label{eq:inconsistent_measure_same_on_down_closure}
    \mu(Y) = \max_{\mb{q} \in \sub{\qcal}} \, \scprod{q}{Y}, \, \forall \, Y \geq 0.
  \end{equation}
\end{proposition}
\begin{proof}{Proof.}
The inequality $ \max_{\mb{q} \in \qcal} \, \scprod{q}{Y} \leq \max_{\mb{q} \in \sub{\qcal}} \, \scprod{q}{Y} $ follows simply because $ \qcal \subseteq \sub{\qcal} $. To prove the reverse, consider any $Y \geq 0$ and let $\mb{q}_1$ be the maximizer of $\max_{\mb{q} \in \sub{\qcal}} \, \scprod{q}{Y} $. By Definition~\ref{def:down_monotone_closure}, there exists $\mb{q}_2 \in \qcal $ such that $\mb{q}_2 \geq \mb{q}_1 \geq 0$. Then:
\begin{align*}
  \max_{\mb{q} \in \qcal} \, \scprod{q}{Y} \geq \mb{q}_2\tr \mb{Y} \geq \mb{q}_1 \tr \mb{Y} = \max_{\mb{q} \in \sub{\qcal}} \, \scprod{q}{Y}~.  \qed
\end{align*}
\end{proof}

In view of this result, one can readily show that testing whether a risk measurement dominates another can be done \emph{equivalently} in terms of the down-monotone closures of the representing sets of measures, as stated in the next result.
\begin{lemma}
  \label{lem:downmonotone_equivalent}
  The inequality $\mu_1(Y) \leq \mu_2(Y), \, \forall \, Y \in \rvspace$ holds if and only if $\sub{\qcal_1} \subseteq \sub{\qcal_2}$.
%$\mu_{1,2}$, with closed, convex sets of measures $\qcal_{1,2} \subseteq \Delta^{|\Omega|}$. Then,
%  \[ \qcal_1 \subseteq \qcal_2 \Leftrightarrow \sub{\qcal_1} \subseteq \sub{\qcal_2}~.\]
\end{lemma}
\begin{proof}{Proof.}
  By Proposition~\ref{prop:inequality_risk_measures}, the above is equivalent to showing \[ \qcal_1 \subseteq \qcal_2 \Leftrightarrow \sub{\qcal_1} \subseteq \sub{\qcal_2}~.\]
  \begin{itemize}
  \item[$(\Rightarrow)$] Consider $\mb{q}_1 \in \sub{\qcal_1}$. Then, by Definition~\ref{def:down_monotone_closure}, there exists $\mb{q}'_1 \in \qcal_1$ such that $\mb{q}'_1 \ge \mb{q}_1$. Since $\qcal_1 \subseteq \qcal_2$, we have $\mb{q}_1'\in\qcal_2$, and therefore $\mb{q}_1 \in \sub{\qcal_2}$.
  \item[$(\Leftarrow)$] Note that $\qcal_i = \sub{\qcal_i} \cap \Delta^{|\Omega|}$ for $i=1,2$. Then:
    \begin{align*}
      \sub{\qcal_1} &\subseteq \sub{\qcal_2} ~ \Rightarrow \\
      \sub{\qcal_1} \cap \Delta^{|\Omega|} &\subseteq \sub{\qcal_2} \cap \Delta^{|\Omega|} ~ \Leftrightarrow \\
      \qcal_1 &\subseteq \qcal_2~.  \quad  \qed
    \end{align*}
  \end{itemize}
\end{proof}

\begin{figure}[h]
  \centering
  \includegraphics[width=0.5\textwidth]{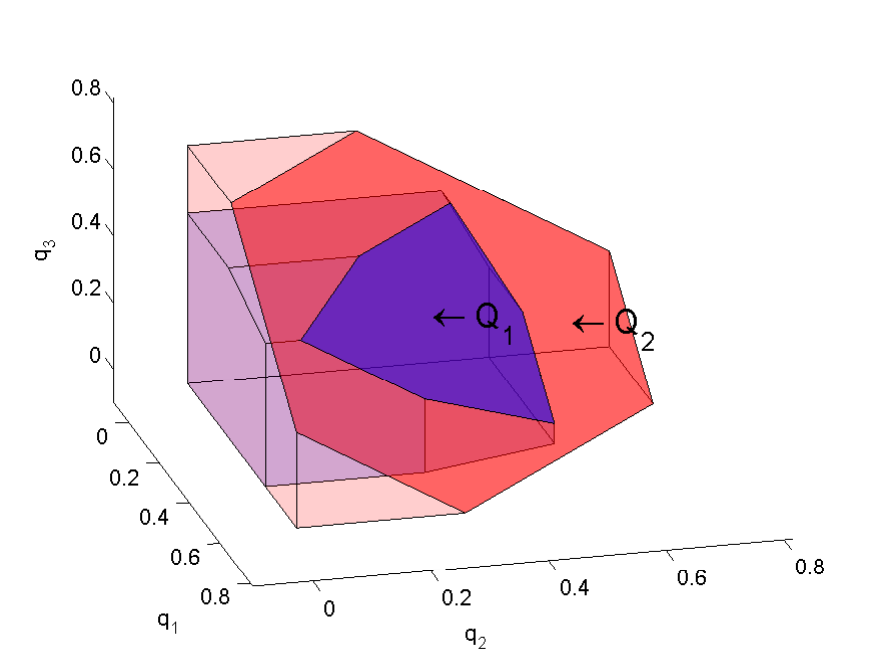}
  \caption{Inclusion relation between $\qcal_1, \, \qcal_2$ (and the corresponding down-monotone closures, $\sub{\qcal_1}$ and $\sub{\qcal_2}$, respectively) that is equivalent to $\mu_1(Y) \leq \mu_2(Y), \, \forall \, Y \in \rvspace_{2}$.}
  \label{fig:qc_subset_qi_subqi_qc_example}
\end{figure}

%From the above lemma and Proposition~\ref{prop:inequality_risk_measures}, one can readily conclude that testing $\mu_1(Y) \leq \mu_2(Y), \, \forall \, Y \in \rvspace$ is also equivalent to $\sub{\qcal_1} \subseteq \sub{\qcal_2}$. %In other words, test whether a risk metric dominates another can be done equivalently in terms of the down-closures of the sets of measures.
Figure~\ref{fig:qc_subset_qi_subqi_qc_example} depicts examples of the sets $\qcal_{1}, \qcal_2$ and their down-monotone closures $\sub{\qcal_1}$ and $\sub{\qcal_2}$, respectively. Note that the conditions provided by Lemma~\ref{lem:downmonotone_equivalent} hold for \emph{any} cost $Y$, i.e., non-negativity is not needed; they may also be more efficient in practice than directly checking $\qcal_1 \subseteq \qcal_2$, in cases when a suitable representation is available for $\sub{\qcal_{1,2}}$, but not for $\qcal_{1,2}$ \citep{Freund_Orlin_85}. 

By considering down-monotone closures and restricting to nonnegative losses, we can also address the second question of interest, namely retrieving the smallest scaling factor $\alpha$ such that $\mu_2(Y) \leq \alpha \cdot \mu_1(Y)$. The following result characterizes any such feasible $\alpha$.

\begin{proposition}
  \label{prop:scaling_property}
  The inequality $\mu_2(Y) \leq \alpha \cdot \mu_1(Y)$ holds for all $Y \geq 0$ if and only if $\sub{\qcal_2} \subseteq \alpha \cdot \sub{\qcal_1}$.
\end{proposition}
\begin{proof}{Proof.}
By Proposition~\ref{prop:can_close}, $\mu_i(Y) = \max_{\mb{q}\in \sub{\qcal_i}} \mb{q}\tr Y$ for $i=1,2$. The claim then follows directly from Lemma~1 in~\citep{Goemans_Hall_1996} (also see Theorem~\ref{thm:goemans_hall_strength_valid_ineq} in the Appendix).
\qed
\end{proof}

In view of this result, the \emph{minimal} $\alpha$ exactly corresponds to the smallest inflation of the down-monotone polytope $\sub{\qcal_1}$ that contains the down-monotone polytope $\sub{\qcal_2}$. This identification leads to the following characterization of the optimal scaling factor. 

\begin{theorem}
 \label{thm:characterization_alpha}
 Let $\alphaopt{\mu_1}{\mu_2}$ denote the smallest value of $\alpha$ such that $\mu_2(Y) \leq \alpha \cdot \mu_1(Y), \, \forall \, Y \geq 0$. 
 \begin{enumerate}
 \item If $\sub{\qcal_1} = \bigl\{ \mb{q} \in \reals^n_{+} \,:\, \mb{a}_i \tr \mb{q} \leq b_i, \, \forall \, i \in \mathcal{I} \bigr\}$, where $\mb{a}_i \geq \mb{0}, \, b_i \geq 0$, then
    \begin{align}
      \alphaopt{\mu_1}{\mu_2} = %\max_{i \in \mathcal{I}} \, \frac{\max_{\mb{x} \in \sub{\qcal_2}} \, \mb{a}_i \tr \mb{x}}{b_i} = 
      \max_{i \in \mathcal{I}} \, \frac{\max_{\mb{q} \in \sub{\qcal_2}} \, \mb{a}_i \tr \mb{q}}{b_i} = \max_{i \in \mathcal{I}} \, \frac{\max_{\mb{q} \in \qcal_2} \, \mb{a}_i \tr \mb{q}}{b_i} = \max_{i \in \mathcal{I}} \, \frac{\mu_2(\mb{a}_i)}{b_i}.
      \label{eq:LP_alphaopt_when_subqcal_available}
    \end{align}
  \item If $\qcal_1 = \bigl\{ \mb{q} \in \reals^n \,:\, A \mb{q} \leq \mb{b} \bigr\}$, then $\alphaopt{\mu_1}{\mu_2}$ is the smallest value $t$ such that the optimal value of the following bilinear program is at most zero:
    \begin{equation}
      \begin{aligned}
        \max_{\mb{q},\mb{\mu}} \quad& ( A \mb{q} - t \mb{b} ) \tr \mb{\mu} \\
        &\mb{q} \in \qcal_2 \\ 
        &\mb{\mu} \geq 0 \\ 
        &A \tr \mb{\mu} \geq 0
      \end{aligned}
      \label{eq:bilinear_alphaopt_when_qcal1_available}
    \end{equation}
  \end{enumerate}
\end{theorem}
\begin{proof}{Proof.}
  The first claim is a known result in combinatorial optimization -- see Theorem~\ref{thm:goemans_hall_strength_valid_ineq} in the Appendix and Theorem~2 in~\citep{Goemans_Hall_1996} for a complete proof.
  
  To argue the second claim, note that the smallest $\alpha$ can be obtained, by definition, as follows:
  \begin{align*}
    \min \Bigl\{ \, t \suchthat \max_{\mb{Y} \geq 0} \Bigl[ \max_{\mb{q} \in \qcal_2} \scprod{Y}{q} - t \cdot \max_{\mb{q} \in \qcal_1} \, \scprod{Y}{q} \Bigr] \leq 0 \, \Bigr\} 
    &=\min \Bigl\{ \, t \suchthat \max_{\mb{Y} \geq 0} \Bigl[ \max_{\mb{q} \in \qcal_2} \scprod{Y}{q} - t \cdot \min_{\begin{smallmatrix} A \tr \mb{\mu} = \mb{Y} \\ \mb{\mu} \geq 0 \end{smallmatrix}} \mb{b} \tr \mb{\mu} \Bigr] \leq 0 \, \Bigr\} \\
    &=\min \Bigl\{ \, t \suchthat 
     \max_{\begin{smallmatrix} \mb{q} \in \qcal_2 \\ \mb{\mu} \geq 0 \\ A \tr \mb{\mu} \geq 0\end{smallmatrix}} \, ( A \mb{q} - t \mb{b} ) \tr \mb{\mu} \leq 0 \, \Bigr\}.
  \end{align*}
  The first equality follows by strong LP duality applied to the maximization over $\mb{q} \in \qcal_1$, which always has a finite optimum since $\qcal_1$ is bounded. The second equality follows by replacing the inner minimization with a maximization, switching the order of the maximizations, and eliminating the variables $\mb{Y}$.
  \qed
\end{proof}

The results in Theorem~\ref{thm:characterization_alpha} give a direct connection between the problem of computing $\alphaopt{\mu_1}{\mu_2}$ and the representations available for the sets $\qcal_i$ and $\sub{\qcal_i}$. More precisely,
\begin{itemize}
\item If a polynomially-sized inequality description is available for $\sub{\qcal_1}$, then $\alphaopt{\mu_1}{\mu_2}$ can be obtained by solving the small number of LPs in~\eqref{eq:LP_alphaopt_when_subqcal_available}. Every such LP essentially entails an evaluation of the risk measure $\mu_2$, leading to an efficient overall procedure.
\item If a compact inequality representation is available for $\qcal_1$, then $\alphaopt{\mu_1}{\mu_2}$ can be found by bisection search over $t \geq 0$, where in each step the bilinear program in~\eqref{eq:bilinear_alphaopt_when_qcal1_available} is solved. Since bilinear programs can be reformulated as integer programs \citep{Horst_Tuy_2003}, for which powerful commercial solvers are available, this approach may lead to a scalable procedure, albeit not one with polynomial-time complexity.
\end{itemize}

  % \dan{Keep this?} For a different interpretation of the latter case, consider the set $K \bydef \{ \mb{\mu} \,:\, \mb{\mu} \geq 0, \, A\tr \mb{\mu} \geq 0 \}$ appearing in~\eqref{eq:bilinear_alphaopt_when_qcal1_available}, and let $K^\polar \bydef \{ \mb{x} \,:\, \scprod{x}{\mu} \leq 0, \, \forall \, \mb{\mu} \in K \}$ denote its polar cone \citep{Rock70}. It can then be readily seen that the result in part~3 is equivalent to
  % \begin{align*}
  %   \alphaopt{\mu_1}{\mu_2} \leq t \quad \Leftrightarrow \quad A \qcal_2 - t \mb{b} \subseteq K^\polar.
  % \end{align*}
  % As such, the complexity of the bilinear program is intrinsically related to the complexity of testing the inclusion of the two polyhedral sets $A \qcal_2 - t \mb{b}$ and $K^\polar$. The former set always has a compact polyhedral representation when $\qcal_2$ does, but in a higher dimensional space; therefore, a compact inequality description in the projected space, i.e., of variables $\mb{y} \bydef A \mb{q} - t \mb{b}, \, \mb{q} \in \qcal_2$, may not be readily available. Similarly, the set $K^\polar$ also has a compact polyhedral representation in a lifted space, so that the representation of the projection may not be compact.

Our observations concerning the complexity of testing $\sub{\qcal_2} \subseteq \alpha \cdot \sub{\qcal_1}$ are summarized in Table~\ref{tab:comput_complexity_table} below. When $\qcal_2$ or $\sub{\qcal_2}$ have polynomially-sized vertex descriptions, the test simply requires checking containment for a finite set of points, and when $\sub{\qcal_2}$ has a polynomially-sized description, the results of Theorem~\ref{thm:characterization_alpha} apply. We conjecture that all the remaining cases are NP-complete, but do not pursue a formal analysis in the present paper. Section~\ref{sec:computational_complexity} revisits the question of computational complexity in the context of {\em comonotonic} risk measures, and argues that the general containment problem is NP-hard.

\begin{table}[h]
%\begin{tabular}{|l|c|c|c|c|c|c}
  \centering
\begin{tabular}{|l|c|c|c|c|c|}
\hline
& Poly $\ext(\qcal_1)$ & Poly $\face(\qcal_1)$ & Poly $\ext(\sub{\qcal_1})$ & Poly $\face(\sub{\qcal_1})$ \\ %& Sub $\qcal_1$ \\
\hline
Poly $\ext(\qcal_2)$ & P & P & P & P \\ %& P  \\
\hline
Poly $\face(\qcal_2)$ & & & & P \\ % & \\
\hline
Poly $\ext(\sub{\qcal_2})$ & P & P & P & P \\ %& P\\
\hline
Poly $\face(\sub{\qcal_2})$ & & & & P \\ %& \\
%\hline
%Sub $\qcal_2$ & & & & P & NP \\
\hline
\end{tabular}
\caption{Computational complexity for determining whether $\mu_2(Y) \le \alpha \cdot \mu_1(Y), \, \forall \, Y \ge \mb{0}$ for a given $\alpha > 0$. ``Poly $\ext$'' and ``Poly $\face$'' denote a polynomially-sized vertex and inequality description, respectively. ``P'' denotes a polynomial-time algorithm is available.}
\label{tab:comput_complexity_table}
\end{table}

We conclude our general discussion by noting that the tightest scaling factor $\alphaopt{\mu_1}{\mu_2}$ can also be used to directly re-examine the first question of interest, namely testing when a given coherent risk measure upper bounds another. This is formalized in the following corollary, which is a direct result of Lemma~\ref{lem:downmonotone_equivalent} and Proposition~\ref{prop:scaling_property}.
\begin{corollary}
  \label{corol:alpha_opt_for_testing_bounds}
  The inequality $\mu_2(Y) \leq \mu_1(Y), \, \forall \, Y \in \rvspace$ holds if and only if $\alphaopt{\mu_1}{\mu_2} \leq 1$.
\end{corollary}

The latter result suggests that characterizing and computing the tightest scaling factor is instrumental in answering \emph{all} questions relating to the approximation of a coherent risk measure by means of another. In particular, given $\mu_1$ and $\mu_2$, by determining the scaling factors $\alphaopt{\mu_1}{\mu_2}$ and $\alphaopt{\mu_2}{\mu_1}$, we can readily test domination and also approximate one measure by the other, as follows:
\begin{itemize}
\item If $\alphaopt{\mu_2}{\mu_1} \leq 1$, then $\mu_1(Y) \in \bigl[ \frac{1}{\alphaopt{\mu_1}{\mu_2}}, \, 1 \bigr] \cdot  \mu_2(Y), \, \forall \, Y \geq 0$. 
\item If $\alphaopt{\mu_1}{\mu_2} \leq 1$, then $\mu_1(Y) \in \bigl[ 1, \, \alphaopt{\mu_2}{\mu_1} \bigr] \cdot \mu_2(Y), \, \forall \, Y \geq 0$. 
\end{itemize}

\subsection{Tightest Time Consistent and Coherent Upper Bound.}
\label{sec:tight-dynam-cons}
The results and exposition in the prior section made no reference to the way in which the coherent risk measures $\mu_{1,2}$ were obtained, as long as the sets of representing measures $\qcal_1, \qcal_2$ were polyhedral. In this section, we discuss some of these results in the context of Section~\ref{sec:cons-meas-scen} -- more precisely, we take $\mu_1$ as the time-inconsistent risk measure $\incons$, while $\mu_2$ denotes the compositional measure $\cons$. 

Our goal is to show that, when $\incons$ is coherent, a complete characterization of the tightest possible uniform upper bound to $\incons$ is readily available, and is given by a popular construction in the literature \citep{Epstein_Schneid_03,Roorda_Schum_Engw_2005_coherent,delbaen_multiperiod_2007,Shapiro_2011}. This not only yields the tightest possible factor $\alphaopt{\incons}{\cons}$, but also considerably simplifies the test $\incons(Y) \leq \cons(Y), \, \forall \, Y$, for any coherent $\cons$. 

The next proposition introduces this construction for an arbitrary coherent measure $\mu$.
\begin{proposition}
  \label{prop:varrho_construction}
  Consider a risk measure $\mu(Y) = \sup_{\mb{q} \in \qcal} \scprod{q}{Y}, \, \forall \, Y \in \rvspace_T$, and define the risk measure $\hat{\mu}(Y) \bydef (\hat{\mu}_{1} \circ \hat{\mu}_{2} \circ \dots \circ \hat{\mu}_{T})(Y)$, where the mappings $\hat{\mu}_{t} \equiv \bigr( \hat{\mu}^i \bigr)_{i \in \nodes{t-1}} : \rvspace_{t} \rightarrow \rvspace_{t-1}$ are given by
    \begin{align}
     \forall \, t \in [1,T], \, \forall \, i \in \nodes{t-1}, ~~ \hat{\mu}^i(Y) &\bydef \sup_{\mb{q} \in \hat{\qcal}^i_\mu} \scprod{q}{Y}, \, \forall \, \mb{Y} \in \reals^{|\child{i}|},
      \label{eq:varrho_def}\\
      \widehat{\qcal}^i_\mu & \bydef \Bigl\{ \mb{q} \in \Delta^{|\child{i}|} \suchthat \exists \, \mb{p} \in \qcal \suchthat q_j = \frac{\mb{p}(\desc{j})}{\mb{p}(\desc{i})}, \, \forall \, j \in \child{i} \, \Bigr\}.
      \label{eq:varrho_sets_of_measures}
    \end{align}
    Then, $\hat{\mu}$ is a time-consistent, coherent risk measure, and $\mu(Y) \leq \hat{\mu}(Y), \, \forall \, Y \in \rvspace_T$.
\end{proposition}

As mentioned, this construction has already been considered in several papers in the literature, and several authors have recognized that it provides an upper bound to $\mu$. It is known that $\hat{\mu}$ is time-consistent, and has a representation of the form $\hat{\mu}(Y) = \sup_{\mb{q} \in \widehat{\qcal}_\mu} \scprod{q}{Y}, \, \forall \, Y \in \rvspace_T$, where the set $\widehat{\qcal}_\mu$ has a \emph{product} or \emph{rectangular} structure. Note that it is obtained by computing products of the sets $\widehat{\qcal}^i_\mu$ of single-step conditional probabilities obtained by marginalization at each node in the tree. Furthermore, $\qcal \subseteq \widehat{\qcal}_\mu$, and therefore $\mu(Y) \leq \hat{\mu}(Y), \, \forall \, Y \in \rvspace_T$ \citep{Epstein_Schneid_03,Roorda_Schum_Engw_2005_coherent,Shapiro_2011}.

\begin{theorem}[Example~\ref{example:naive_inconsistencies} Revisited.]
  To understand the construction, consider again Example~\ref{example:naive_inconsistencies}. The set $\qcal$ yielding the inconsistent measure $\mu$ at the root node R is given by two probabilities, corresponding to $p = 0.4$ and $p = 0.6$, i.e., 
  \begin{align*}
    \qcal = \{ \, (0.16, \, 0.24, \, 0.24, \, 0.36), \, (0.36, \, 0.24, \, 0.24, \, 0.16) \, \}.
  \end{align*}
  The sets of conditional one-step probabilities corresponding to nodes U, D, and R are then:
  \begin{align*}
    \widehat{\qcal}^U_\mu = \widehat{\qcal}^D_\mu = \widehat{\qcal}^R_\mu = \{ (0.4, \, 0.6), \, (0.6, \, 0.4) \}.
  \end{align*}
  This yields a set $\widehat{\qcal}_\mu$ containing eight different probability measures, for all possible products of one-step measures chosen from $\widehat{\qcal}^U_\mu,\, \widehat{\qcal}^D_\mu,\, \widehat{\qcal}^R_\mu$. More precisely,
  \begin{align*}
    \widehat{\qcal}_\mu = & 
    \bigl\{ \, ( 0.16, \, 0.24, \, 0.24, \, 0.36 ), \, 
    (0.16, \, 0.24, \, 0.36, \, 0.24 ), \, 
    (0.24,\, 0.16,\, 0.24, \, 0.36), \, 
    (0.24,\, 0.16,\, 0.36, \, 0.24), \,\\
    & ~~ (0.24,\, 0.36,\,	0.16,\,	0.24), \,
    (0.24,\, 0.36,\,	0.24,\,	0.16), \,
    (0.36,\, 0.24,\,	0.16,\,	0.24),\,
    (0.36,\, 0.24,\,	0.24,\,	0.16) \, \bigr\}.
  \end{align*}
\end{theorem}

In this context, we claim that $\widehat{\incons}$ actually represents the \emph{tightest} upper bound for $\incons$, among all possible coherent and time-consistent upper bounds. This is formalized in the following result.%, which we prove in Section~\ref{sec:technical-proofs} of the paper's Appendix.

\begin{lemma}
  \label{lem:varrho_best_upper_bound}
  Consider any risk measure $\incons(Y) = \sup_{\mb{q} \in \qincons} \scprod{q}{Y}, \, \forall \, Y \in \rvspace_T$, and let $\widehat{\incons}$ be the corresponding risk measure obtained by the construction in Proposition~\ref{prop:varrho_construction}. Also, consider any time-consistent, coherent risk measure $\cons(Y) \bydef (\rho_1 \circ \rho_2 \dots \circ \rho_T)(Y)$, where $\rho_t \equiv \bigl( \rho_t^i \bigr)_{i \in \nodes{t-1}} : \rvspace_t \rightarrow \rvspace_{t-1}$ are given by
  \begin{align*}
    \rho_t^i(Y) = \max_{\mb{q} \in \qcal^i_\rho} \scprod{q}{Y}, \, \forall\, \mb{Y} \in \reals^{|\child{i}|},
  \end{align*}
for some closed and convex sets $\qcal^i_\rho \subseteq \Delta^{|\child{i}|}$. Then, the following results hold:
\begin{enumerate}
%\item $\cons(Y) \geq \incons(Y), \, \forall\, Y \in \rvspace_T$ holds if and only if $\cons(Y) \geq \widehat{\incons}(Y), \, \forall \, Y \in \rvspace_T$. 
%\item If $\cons(Y) \geq \incons(Y), \, \forall\, Y \in \rvspace_T$, then $\alphaopt{\incons}{\widehat{\mu}} \leq \alphaopt{\incons}{\cons}$.
\item If $\cons(Y) \geq \incons(Y), \, \forall\, Y \in \rvspace_T$, then 
  \begin{align}
    \cons(Y) \geq \widehat{\incons}(Y), \, \forall \, Y \in \rvspace_T \qquad \textup{and} \qquad \alphaopt{\incons}{\widehat{\mu}} \leq \alphaopt{\incons}{\cons}.
    \label{eq:cons_dominates_varrho}
  \end{align}
%$\cons(Y) \geq \widehat{\incons}(Y), \, \forall \, Y \in \rvspace_T$, and $\alphaopt{\incons}{\widehat{\mu}} \leq \alphaopt{\incons}{\cons}$.
\item $\cons(Y) \geq \incons(Y), \, \forall\, Y \in \rvspace_T$ holds if and only if 
  \begin{align}
    \widehat{\qcal}_{\incons}^i \subseteq \qcal^i_\rho, \, \forall \, i \in \nodes{t-1}, \, \forall \, t \in [1,T] .
    \label{eq:cons_dominates_incons_using_varrho}
  \end{align}
\end{enumerate}
\end{lemma}
\begin{proof}{Proof.}
  [1] Since $\cons$ is a coherent risk measure, it can always be written as $\cons(Y) = \max_{\mb{q} \in \qcons} \scprod{q}{Y}$. Furthermore, it is known that the set of representing measures  $\qcons$ is obtained by taking products of the sets $\qcal_\rho^i$ (see, e.g., \citet{Roorda_Schum_Engw_2005_coherent} or \citet{Follmer_Schied}). Due to this property, $\qcons$ is closed under the operation of taking marginals and computing the product of the resulting sets of conditional one-step measures \citep{Epstein_Schneid_03,Roorda_Schum_Engw_2005_coherent,delbaen_multiperiod_2007}, i.e.,
  \begin{align}
    \qcal^i_\rho = \Bigl\{ \mb{q} \in \Delta^{|\child{i}|} \suchthat \exists \, \mb{p} \in \qcons \suchthat q_j = \frac{\mb{p}(\desc{j})}{\mb{p}(\desc{i})}, \, \forall \, j \in \child{i} \Bigr\}.% \bydef \widehat{\qcal}^i_{\cons}.
    \label{eq:qcalrho_closed_under_pasting}
  \end{align}
  Since $\cons(Y) \geq \incons(Y)$, we must have $\qincons \subseteq \qcons$. But then, from~\eqref{eq:varrho_sets_of_measures} and~\eqref{eq:qcalrho_closed_under_pasting}, we obtain that $\hat{Q}^i_{\incons} \subseteq \qcal^i_\rho, \, \forall \, i \in \nodes{t-1}, \, \forall \, t$. This readily implies that $\widehat{\qcal}_{\incons} \subseteq \qcons$, and hence $\widehat{\incons}(Y) \leq \cons(Y), \, \forall \, Y$. The inequality for the multiplicative factors $\alphaopt{\incons}{\cdot}$ follows from the definition.

  [2] For the second result, note that the ``$\Rightarrow$'' implication has already been proved in the first part. The reverse direction follows trivially since $\widehat{\qcal}_{\incons}^i \subseteq \qcal^i_\rho$ implies that $\widehat{\qcal}_{\incons} \subseteq \qcons$, and, since $\incons(Y) \leq \widehat{\incons}(Y)$, we have $\incons(Y) \leq \cons(Y), \, \forall \, Y$.
  \qed
\end{proof}

%The proof follows quite directly from the construction and several result boun s known in the literature. Due to space considerations, we choose to relegate it to the online supplement of the paper.

The result above is useful in several ways. First, it suggests that the tightest time-consistent, coherent upper bound for a given $\incons$ is $\widehat{\incons}$. This not only yields the smallest possible multiplicative factor $\alphaopt{\incons}{\cdot}$, but the upper-bound is \emph{uniform}, i.e., for any loss $Y$. Also, $\alphaopt{\incons}{\widehat{\incons}}$ is a lower bound on the best possible $\alphaopt{\incons}{\cons}$ when the consistent measures $\cons$ are further constrained, e.g., to be \emph{comonotonic}. 

The conditions~\eqref{eq:qcalrho_closed_under_pasting} also prescribe a different way of testing $\incons \leq \cons$, by examining several smaller-dimensional tests involving the sets $\widehat{\qcal}^i_{\incons}, \qcal^i_\rho \subseteq \Delta^{|\child{i}|}$. This will also prove relevant in our subsequent analysis of the case of comonotonic risk measures.

\subsection{The Comonotonic Case.}
\label{sec:charact_qc_qi_subqc_subqi}
% The results and exposition in the prior section made no reference to the way in which the coherent risk measures $\mu_{1,2}$ were obtained, as long as the sets of representing measures were polyhedral. In this section, we particularize the results to the setting described in Section~\ref{sec:cons-meas-scen}, where dynamic risk is measured by means of comonotonic risk measures. More precisely, we take $\mu_1$ as the time-inconsistent risk measure $\incons$, while $\mu_2$ denotes the consistent, compositional risk measure $\cons$. 

The results introduced in Section~\ref{sec:find-optim-alpha} and Section~\ref{sec:tight-dynam-cons} become more specific when the risk measures $\qincons$ and $\qcons$ are further restricted to be comonotonic. We discuss a model with $T=2$, but the approach and results readily extend to a finite number of time periods, a case which we revisit in Section~\ref{sec:multi-stage-extens}.

We start by characterizing $\incons$, with its set of representing measures $\qincons$ and its down closure $\sub{\qincons}$. %As mentioned above, the set of measures characterizing a single distortion risk metric is the \emph{base polytope} corresponding to a particular rank function $c$, and the down-monotone closure of this set is the polymatroid corresponding to $c$.
The central result here, formalized in the next proposition, is the identification of $\qincons$ with the base polytope corresponding to a particular Choquet capacity $c$. This analogy proves very useful in our analysis, since it allows stating all properties of $\qincons$ by employing known results for base polytopes of polymatroid rank functions\footnote{We note that, with the exception of the normalization requirement $c(\Omega) = 1$ that is unimportant for analyzing fundamental structural properties, the definition of a \emph{Choquet capacity} is identical to that of a \emph{rank function of a polymatroid} \citep[Chapter 2]{Fujishige_book_05}. Therefore, we use the two names interchangeably throughout the current paper.}, a concept studied extensively in combinatorial optimization (see Section~\ref{sec:subm-polyh-polym} of the Appendix for all the results relevant to our treatment, and \citep{Fujishige_book_05, Schrijver_Alex_book_2003} for a comprehensive review).

\begin{proposition}
  \label{prop:incon_rep_thm}
  Consider a na\"{i}ve dynamic comonotonic risk measure $\incons : \rvspace_2 \rightarrow \reals$, with $\incons(Y) \bydef \max_{\mb{q} \in \qincons} \, \scprod{q}{Y}, \, \forall \, Y \in \rvspace_2$. Then, there exists a Choquet capacity $c : 2^{|\nodes{2}|} \rightarrow \reals$ such that
  \begin{enumerate}
  \item The set of measures $\qincons$ is given by the \emph{base polytope} corresponding to $c$, i.e.,
    \begin{align}
      \qincons \equiv \basepoly{c} \bydef \bigl\{~ \mb{q} \in \reals^{|\nodes{2}|} \suchthat \mb{q}(S) \leq c(S),  \, \forall \, S \subseteq \nodes{2}, \, \mb{q}(\nodes{2}) = c(\nodes{2})~ \bigr\}.
      \label{eq:inconsistent_measure_set}
    \end{align}
    % where $c : 2^{|\nodes{2}|} \rightarrow \reals$ is a Choquet capacity,
    %and $\basepoly{c}$ is the base polytope corresponding to $c$.
  \item The down-monotone closure of $\qincons$ is given by the \emph{polymatroid} corresponding to $c$, i.e.,
    \begin{align}
      \sub{\qincons} \equiv \polymat{c} \bydef \bigl\{\, \mb{q} \in \spacenodes{2}_{+} \,:\, \mb{q}(S) \leq c(S), \, \forall \, S \subseteq \nodes{2} \, \bigr\}.
      \label{eq:down_closure_inconsist_set}
    \end{align}
    %where $\polymat{c}$ is the polymatroid corresponding to $c$.
  \end{enumerate}
\end{proposition}
\begin{proof}{Proof.}
  %The first claim follows directly from Theorem~\ref{thm:representation_comonotonic_rm} for distortion risk measures (also see the discussion in Section~\ref{sec:background_risk_theory} and Theorem 4.88 in \citep{Follmer_Schied}).
  By Theorem~\ref{thm:representation_comonotonic_rm} for comonotonic risk measures, there exists a Choquet capacity $c$ such that $\qincons = \bigl\{~ \mb{q} \in \Delta^{|\nodes{2}|} \suchthat \mb{q}(S) \leq c(S),  \, \forall \, S \subseteq \nodes{2} \bigr\}$. Since $c(\nodes{2}) = 1$, this set can be rewritten equivalently as the base polytope corresponding to $c$ (also refer to Corollary~\ref{corol:nonnegative_base_poly} of the Appendix for the argument that $\basepoly{c} \subset \reals^{|\nodes{2}|}_+$).
  For the second claim, we can invoke a classical result in combinatorial optimization, that the downward monotone closure of the base polytope $\basepoly{c}$ is exactly given by the \emph{polymatroid} corresponding to the rank function $c$, i.e., $\polymat{c}$ (see Theorem~\ref{thm:properties_base_poly} in Section~\ref{sec:subm-polyh-polym}).
  \qed
\end{proof}

In particular, both sets $\qincons$ and $\sub{\qincons}$ are polytopes contained in the non-negative orthant, and generally described by exponentially many inequalities, one for each subset of the ground set $\nodes{2}$. However, evaluating the risk measure $\incons$ for a given $Y \in \rvspace_{2}$ can be done in time polynomial in $|\nodes{2}|$, by a simple Greedy procedure (see Theorem~\ref{thm:greedy_evaluation_comonotonic_case} in the Appendix or Lemma~4.92 in~\citep{Follmer_Schied}).

In view of the results in Section~\ref{sec:tight-dynam-cons}, one may also seek a characterization of the tightest upper bound to $\incons$, i.e., $\widehat{\incons}$, or of its set of representing measures $\widehat{\qcal}_{\incons}$. Unfortunately, this seems quite difficult for general Choquet capacities $c$ -- a particular case when it is possible is when $\incons$ is given by $\cvar_\lev$, a case discussed in our companion paper \citet{Huang_I_Petrik_Subraman_cvar_2012}. However, the result in Lemma~\ref{lem:varrho_best_upper_bound} nonetheless proves useful for several of the results in this section. 

The following result provides a characterization for the time-consistent and comonotonic risk measure  $\cons = \mu_1 \circ \mu_2$ as a \emph{coherent} risk measure, by describing its set of representing measures $\qcons$ and its down-monotone closure $\sub{\qcons}$.
\begin{proposition}
  \label{prop:rep_comp_measure}
  Consider a two-period consistent, comonotonic risk measure $\cons(Y) = \mu_1 \circ \mu_2$, where $\mu_t : \rvspace_{t} \rightarrow \rvspace_{t-1}$. Then,
  \begin{enumerate}
  \item There exists $\qcons \subseteq \Delta^{|\nodes{2}|}$ such that $\cons(Y) \bydef \max_{\mb{q} \in \qcons} \, \scprod{q}{Y}, \, \forall \, Y \in \rvspace_2$.
  \item  The set of measures $\qcons$ is given by
    \begin{align*}
      \qcons &\bydef \biggl\{ \mb{q} \in \Delta^{|\nodes{2}|} \suchthat \exists \, \mb{p} \in \Delta^{|\nodes{1}|}, \quad
      \begin{aligned}
        \mb{p}(S) &\leq c_1(S), \, \forall \, S \subseteq \nodes{1} \\
        \mb{q}(U) & \leq p_i \cdot c_{2|i}(U), \, \forall \, U \subseteq \child{i}, \, \forall \, i \in \nodes{1}
      \end{aligned}
      \biggr\}  \\
      & \equiv \Bigl\{ \, \mb{q} \in \Delta^{|\nodes{2}|} \suchthat \exists \, \mb{p} \in \basepoly{c_1} \suchthat \mb{q}\vert_{\child{i}} \in \basepoly{p_i \cdot c_{2|i}}, \, \forall \, i \in \nodes{1} \, \Bigr\},
    \end{align*}
    where $c_1 : 2^{|\nodes{1}|} \rightarrow \reals$ and $c_{2|i} : 2^{|\child{i}|} \rightarrow \reals, \, \forall \, i \in \nodes{1}$ are Choquet capacities, and $\basepoly{c_1}, \basepoly{c_{2|i}}$ are the base polytopes corresponding to $c_1$ and $c_{2|i}$, respectively.
  \item The downward monotone closure of $\qcons$ is given by
    \begin{align*}
      \sub{\qcons} & \bydef \biggl\{ \mb{q} \in \spacenodes{2}_+ \,:\, \exists \, \mb{p} \in \spacenodes{1}_+, \quad
      \begin{aligned}
        \mb{p}(S) &\leq c_1(S), \, \forall \, S \subseteq \nodes{1}, \\
        \mb{q}(U) & \leq p_i \cdot c_{2|i}(U), \, \forall \, U \subseteq \child{i}, \, \forall \, i \in \nodes{1}
      \end{aligned}
      \biggr\} \\
      &= \Bigl\{ \mb{q} \in \spacenodes{2}_+ \,:\, \exists \, \mb{p} \in \polymat{c_1} \suchthat \mb{q}\vert_{\child{i}} \in \polymat{p_i \cdot c_{2|i}}, \, \forall \, i \in \nodes{1} \Bigr\}
    \end{align*}
    where $\polymat{c_1}$ and $\polymat{p_i c_{2|i}}$ are the polymatroids associated with $c_1$ and $p_i c_{2|i}$, respectively.
  \end{enumerate}
\end{proposition}
\begin{proof}{Proof.}
The proof is technical, and involves a repeated application of ideas similar to those in the proof of Proposition~\ref{prop:incon_rep_thm}. Therefore, we relegate it to Section~\ref{sec:technical-proofs} of the Appendix.
\qed
\end{proof}

As expected, the set of product measures $\qcons$ and its down-monotone closure $\sub{\qcons}$ have a more complicated structure than $\qincons$ and $\sub{\qincons}$, respectively. However, they remain polyhedral sets, characterized by the base polytopes and polymatroids associated with particular Choquet capacities $c_1$ and $c_{2|i}$. The inequality descriptions of $\qcons$ and $\sub{\qcons}$ involve exponentially many constraints, but evaluating $\cons(Y)$ at a given $Y \in \rvspace_{2}$ can still be done in time polynomial in $|\nodes{2}|$, by using the Greedy procedure suggested in Theorem~\ref{thm:greedy_evaluation_comonotonic_case} in a recursive manner.

Because $\qincons$ and $\qcons$ are polytopes, they can also be described in terms of their extreme points. The description of the vertices of polymatroids and base polytopes has been studied extensively in combinatorial optimization (see Theorem~\ref{thm:extreme_points_base_polytope} in the Appendix or \citep{Schrijver_Alex_book_2003} for details). Here, we apply the result for the case of $\qincons$, and extend it to the special structure of the set $\qcons$.
\begin{proposition}
  \label{prop:extreme_points}
  Consider two risk measures $\incons$ and $\cons$, as given by Proposition~\ref{prop:incon_rep_thm} and Proposition~\ref{prop:rep_comp_measure}. Then,
  \begin{enumerate}
  \item  The extreme points of $\qincons$ are given by
    \begin{align*}
      q_{\sigma(i)} = c\bigl( \cup_{k=1}^{i} \sigma(k) \bigr) - c\bigl( \cup_{k=1}^{i-1} \sigma(k) \bigr), \, i \in [1,|\nodes{2}|],
    \end{align*}
    where $\sigma \in \Pi(\nodes{2})$ is any permutation of the elements of $\nodes{2}$.
  \item The extreme points of $\qcons$ are given by
    \begin{align*}
      q_{\sigma_{\ell}(i)} = \Bigl[ c_1 \bigl(\cup_{k=1}^\ell \pi(k) \bigr) - c_1\bigl(\cup_{k=1}^{\ell-1} \pi(k) \bigr) \Bigr] \cdot \Bigl[ c_{2|\ell} \bigl(\cup_{k=1}^i \sigma_\ell(k) \bigr) - c_{2|\ell}\bigl(\cup_{k=1}^{i-1} \sigma_\ell(k) \bigr) \Bigr] , \, \forall \, i \in [1,|\child{\ell}|], \forall \, \ell \in\nodes{1},
    \end{align*}
    where $\pi \in \Pi(\nodes{1})$ is any permutation of the elements of $\nodes{1}$, and $\sigma_\ell \in \Pi(\child{\ell})$ is any permutation of the elements of $\child{\ell}$ (for each $\ell \in \nodes{1}$).
  \end{enumerate}
\end{proposition}
\begin{proof}{Proof.}
Part (1) follows directly from the well-known characterization of the extreme points of an extended polymatroid, summarized in Theorem~\ref{thm:extreme_points_base_polytope}.

Part (2) follows by a repeated application of Theorem~\ref{thm:extreme_points_base_polytope} to both $\mb{p}$ and $\mb{q}$ in the description of $\qcons$ of Proposition~\ref{prop:rep_comp_measure}. In particular, any value of $\mb{p}$ can be expressed as a convex combination of the extreme points $\mb{p}^\pi$ of $\basepoly{c_1}$  such that $ \mb{p} = \sum_{\pi \in \Pi(\nodes{1})} \lambda_{\pi} \mb{p}^{\pi}$ for appropriate  $\{\lambda_{\pi}\}_{\pi \in \Pi(\nodes{1})}$. Now, for each $\ell \in \nodes{1}$ the value $\mb{q}\vert_{\child{\ell}} \in \polymat{p_\ell \cdot c_{2|\ell}}$ can be similarly expressed as a convex combination of the extreme points $\mb{q}_\ell^\sigma$ for an appropriate set of convex weights $\{\xi_{\sigma}\}_{\sigma \in \Pi(\child{\ell})}$, such that
\begin{align*}
 \mb{q} \vert_{\child{\ell}} &= p_\ell \cdot \sum_{\sigma\in\Pi(\child{\ell})} \xi_{\sigma} \mb{q}_\ell^{\sigma} 
 = \sum_{\pi \in \Pi(\nodes{1})} \sum_{\sigma \in \Pi(\child{l})} \lambda_{\pi} \xi_{\sigma} p_\ell^{\pi} \mb{q}_\ell^{\sigma}
= \sum_{\pi \in \Pi(\nodes{1})} \sum_{\sigma \in \Pi(\child{\ell})} \chi_{\pi,\sigma} p_\ell^{\pi} \mb{q}_\ell^{\sigma} ~.
\end{align*}
The proposition then follows directly from the fact that $\chi_{\pi,\sigma}$ are themselves convex combination coefficients, and $p_\ell^{\pi} \mb{q}_\ell$ are extreme points.
\qed
\end{proof}

\subsection{Computing the Optimal Bounds $\alphaopt{\cons}{\incons}$ and $\alphaopt{\incons}{\cons}$.}
\label{sec:basic bounds}
With the representations provided above, we now derive our main technical result, establishing a method for computing the tightest multiplicative bounds for a pair of consistent and inconsistent comonotonic risk measures. The following theorem summarizes the result.

\begin{theorem}
  \label{thm:optimal_alpha}
  For any pair of risk measures $\incons$ and $\cons$ as introduced in Section~\ref{sec:charact_qc_qi_subqc_subqi},
  \begin{align}
    \label{eq:optimal_alpha_formula}
    \alphaopt{\cons}{\incons} &= \max_{\mb{q} \in \sub{\qincons}} \, \max_{S \subseteq \nodes{1}} \, \frac{\sum_{i \in S} \max_{U \subseteq \child{i}} \frac{\mb{q}(U)}{c_{2|i}(U)} }{c_1(S)} ~ \\
    \label{eq:optimal_beta_formula}
    \alphaopt{\incons}{\cons} &= \max_{\mb{q} \in \sub{\qcons}} \, \max_{S \subseteq \nodes{2}} \frac{\mb{q}(S)}{c(S)} ~.
  \end{align}
  Furthermore, the value for $\alphaopt{\cons}{\incons}$ remains the same if the outer maximization over $\mb{q}$ is done over $\qincons, \ext(\qincons)$ or $\ext(\sub{\qincons})$, and corresponding statements hold for $\alphaopt{\incons}{\cons}$.
\end{theorem}
\begin{proof}{Proof.}
  To prove the first result, recall from Proposition~\ref{prop:scaling_property} that for any $Y \geq 0$,
  \begin{align*}
     \incons(Y) \leq \alpha \cdot \cons(Y) ~ \Leftrightarrow ~ \sub{\qincons} \subseteq \alpha \cdot \sub{\qcons} ~.
  \end{align*}
 Consider an arbitrary $\mb{q} \in \sub{\qincons}$. Any feasible scaling $\alpha > 0$ must satisfy that $\frac{1}{\alpha} \mb{q} \in \sub{\qcons}$. Using the representation for $\sub{\qcons}$ in Proposition~\ref{prop:rep_comp_measure}, this condition yields
  \begin{align*}
    \frac{1}{\alpha} \mb{q} \in \sub{\qcons} & \Leftrightarrow
   \exists \, \mb{p} \in \spacenodes{1}_{+} \,:\, \, \left\{
   \begin{aligned}
     \mb{p}(S) &\leq c_1(S), \, \forall \, S \subseteq \nodes{1} \\
     \frac{1}{\alpha} \, \mb{q}(U) &\leq p_i \cdot c_{2|i}(U), \, \forall \, U \subseteq \child{i}, \, \forall \, i \in \nodes{1}.
   \end{aligned}
   \right.
  \end{align*}
  The second set of constraints implies that any feasible $\mb{p}$ satisfies $p_i \geq \frac{1}{\alpha} \max_{U \subseteq \child{i}}  \frac{\mb{q}(U)}{c_{2|i}(U)}, \, \forall \, i \in \nodes{1}$. Corroborated with the first set of constraints, this yields
  \begin{align*}
    \frac{1}{\alpha} \sum_{i \in S} \max_{U \subseteq \child{i}} \frac{\mb{q}(U)}{c_{2|i}(U)} &\leq \sum_{i \in S} p_i \leq c_1(S), \, \forall \, S \subseteq \nodes{1} \quad \Leftrightarrow \\
    \alpha &\geq \max_{S \subseteq \nodes{1}} \, \frac{\sum_{i \in S} \max_{U \subseteq \child{i}} \frac{\mb{q}(U)}{c_{2|i}(U)} }{c_1(S)}.
  \end{align*}
  Since this must be true for any $\mb{q} \in \sub{\qincons}$, the smallest possible $\alpha$ is given by maximizing the expression above over $\mb{q} \in \sub{\qincons}$, which leads to the result~\eqref{eq:optimal_alpha_formula}.

  The expression for $\alphaopt{\incons}{\cons}$ is a direct application of the second part of Theorem~\ref{thm:characterization_alpha}, by identifying $\sub{\qcal_1}$ with $\sub{\qincons}$ and using the compact representation for $\sub{\qincons}$ from Proposition~\ref{prop:incon_rep_thm}. 

  The claim concerning the alternative sets  follows by recognizing that the function maximized is always nondecreasing in the components of $\mb{q}$, so that $\sub{\qcal}$ can be replaced with $\qcal$, and it is also convex in $\mb{q}$, hence reaching its maximum at the extreme points of the feasible set.  \qed
\end{proof}

From Theorem~\ref{thm:optimal_alpha}, it can readily seen that, when $\cons \leq \incons$, the optimal $\alphaopt{\cons}{\incons}$ will always be at least $1$, and can be $+\infty$ whenever the dimension of the polytope $\qcons$ is strictly smaller than that of $\qincons$. Similarly, when $\incons \leq \cons$, the optimal $\alphaopt{\incons}{\cons}$ is always at least $1$, and can be $+\infty$ when the dimension of the polytope $\qincons$ is smaller than $\qcons$. To avoid the cases of unbounded optimal scaling factors, one can make the following assumption about the Choquet capacities.
\begin{assumption}[Relevance]
\label{assum:nontrivial_choquet_capacity}
The Choquet capacities $c, c_1, \, c_{2|i}$ appearing in the representations for $\incons$ and $\cons$ (Proposition~\ref{prop:incon_rep_thm} and Proposition~\ref{prop:rep_comp_measure}) satisfy the properties
\begin{align*}
     c(\{k\}) &> 0, \, \forall \, k \in \nodes{2} \\
     c_1(\{i\}) &> 0, \, \forall \, i \in \nodes{1} \\
     c_{2|i}(\{j\}) & > 0, \, \forall \, i \in \nodes{1}, \, \forall \, j \in \child{i}.
\end{align*}
\end{assumption}
This ensures that both risk measures consider all possible outcomes in the scenario tree, and is in line with the original requirement of \emph{relevance} in \citep{Artzner_Delbaen_1999}, which states that, for any random cost $Y$ such that $Y \geq 0$ and $Y \neq 0$, any risk measure $\mu$ should satisfy $\mu(Y) > 0$. In this case, the polytopes $\qincons$ and $\qcons$ are both full-dimensional (see \citep{Balas_Fischetti_97} and Appendix A), which leads to finite minimal scalings.

As suggested in our general exposition at the beginning of Section~\ref{sec:find-optim-alpha}, determining the optimal scaling factors $\alphaopt{\cons}{\incons}$ and $\alphaopt{\incons}{\cons}$ also leads to direct conditions for determining whether $\cons$ lower bounds $\incons$ or viceversa. The following corollary states these in terms of optimization problems.

\begin{corollary} \label{cor:lower_upper_bounds}
For any pair of risk measures $\incons$ and $\cons$ as introduced in Section~\ref{sec:charact_qc_qi_subqc_subqi},
  \begin{enumerate}
  \item The inequality $\cons(Y) \leq \incons(Y), \, \forall \, Y \in \rvspace_{2}$ holds if and only if
    \[
    \max_{\mb{q} \in \ext(\qcons)} \, \max_{S \subseteq \nodes{2}} \bigl[ \, \mb{q}(S) - c(S) \, \bigr] \leq 0.
    \]
  \item The inequality $\incons(Y) \leq \cons(Y), \, \forall \, Y \in \rvspace_{2}$ holds if and only if
    \[
    \max_{\mb{q} \in \ext(\qincons)} \, \max_{S \subseteq \nodes{1}} \, \biggl[\, \sum_{i \in S} \max_{U \subseteq \child{i}} \frac{\mb{q}(U)}{c_{2|i}(U)} - c_1(S) \, \biggr] \leq 0.
    \]
  \end{enumerate}
\end{corollary}
\begin{proof}{Proof.}
  The proof follows from Corollary~\ref{corol:alpha_opt_for_testing_bounds}, by recognizing that the condition $\cons \leq \incons$ is equivalent to setting $\alphaopt{\incons}{\cons} \leq 1$ (and similarly for the reverse inequality and $\alphaopt{\cons}{\incons}$). The formulas in Theorem~\ref{thm:optimal_alpha} then immediately yield the desired conclusions.
  \qed
\end{proof}

The results in Corollary~\ref{cor:lower_upper_bounds} are stated in terms of non-trivial optimization problems. It is also possible to write out the conditions in a combinatorial fashion, using the analytical description of the extreme points of $\qcons$ and $\qincons$, as summarized in the following corollary.

\begin{corollary} 
  \label{cor:lower_upper_bounds_explicit}
  For any pair of risk measures $\incons$ and $\cons$ as introduced in Section~\ref{sec:charact_qc_qi_subqc_subqi},
  \begin{enumerate}
  \item The inequality $\cons(Y) \leq \incons(Y), \, \forall \, Y \in \rvspace_2$ holds if and only if
    \begin{align*}
      \sum_{j = 1}^{|\nodes{1}|} \, \Bigl[ c_1\bigl(\cup_{k=1}^j s_k \bigr) - c_1\bigl(\cup_{k=1}^{j-1} s_k \bigr) \Bigr] \cdot c_{2|s_j}(U_{s_j}) \leq c \bigl( \cup_{i \in \nodes{1}} U_i \bigr),
    \end{align*}
     where $(s_1,\dots,s_{|\nodes{1}|})$ denotes any permutation of the elements of $\nodes{1}$, and $U_i \subseteq \child{i}$ for any $i \in \nodes{1}$.
 \item The inequality $\incons(Y) \leq \cons(Y), \, \forall \, Y\in\rvspace_{2}$ holds if and only if
    \begin{align*}
      c\bigl( \cup_{i \in S} \child{i} \bigr) &\leq c_1(S), ~~ \forall \, S \subseteq \nodes{1}, \\
      \frac{c(U)}{c(U) + 1 - c(\nodes{2} \setminus \child{i} \cup U)} &\leq c_{2,i}(U), \, \, \forall \, U \subseteq \child{i}, \, \forall \, i \in \nodes{1}.
    \end{align*}
  \end{enumerate}
\end{corollary}
\begin{proof}{Proof.}
  The proof is slightly technical, so we defer it to Section~\ref{sec:technical-proofs} of the Appendix.
  \qed
\end{proof}

The above conditions are explicit, and can always be checked when oracles are available for evaluating the relevant Choquet capacities. The main shortcoming of that approach is that the number of conditions to test is generally exponential in the size of the problem, even for a fixed $T$: $\mathcal{O}\bigl( (|\nodes{1}|!) \cdot 2^{|\nodes{2}|} \bigr)$ for $\cons \leq \incons$, and $\mathcal{O}( |A| \cdot 2^{\max_{i \in A} |\child{i}|} )$ for $\incons \leq \cons$, respectively, where $A \bydef \cup_{t \in [0,T-1]} \nodes{t}$. However, under additional assumptions on the Choquet capacities or the risk measures, it is possible to derive particularly simple polynomially-sized tests. We refer the interested reader to the discussion in Section~\ref{sec:computational_complexity} and the example in Section~\ref{sec:case-average-value-two-period}.

We note that the reason the conditions for $\incons \leq \cons$ take a decoupled form and result in a smaller overall number of inequalities is directly related to the results of Lemma~\ref{lem:varrho_best_upper_bound}, which argues that testing $\incons \leq \cons$ can be done by separately examining conditions at each node of the scenario tree.

\subsection{Multi-stage Extensions.}
\label{sec:multi-stage-extens}
Although we focused our discussion thus far on a setting with $T = 2$, the ideas can be readily extended to an arbitrary, finite number of periods. We briefly outline the most relevant results in this section, but omit including the proofs, which are completely analogous to those for $T=2$. 

In a setting with general $T$, our goal is to compare a comonotonic $\incons$ with a time-consistent, comonotonic $\cons \bydef \mu_{1} \circ \mu_{2} \circ \dots \circ \mu_T$. The former is exactly characterized by Proposition~\ref{prop:incon_rep_thm}, while the representation for the latter can be summarized in the following extension of Proposition~\ref{prop:rep_comp_measure}.
\begin{proposition}
  \label{prop:rep_comp_measure_multi_period}
  Consider a time-consistent, comonotonic risk measure $\cons$. Then,
  \begin{enumerate}
  \item There exists $\qcons \subseteq \Delta^{|\nodes{T}|}$ such that $\cons(Y) \bydef \max_{\mb{q} \in \qcons} \, \scprod{q}{Y}, \, \forall \, Y \in \rvspace_2$.
  \item  The set of measures $\qcons$ is given by
    \begin{align}
      \qcons & \bydef \bigl\{ \, \mb{p}_T \in \Delta^{|\nodes{T}|} \suchthat \exists \, \{\mb{p}_t \in \Delta^{|\nodes{t}|}\}_{t \in [1,T-1]}, %\\ & \qquad \qquad \qquad \qquad \qquad \qquad \qquad \quad
      ~ \mb{p}_{t}(U) \leq \mb{p}_{t-1}(\{i\}) \cdot c_{t|i}(U), \, \forall \, U \subseteq \child{i}, \, \forall \, i \in \nodes{t-1}, \, \forall \, t \in [1,T] \, \bigr\}  \nonumber \\
      & \equiv \bigl\{ \, \mb{p}_T \in \Delta^{|\nodes{T}|} \suchthat \exists \, \{\mb{p}_t \in \Delta^{|\nodes{t}|}\}_{t \in [1,T-1]}, ~ \mb{p}_{t}\vert_{\child{i}} \in \basepoly{\mb{p}_{t-1}(\{i\}) \cdot c_{t|i}}, \, \forall \, i \in \nodes{t-1}, \, \forall \, t \in [1,T] ~ \bigr\},
      \label{eq:qcons_multiperiod}
    \end{align}
    where $c_{t|i} : 2^{|\child{i}|} \rightarrow \reals$ are Choquet capacities with corresponding base polytopes $\basepoly{c_{t|i}}$, for every $t \in [1,T]$ and for every $i \in \nodes{t-1}$.
  \item The downward monotone closure $\sub{\qcons}$ of $\qcons$ is obtained by replacing $\Delta^{|\nodes{t}|}$ with $\reals^{|\nodes{t}|}_+$ and $\basepoly{\mb{p}_{t-1}(\{i\}) \cdot c_{t|i}}$ with the polymatroid $\polymat{\mb{p}_{t-1}(\{i\}) \cdot c_{t|i}}$ in equation~\eqref{eq:qcons_multiperiod}.
  \end{enumerate}
\end{proposition}

The proof exactly parallels that of Proposition~\ref{prop:rep_comp_measure}, and is omitted due to space considerations. With this result, we can now extend our main characterization in Theorem~\ref{thm:optimal_alpha} for the optimal multiplicative factors to a multi-period setting, as follows.
\begin{theorem}
  \label{thm:optimal_alpha_multi_period}
  For any comonotonic measure $\incons$ and time-consistent comonotonic measure $\cons$,
  \begin{align}
    \label{eq:optimal_alpha_multi_period}
    \alphaopt{\cons}{\incons} &= \max_{\mb{q} \in \sub{\qincons}}  \max_{S \subseteq \nodes{1}} \, \frac{ \sum_{i \in S} z_1(i,\mb{q})}{c_{1}(S)},
  \end{align}
  where $\mb{z}_T(i,\mb{q}) \bydef q_i, \, \forall \, i \in\nodes{T}$, and $z_t(i,\mb{q}) \bydef \max_{U \subseteq \child{i}} \frac{\sum_{i\in U} z_{t+1} (i,\mb{q})}{c_{t+1|i}(U)}, \, \forall \, t \in [1,T-1], \, \forall \, i \in \nodes{t}$. Also,
  \begin{align}
    \label{eq:optimal_beta_multi_period}
    \alphaopt{\incons}{\cons} &= \max_{\mb{q} \in \sub{\qcons}} \, \max_{S \subseteq \nodes{T}} \frac{\mb{q}(S)}{c(S)}.
  \end{align}
  Furthermore, the value for $\alphaopt{\cons}{\incons}$ would remain the same if the outer maximization were taken over $\qincons, \ext(\qincons)$ or $\ext(\sub{\qincons})$. Corresponding statements hold for $\alphaopt{\incons}{\cons}$. 
\end{theorem}
The proof follows analogously to that of Theorem~\ref{thm:optimal_alpha}, by using the expressions for $\sub{\qincons}$ and $\sub{\qcons}$ provided by Proposition~\ref{prop:incon_rep_thm} and Proposition~\ref{prop:rep_comp_measure_multi_period}, respectively, to analyze the conditions $\sub{\qincons} \subseteq \alpha \cdot \sub{\qcons}$ or vice-versa. We omit it for brevity.

By comparing~\eqref{eq:optimal_alpha_multi_period} and~\eqref{eq:optimal_beta_multi_period} with their two-period analogues in~\eqref{eq:optimal_alpha_formula} and~\eqref{eq:optimal_beta_formula}, respectively, it is interesting to note that the complexity of the formulation for $\alphaopt{\incons}{\cons}$ remains the same, while the optimization problems yielding $\alphaopt{\cons}{\incons}$ get considerably more intricate. Section~\ref{sec:discussion} contains a detailed analysis of the computational complexity surrounding these problems.

For completeness, we remark that direct multi-period counterparts for Corollary~\ref{cor:lower_upper_bounds} and Corollary~\ref{cor:lower_upper_bounds_explicit} can be obtained, by recognizing that $\cons(Y) \leq \incons(Y), \, \forall \, Y$ is equivalent to $\alphaopt{\incons}{\cons} \leq 1$, and by using the results in Theorem~\ref{thm:optimal_alpha_multi_period} and Lemma~\ref{lem:varrho_best_upper_bound} to simplify the latter conditions. We do not include these extensions due to space considerations. 

\section{Discussion of the Results.} 
\label{sec:discussion}
%In this section, we discuss the results of Section~\ref{sec:find-optim-alpha} in more detail, and suggest examples of how the various conditions can be used. In particular, Section~\ref{sec:computational_complexity} analyzes the complexity of computing the optimal scalings $\alphaopt{\qincons}{\qcons}$ and $\alphaopt{\qcons}{\qincons}$, showing that these problems are NP complete in general \dan{are we sure they are NP complete instead of NP hard?}, even when restricting attention to a subclass of comonotonic risk measures. However, we suggest particular cases when polynomial-time algorithms are nonetheless possible. Section~\ref{sec:examples} then demonstrates how to derive simple, tractable, dynamically consistent risk measure bounds from a given inconsistent risk measure using the optimal scalings  $\alphaopt{\qincons}{\qcons}$ and $\alphaopt{\qcons}{\qincons}$. \dan{Include $\cvar$ results.}

In view of the results in the previous section, several natural questions emerge. What is the computational complexity of determining the optimal scaling factors $\alphaopt{\incons}{\cons}$ and $\alphaopt{\cons}{\incons}$ for coherent/comonotonic risk measures? If this is generally hard, are there special cases that are easy, i.e., admitting polynomial-time algorithms? What examples of time-consistent risk measures can be derived starting with a given $\incons$, and how closely do they approximate the original measure? 

The goal of the present section is to address these questions in detail. As we argue in Section~\ref{sec:computational_complexity}, computing the scaling factors is hard even when restricting attention to \emph{distortion} risk measures -- a proper subclass of comonotonic measures. However, several relevant cases are nonetheless tractable. Section~\ref{sec:examples} introduces examples obtained by composing $\incons$ with the conditional expectation operator ``$\E$'' or conditional worst-case operator ``$\max$'', and compares them in terms of their approximation strength. Section~\ref{sec:case-average-value-two-period} then summarizes the case when $\incons$ and $\cons$ correspond to the $\cvar$ risk measure, and shows how many of the results drastically simplify.

\subsection{Computational Complexity.} 
\label{sec:computational_complexity}
%\dan{Either here, or in Section~\ref{sec:find-optim-alpha}, we should have the discussion about the hardness for computing $\alpha^\star$ when the risk measures are given by $\qincons$ and $\qcons$ (essentially, the complexity table Marek was drawing up). I believe that testing $Q \subseteq P$ is hard even when $P, Q$ are given as projections of some polytopes with compact inequality representations. To see that, supposed we could always test efficiently whether $Q \subseteq P$, when $P, Q$ are given via projections of some $\hat{P}, \hat{Q}$ H-representable polyhedra. Then, we can solve $Q \subseteq P$, when $Q$ is H-representable and $P$ is V-representable (which is known to be hard!) The reason is that a V-representation is basically the same as $P = \{ x \,:\, \exists \, \lambda \geq 0, \, 1 \tr \lambda = 1, \, x = V \lambda \}$, where $V$ has the extreme points of $Q$ on columns.}

As argued in Section~\ref{sec:find-optim-alpha}, computing the optimal scaling factors $\alphaopt{\incons}{\cons}$ and $\alphaopt{\cons}{\incons}$ entails solving the optimization problems in~\eqref{eq:optimal_alpha_formula} and~\eqref{eq:optimal_beta_formula}. We now show that this is NP hard even for a problem with $T=1$, and even when only examining \emph{distortion} risk measures. We use a reduction from the SUBSET-SUM problem, which is NP hard~\citep{CLRS01} and is defined as follows.
\begin{definition} [SUBSET-SUM] Given a set of integers $\{k_1,k_2,\ldots,k_m\}$, is there a subset that sums to $s$?
\end{definition}

This following result is instrumental in showing the complexity of computing the optimal scalings $\alphaopt{\incons}{\cons}$ and $\alphaopt{\cons}{\incons}$. 
\begin{theorem} 
  \label{thm:complexity_general}
  Consider two arbitrary distortion risk measures $\mu_{1,2} : \rvspace_{1} \rightarrow \reals$. Then, it is NP-hard to decide if $\alphaopt{\mu_2}{\mu_1} \ge \gamma$, for any  $\gamma \ge 0$. The problem remains NP-hard even when $\mu_2(Y) \leq \mu_1(Y)$, for all $Y \in \rvspace_1 \, (Y \geq 0)$.
\end{theorem}
\begin{proof}{Proof.}
  We use the representation of distortion risk measures to show the reduction from the SUBSET-SUM problem. By the representation Theorem~\ref{thm:representation_comonotonic_rm} written for the specific case of distortion measures yields, $\mu_i(Y) = \max_{\mb{q} \in \qcal_i} \, \scprod{q}{Y}$, where 
  \begin{equation*}
    \qcal_i = \Bigl\{~ \mb{q} \in \Delta^{|\nodes{2}|} \suchthat \mb{q}(S) \leq c_i(S),  \, \forall \, S \subseteq \nodes{1}~ \Bigr\}, ~~ \forall \, i \in \{1,2\},
  \end{equation*}
  and $c_i(S) = \Psi_i\bigl( \Pr(S) \bigr)$, where $\Psi_i : [0,1] \rightarrow [0,1]$ are concave, increasing functions satisfying $\Psi_i(0) = 0, \, \Psi_i(1) = 1$. Because both $\sub{\qcal_1}$ and $\sub{\qcal_2}$ are polymatroids and downward monotone, the second result in Theorem~\ref{thm:optimal_alpha} can be further simplified to:
  \begin{equation} 
    \label{eq:optimization_goemans_reduced}
    \alphaopt{\mu_2}{\mu_1} = \max_{S \subseteq \nodes{1}} \frac{c_1(S)}{c_2(S)}.
\end{equation}
Now, consider a SUBSET-SUM problem with values $k_1, k_2 \ldots k_m$ and a value $s$ such that $1 \le s < K$, where $K = \sum_{j=1}^m k_j$. Construct the functions $c_1$ and $c_2$ as follows:
\begin{align*}
\mathbb{P}(s_i) &= k_i / K &
c_1(S) &= \min\Bigl\{ \bigl( \mathbb{P}(S) \cdot K \bigr) /s , 1  \Bigr\} &
c_2(S) &= \min\Bigl\{c_1(S), \sqrt{\mathbb{P}(S)} \Bigr\}
\end{align*}
Since both $c_1$, $c_2$ satisfy the conditions of distortion risk measures, any SUBSET-SUM problem can be reduced to the problem of computing the optimal scale of two distortion risk measures.

Now, the optimal value of \eqref{eq:optimization_goemans_reduced} is upper bounded as:
\[\max_{S \subseteq \nodes{1}} \frac{c_1(S)}{c_2(S)} \le  \sqrt{\frac{K}{s}}~.  \]
The maximum is achieved when there exists $S$ such that $\mathbb{P}(S) = s/K$. To show this, consider $c_1(S)/c_2(S)$ as a function of $\mathbb{P}(S)$. This function is: (1) non-decreasing on the interval $[0,s/K)$ and non-increasing on the interval $(s/K,1]$, (2) strictly greater than one for $\mathbb{P}(S) = s/K$, (3) equal to 1 for $\mathbb{P}(S) \in \{0,1\}$, and (4) continuous. Therefore, the SUBSET-SUM problem has a subset that sums to $s$ if and only if the optimal value of \eqref{eq:optimization_goemans_reduced} is $\sqrt{K/s}$. Finally, the result also holds when $\mu_2(Y) \leq \mu_1(Y)$, since our choice already has $c_2(S) \le c_1(S)$ for all $S \subseteq \nodes{1}$, which implies $\mu_2(Y) \leq \mu_1(Y)$. \qed
\end{proof}

The complexity of computing the optimal scalings $\alphaopt{\incons}{\cons}$ and $\alphaopt{\cons}{\incons}$ readily follows as a direct corollary of Theorem~\ref{thm:complexity_general}.
\begin{corollary} 
  \label{cor:complexity_alpha}
  Under a fixed $T \geq 1$ and for any given $\gamma \geq 0$, it is NP-complete to decide whether $\alphaopt{\cons}{\incons} \geq \gamma$ for an arbitrary inconsistent distortion measure $\incons$ and a consistent distortion measure $\cons$. The result remains true even when $\cons$ and $\incons$ are such that $\cons(Y) \leq \incons(Y)$ for all $Y \in \rvspace_T \, (Y \geq 0)$. Similarly, it is NP-complete to decide whether $\alphaopt{\incons}{\cons} \geq \gamma$, and the result remains true even when $\incons(Y) \leq \cons(Y), \, \forall \, Y \in \rvspace_T \, (Y \geq 0)$.
\end{corollary}
\begin{proof}{Proof.}
  First, note that finding the scaling factors for any $T > 1$ is at least as hard as for $T=1$. This can be seen by setting  $|\nodes{t}| = 1$ for all $t\in [2, T-1]$. The NP-hardness then follows from Theorem~\ref{thm:complexity_general} by setting $\mu_2 = \cons$ and $\mu_1 = \incons$. The membership in NP follows by checking the inequality \eqref{eq:optimal_alpha_formula} for every extreme point $\mb{q}$, subset $S$, and the appropriate subsets $U$. The second result follows analogously.\qed
\end{proof}

Corollary~\ref{cor:complexity_alpha} argues that computing the optimal scaling factors for arbitrary distortion risk measures cannot be done in polynomial time. While the NP-hardness may be somewhat disappointing, solving the two optimization problems in Theorem~\ref{thm:optimal_alpha} is nonetheless clearly preferable to simply examining all possible values of $Y$. 

%\subsection{Cases Allowing Polynomial-Time Algorithms.}
%\label{sec:cases-allow-polyn}
While the problem of computing the scaling factors is hard for general distortion measures, polynomial-time algorithms are possible when the representations of $\qincons, \qcons$ or $\sub{\qincons}, \sub{\qcons}$ fall in the tractable cases discussed in Table~\ref{tab:comput_complexity_table} of Section~\ref{sec:find-optim-alpha}. 

In fact, some of the results of Table~\ref{tab:comput_complexity_table} can even be strengthened - one such case is when a vertex description for the polytope $\qincons$ is available, and problem~\eqref{eq:optimal_alpha_multi_period} can be solved in time polynomial in $|\nodes{T}|$, under oracle access to the Choquet capacities $c_{t|i}$ yielding the measure $\cons$.
\begin{lemma}
  If the polytope $\qincons$ is specified by a polynomial number of extreme points, then $\alphaopt{\cons}{\incons}$ can be computed in time polynomial in $|\nodes{T}|$.
\end{lemma}
\begin{proof}{Proof.}
  Consider the specialization of~\eqref{eq:optimal_alpha_multi_period} for a fixed $\mb{q}\in \qincons$: 

  \begin{align}
    \label{eq:optimal_alpha_multi_period}
    \alphaopt{\cons}{\incons} &= \max_{\mb{q} \in \sub{\qincons}}  \max_{S \subseteq \nodes{1}} \, \frac{ \mb{z}_1(U,\mb{q})}{c_{1}(S)},
  \end{align}
  where $\mb{z}_T(i,\mb{q}) \bydef q_i, \, \forall \, i \in\nodes{T}$, and $z_t(i,\mb{q}) \bydef \max_{U \subseteq \child{i}} \frac{\mb{z}_{t+1} (U,\mb{q})}{c_{t+1|i}(U)}, \, \forall \, t \in [1,T-1], \, \forall \, i \in \nodes{t}$. 

  Note that each value $z_t(i,\mb{q})$ and also $\alphaopt{\cons}{\incons}$ can be written as:
  \[ z_t(i,\mb{q}) = \max_{U \subseteq \child{i}} \frac{\mb{z}_{t+1} (U,\mb{q})}{c_{t+1|i}(U)} ~= \min \Bigl\{ l \in \reals \suchthat l \cdot c_{t+1|i}(U) - \mb{z}_{t+1} (U,\mb{q}) \ge 0, \; \forall \, U \subseteq \child{i} \Bigr\} \,.\]
  %This latter optimization can be solved in polynomial time by bisection on $l$. 
  For any $l$, the constraint $l \cdot c_{t+1|i}(U) - \mb{z}_{t+1}(U,\mb{q}) \ge 0,\, \forall \, U \subseteq \child{i}$ can be checked in polynomial time, since the set function on the left-hand side is submodular in $U$, and can be minimized with a polynomial number of function evaluations~\citep{Schrijver_Alex_book_2003}. %The problem $\max_{S \subseteq \nodes{1}} \, {\sum_{i \in S} z_i}/{c_1(S)}$ can then be solved similarly using the linearity of $\sum_{i \in S} z_i$ with respect to $S$. 
  \qed
\end{proof}

The result above is slightly stronger than what Table~\ref{tab:comput_complexity_table} suggests, since the representation of $\qcons \equiv \qcal_1$ can still be exponential both in terms of extreme points and vertices, as long as oracle access to $c_{t|i}$ is available.

%
%  Examples with rho(E(.)) and E(rho(.))
%
\subsection{Examples.}
\label{sec:examples}
To see how our results can be used to examine the tightness of particular dynamically consistent risk measures, we now consider several constructions suggested in the literature. The starting point is typically a single \emph{distortion} risk measure $\incons : \rvspace_2 \rightarrow \reals$, denoting the inconsistent evaluation. This is then composed with other suitable measures (for instance, with itself, with the conditional expectation and/or the conditional worst-case operator), to obtain time-consistent risk measures that are derived from $\incons$. The questions we would like to address here is which of these measures lower-bound or upper-bound the inconsistent evaluation $\incons$, and what can be said about the relative tightness of the various formulations.

In order to construct dynamically-consistent measures by composing $\incons$, we must first specify the conditional one-step risk mappings corresponding to $\incons$, formally denoted by $\incons^1 : \rvspace_1 \rightarrow \reals$ and $\incons^{2} : \rvspace_2 \rightarrow \rvspace_1$. When $\incons$ is a \emph{distortion} risk measure, this can be done in a natural way in terms of the corresponding concave distortion function. To this end, recall that, by the representation Theorem~\ref{thm:representation_comonotonic_rm}, any distortion measure $\incons$ is uniquely specified by the concave function $\Psi$ yielding its set of representing measures, through the Choquet capacity $c(S) = \Psi(\Pr(S)), \, \forall \, S \subseteq \Omega_2$. The conditional one-step risk mappings $\incons^1$ and $\incons^2 \equiv (\incons^{2|i})_{i \in \nodes{1}}$ are then obtained by applying the same distortion function $\Psi$ to suitable conditional probabilities. More precisely, $\incons^1$ and $\incons^{2|i}$ are the distortion risk measures corresponding to the Choquet capacities:
  \begin{align*}
    c_1 : 2^{\nodes{1}} \rightarrow \reals, \, c_1(S) &= \Psi\Bigl(\, \sum_{i \in S} \Pr(\child{i}) \Bigr), \, \forall \, S \subseteq \nodes{1} \\
    c_{2|i} : 2^{\child{i}} \rightarrow \reals, \, c_{2|i}(U_i) &= \Psi\Bigl( \frac{\Pr(U_i)}{\Pr(\child{i})} \Bigr), \, \forall \, U_i \subseteq \child{i}, \, \forall \, i \in \nodes{1}.
  \end{align*}
%Note that the same distortion function $\Psi$ (yielding $\incons$) is applied in the first stage, but to the appropriate conditional probability measure. In the second stage, the probability measure is conditioned on the first-stage outcome.

The conditional risk mappings $\incons^1$ and $\incons^{2}$ can be used to define dynamic time-consistent risk measures, either alone or by composition with other conditional risk mappings. In particular, all of the following dynamic time-consistent risk measures have been considered in the literature:
\[ \E \circ \incons^2 \qquad \incons^1 \circ \E \qquad \incons^1 \circ \incons^2 \qquad \incons^1 \circ \max \qquad \max \circ \incons^2,\]
where $\E$ denotes the conditional expectation operator, and $\max$ is the conditional worst-case operator. Whenever the meaning is clear from context, we sometimes omit the time-subscript, and use shorthand notation such as $\E \circ \incons, \, \incons \circ \E$, $\incons \circ \incons$,  etc., although we are formally referring to compositions with $\incons^1$ and/or $\incons^2$.

\subsubsection{Time-Consistent Lower Bounds Derived From a Given $\incons$.}
\label{sec:time-cons-lower}
We begin by discussing two choices for lower-bounding consistent risk measures derived from $\incons$. The following proposition formally establishes the first relevant result.
\begin{proposition}
  \label{prop:lower_bounds_by_cond_expectation}
  Consider any \emph{distortion} risk measure $\incons : \rvspace_2 \rightarrow \reals$, and the time-consistent, comonotonic measures $\incons \circ \E$ and $\E \circ \incons$. Then, for any cost $Y \in \rvspace_{2}$,
  \begin{align*}
    (\incons \circ \E)(Y) \leq \incons(Y) \quad \textup{and} \quad (\E \circ \incons)(Y) \leq \incons(Y).
  \end{align*}
\end{proposition}
\begin{proof}{Proof.}
  The proof entails directly checking the conditions in Corollary~\ref{cor:lower_upper_bounds_explicit}. A complete derivation is included in Section~\ref{sec:technical-proofs} of the Appendix.
\end{proof}

This is not a surprising result, since the $\E$ operator is known to be a uniform lower bound for any static coherent risk measure \citep{Follmer_Schied}. We confirm that the same remains true in dynamic settings, provided that the risk measure $\incons$ is applied in a single time step, and conditional expectation operators are applied in other stages.

Since both $\incons \circ \E$ and $E \circ \incons$ are lower bounds for $\incons$, a natural question is whether one provides a ``better'' approximation than the other. More precisely, the following are questions of interest:
\begin{enumerate}
\item For a given $\incons$, is it true that
  $ (\incons \circ \E)(Y) \leq (\E \circ \incons)(Y), \, \forall \, Y \in \rvspace_2$
  (or vice-versa)?
\item Is it true that $\alphaopt{\incons \circ \E}{\incons} \geq \alphaopt{\E \circ \incons}{\incons}$ \emph{for any} distortion measure $\incons$ (or vice-versa)?
\end{enumerate}
Clearly, a positive answer to the first question would provide a very strong sense of tightness of approximation. However, as the following example shows, neither inequality holds in general.
\begin{example}
  Consider a scenario tree with $T=2$, $|\nodes{1}| = 2, \, |\child{i}| = 2, \, \forall \, i \in \nodes{1}$, under uniform reference measure. Introduce the following two random costs $X, Y$ (specified as vectors in $\reals^{|\nodes{2}|}$):
  \begin{gather*}
    X\vert_{\child{1}} = M \cdot \oneN{}, ~~ X\vert_{\child{2}} = \mb{0} \\
    Y\vert_{\child{1}} = Y\vert_{\child{2}} = [M, ~ 0]\tr.
  \end{gather*}
  With $M > 0$, and $\incons \equiv \cvar_{1/2}$, it can be checked that $(\incons \circ \E)(X) = M > (\E \circ \incons)(X) = \frac{M}{2}$, while $(\incons \circ \E)(Y) = \frac{M}{2} < (\E \circ \incons)(Y) = M$.
\end{example}

Insofar as the second question is concerned, we note that it can always be answered for a \emph{specific} distortion measure $\qincons$, by calculating the optimal scalings, so that it really makes sense when posed for \emph{all} risk measures. Unfortunately, our computational experiments show that counterexamples can be constructed for this claim, as well, and that any one of the scaling factors can be better than the other. However, it would be very interesting to characterize conditions (on the risk measures, the underlying probability space, or otherwise) under which a particular compositional form always results in a smaller scaling factor. The following result, which we prove in the Appendix, is a potential first step in this direction, suggesting that the two lower bounds can result in \emph{equal} tightness of approximation in certain cases of interest.
\begin{theorem}
  \label{thm:alpha_rhoE_equal_Erho_uniform_rm}
  Consider a uniform scenario tree, i.e., $|\nodes{1}| = N, \, |\child{i}| = N, \, \forall \, i \in \nodes{1}$, under a uniform reference measure. Then, for any distortion risk measure $\incons$, we have 
  \begin{equation*}
    \alphaopt{\incons \circ \E}{\incons} = \alphaopt{\E \circ \incons}{\incons} = N \cdot \max \Bigl\{ \frac{\Psi( 1/N^2)}{\Psi(1/N)}, \, \frac{\Psi( 2/N^2)}{\Psi(2/N)}, \dots, \Psi( 1/N) \Bigr\}.
  \end{equation*}
\end{theorem}

\subsubsection{Time-Consistent Upper Bounds Derived From a Given $\incons$.}
\label{sec:time-cons-upper}

In an analogous fashion to the previous discussion, one can ask what time-consistent \emph{upper bounds} can be derived from a distortion measure $\incons$. In particular, a natural supposition, analogous to the results of Section~\ref{sec:time-cons-lower}, may be that $\incons \circ \max$ and $\max \circ \incons$ are upper bounds to $\incons$, since $\max$ is the most conservative risk mapping possible \citep{Follmer_Schied}. The following result shows that, unlike in the lower bound setting, \emph{only one} of the two composed measures is a valid upper bound.
\begin{proposition}
  \label{prop:upper_bounds_by_max}
  Consider any distortion risk measure $\incons$, and the time-consistent, comonotonic measures $\incons \circ \max$ and $\max \circ \incons$, where $\max$ denotes the conditional worst-case operator. Then:
  \begin{enumerate}
  \item[(i)] For any cost $Y \in \rvspace_{2}$, $\incons(Y) \leq (\incons \circ \max)(Y)$.
  \item[(ii)] There exists a choice of $\incons$ and of random costs $Y_{1,2} \in \rvspace_2$ such that
  $(\max \circ \incons)(Y_1) < \incons(Y_1)$ and $(\max \circ \incons)(Y_2) > \incons(Y_2)$.
  \end{enumerate}
\end{proposition}
\begin{proof}{Proof.}
  The proof for Part~(i) entails checking the conditions of Corollary~\ref{cor:lower_upper_bounds_explicit}. Since it is rather technical in nature, we leave it for Section~\ref{sec:technical-proofs} of the Appendix of the paper. 

  To show Part~(ii), consider a uniform scenario tree with $|\Omega_1| = |\child{i}| = 2, \, \forall \, i \in \nodes{1}$, and let the reference measure be $\refmeas = [0.1,~0.5,~0.2,~0.3]\tr$. For simplicity, assume the first two components of $\refmeas$ correspond to nodes in the same child. Then, for the risk measure $\incons = \cvar_{1/2}$, and the costs $\mb{Y}_1 = [1, ~0, ~0, ~0.4]\tr$ and $\mb{Y}_2 = [0, ~0, ~0, ~1]\tr$, it can be checked that $\incons(Y_1) = 0.44 > (\max \circ \incons)(Y_1) = 0.4$, but $\incons(Y_2) = 0.3 < (\max \circ \incons)(Y_2) = 1$. 
  \qed
\end{proof}

The result in Proposition~\ref{prop:upper_bounds_by_max} also suggests that upper bounds to $\incons$ can be derived by composing $\incons$ with more conservative mappings in later time periods. This intuition is sharpened in Section~\ref{sec:case-average-value-two-period} and our companion paper~\citep{Huang_I_Petrik_Subraman_cvar_2012}, which show that, when $\incons = \cvar_\lev$, all upper bounds of the form $\cvar_\lev \circ \cvar_\gamma$ must have $\gamma \leq \lev$, and, in many practical settings, $\gamma = 0$, i.e., worst-case as the second-stage evaluation.

Since $\incons \circ \max$ is an upper bound for a given $\incons$, one can also turn to the question of comparing the resulting scaling factor $\alphaopt{\incons}{\incons \circ \max}$ with the factors of the previous section, namely $\alphaopt{\incons \circ \E}{\incons}$ or $\alphaopt{\E \circ \incons}{\incons}$. Our computational tests show that there is no general relation between these, even when the scenario tree and the reference measure are uniform, a claim due to the following result, whose proof is included in the paper's Appendix.
\begin{proposition}
  \label{prop:beta_star_mu_of_max}
  Consider a uniform scenario tree, i.e., $|\nodes{1}| = N, \, |\child{i}| = N, \, \forall \, i \in \nodes{1}$, under a uniform reference measure. Then, for any distortion risk measure $\incons$, we have 
  \begin{equation*}
    \alphaopt{\incons}{\incons \circ \max} = \max \Bigl\{ \frac {\Psi(1/N)}{\Psi( 1/N^2)}, \, \frac {\Psi(2/N)}{\Psi( 2/N^2)}, \dots, \frac{1}{\Psi( 1/N)} \Bigr\}.
  \end{equation*}
\end{proposition}
Corroborating this result with the expression in Theorem~\ref{thm:alpha_rhoE_equal_Erho_uniform_rm} for $\alphaopt{\incons \circ \E}{\incons}$, one can readily find simple examples of distortions $\Psi$ such that either the latter or the former scaling factor is smaller.

An opinion often held among practitioners, and informally argued in the literature \citep{Roorda_Schumach_2007_tail_var,Roorda_Schumacher_2008} is that composing a risk measure with itself would compound the losses, resulting in a larger evaluation of risk, i.e., that $\incons \circ \incons$ should over-bound $\incons$. For instance, if $\incons = \cvar$ -- the case considered in \citep{Roorda_Schumach_2007_tail_var} -- the compositional measure corresponds to the so-called ``iterated tail-CTE'', which takes tail conditional expectations of quantities that are already tail conditional expectations. We show by means of an example that this informal belief is actually \emph{not} true, even in the case of $\cvar$.

\begin{example}[Iterated $\cvar$]
  \label{examp:iterated_cvar}
  Consider a uniform scenario tree (i.e., $|\nodes{1}| = |\child{i}| = 4, \, \forall \, i \in \nodes{1}$), and a uniform reference measure. Furthermore, consider the risk measure
  % \begin{align*}
  %   \incons(Y) &= \cvar_{3/4}(Y), \, \forall \, Y \in \rvspace_2\\
  %   \cons(Y) &= (\mu_1 \circ \mu_2)(Y), \forall \, Y \in \rvspace_2, ~ \textup{where}~\\
  %   \mu_1(Y) &= \cvar_{3/4}(Y), \, \forall \, Y \in \rvspace_1 \\
  %   \mu_2& = ( \mu^i )_{i \in \nodes{1}}, \, \mu^i(Y) = \cvar_{3/4}(Y), \, \forall \, Y \in \reals^{|\child{i}|}, \, \forall \, i \in \nodes{1}.
  % \end{align*}
  $\incons \equiv \cvar_{3/4}$, and the following two costs (specified as real vectors in $\reals^{|\nodes{2}|}$, with components split in the four sub-trees of stage $T=2$):
  \begin{align*}
    & X\vert_{\child{1}} = X\vert_{\child{2}} = \oneN{}, \quad X\vert_{\child{3}} = X\vert_{\child{4}} = [1,\, 1,\, -M, \, -M]\tr \\
    & Y\vert_{\child{1}} = Y\vert_{\child{2}} = Y\vert_{\child{3}} = [1,\, 1,\, 1, \, -M]\tr, \quad Y\vert_{\child{4}} = -M \cdot \oneN{}.
  \end{align*}
When $M > -1$, it can be readily checked\footnote{For the case of discrete probability measures, one has to be careful in defining $\cvar_\lev$, since it is no longer exactly given by the conditional expectation of the loss exceeding $\var_\lev$. The precise concepts are presented and discussed at length in~\citep{Rock_Uryas_2002}, which we follow here.} that $\incons(X) = 1 > (\incons \circ \incons)(X) = \frac{8-M}{9}$, while $(\incons \circ \incons)(Y) = 1 > \incons(Y) = \frac{3-M}{4}$.
\end{example}

The example shows that the iterated $\cvar$ is \emph{neither an upper nor a lower bound} to the static $\cvar$. We direct the interested reader to our companion paper \citep{Huang_I_Petrik_Subraman_cvar_2012}, which is focused specifically on the $\cvar$ case, and discusses the exact necessary and sufficient conditions for when one of the two dominates the other.

\subsubsection{The Tightest Possible Time-Consistent Upper-bound.}
\label{sec:risk-measure-cons}

A natural time-consistent upper bound to a given $\incons$ is the measure $\widehat{\incons}$, obtained by the rectangularization procedure in Proposition~\ref{prop:varrho_construction}. It is the tightest possible coherent upper bound to $\incons$, both in a uniform and multiplicative-alpha sense. The main potential drawback in using $\widehat{\incons}$ is that it may not satisfy additional axiomatic properties, and it typically bears no interpretation in terms of $\incons$. For instance, starting with a comonotonic $\incons$ does \emph{not} generally result in a comonotonic $\widehat{\incons}$, and $\widehat{\incons}$ is usually not given by compositions of one-step risk measures that correspond to $\incons$. Determining conditions that guarantee the latter two properties is an interesting question, which we do not pursue further in the present paper. However, we note that this is possible in at least one case of practical interest: when $\incons = \cvar_\lev$, one can show that $\widehat{\incons}$ always corresponds to a composition of one-step $\cvar$ measures, at appropriate levels -- see our discussion in Section~\ref{sec:case-average-value-two-period} and the detailed treatment in our companion paper \citep{Huang_I_Petrik_Subraman_cvar_2012}. 

For completeness, we also note that the ordering relation between the scaling factor $\alphaopt{\incons}{\hat{\mu}_I}$ and scalings $\alphaopt{\cons}{\incons}$ derived from lower-bounding measures $\cons$ is generally not obvious: our computational experiments suggest that either one could dominate the other. However, more can be said in particular settings, such as the case of $\cvar$, which we discuss next.

\subsection{The Case of $\cvar$.}
\label{sec:case-average-value-two-period}
In this section, we discuss how several of the results introduced throughout the paper can be considerably simplified when the risk measures in question correspond to $\cvar$. In particular, analytical expressions or polynomial-time procedures can be derived for computing $\alphaopt{\cons}{\incons}$ and $\alphaopt{\incons}{\cons}$ and for testing $\incons(Y) \leq \cons(Y)$ or viceversa. Furthermore, one can consider designing the risk measures $\cons$ that provide the tightest possible lower or upper approximations to a given $\incons$. 

The case is discussed at length in our companion paper \citep{Huang_I_Petrik_Subraman_cvar_2012}, to which we direct the interested reader for any technical details and proofs. Our goal for the remainder of the section is to outline the main results, and briefly discuss the implications.

To start, we consider a uniform scenario tree under uniform reference measure ($|\child{i}| = N, \, \forall \, i \in \cup_{t=0}^{T-1} \nodes{t}$, and $\Pr = \frac{\oneN{}}{N^T}$), and the following choice of risk measures:
\begin{subequations}
  \begin{align}
    \incons &= \cvar_\lev, && \lev \in [1/N^T,\, 1] 
    \label{eq:inconsistent_avar} \\
    \cons &= \cvar_{\lev_1} \circ \cvar_{\lev_2} \circ \dots \circ \cvar_{\lev_T}, && \lev_{t} \in [1/N, \, 1], \, \forall \, t \in [1,T].
    \label{eq:consistent_avar}
  \end{align}
\end{subequations}
Note that the restriction on $\lev$ and $\lev_t$ is without loss of generality, since $\cvar_\lev$ with $\lev \le \frac{1}{N^T}$ is identical to the worst-case risk measure, rendering the case $\lev \in [0, \frac{1}{N^T})$ analogous to $\lev = \frac{1}{N^T}$.

In this setup, we can revisit our main results in Theorem~\ref{thm:optimal_alpha}, and provide the following expressions for the tightest factors $\alphaopt{\cons}{\incons}$ and $\alphaopt{\incons}{\cons}$ for the case $T=2$.

%
%  optimal alpha and beta
%
\begin{theorem}
\label{thm:alpha_star_bound_thm2}
Consider a case $T=2$, and the pair of risk measures in~\eqref{eq:inconsistent_avar} and \eqref{eq:consistent_avar}. Then,
\begin{subequations}
  \begin{align}
    \alphaopt{\cons}{\incons} &=
    \begin{cases}
      \max \big\{ N \lev_1,  \frac{\lev_1 \lev_2}{\lev}, N \lev_2 \big\}, &~~ \lev \leq \frac{1}{N} \\
      \max \big\{ \frac{\lev_1}{\lev}, f(N, \lev, \lev_2)  \big\}, & ~~\lev > \frac{1}{N},
  \end{cases}
  \label{eq:optimal_alpha_exactlong_expression} \\
  \label{eq:optimal_beta_analytical}
  \alphaopt{\incons}{\cons} &= \max \Bigl\{1, \, \frac{\lev}{\lev_1 \, \lev_2} \Bigr\},
\end{align}
\end{subequations}
where $f(N, \lev, \lev_2)$ is an explicit analytical function. Furthermore, the result for $\alphaopt{\incons}{\cons}$ remains true under an \emph{arbitrary} scenario tree and reference measure $\Pr$.  
\end{theorem}

Note that the above result has several immediate implications. First, it readily allows checking whether $\cons(Y) \leq \incons(Y), \, \forall \, Y$ (or vice-versa), since the latter conditions are equivalent to $\alphaopt{\incons}{\cons} \leq 1$ (respectively, $\alphaopt{\cons}{\incons} \leq 1$). This leads to the following simple tests.
\begin{corollary}
\label{corol:qc_qi_epsilons}
Consider the pair of risk measures in~\eqref{eq:inconsistent_avar} and~\eqref{eq:consistent_avar}. Then, 
\begin{enumerate}
\item the inequality $\cons(Y) \leq \incons(Y), \, \forall \, Y \in \rvspace_2$ holds if and only if 
  \begin{align}
    \lev_1 \, \lev_2 \geq \lev,
    \label{eq:conditions_incons_dominates_cons_cvar}
  \end{align}
\item the inequality $\incons(Y) \leq \cons(Y), \, \forall \, Y \in \rvspace_2$ holds if and only if 
  \begin{align}
    \lev_1 \leq \max \Bigl( \frac{1}{N}, \lev \Bigr) \quad \textup{and} \quad \lev_2 \leq \max \Bigr(\frac{1}{N}, N \lev - N + 1 \Bigr).
    \label{eq:conditions_cons_dominates_incons_cvar}
  \end{align}
  Furthermore, ~\eqref{eq:conditions_incons_dominates_cons_cvar} remains true under an \emph{arbitrary} scenario tree and reference measure $\Pr$.
\end{enumerate}
\end{corollary}

The latter result confirms the observation in Example~\ref{examp:iterated_cvar} that the iterated $\cvar$, i.e., $\cons = \cvar_\lev \circ \cvar_\lev$, is generally neither an upper nor a lower bound to the inconsistent choice $\incons = \cvar_\lev$. By~\eqref{eq:conditions_cons_dominates_incons_cvar}, $\lev_1 = \lev$ is always a feasible option, but one must take $\lev_2 \leq \max( 1/N, \, N \lev - N + 1)$. In fact, as argued in \citep{Huang_I_Petrik_Subraman_cvar_2012}, most relevant choices of $\lev$ would actually lead to taking $\lev_2 = 1/N$, i.e., the worst-case operator in the second stage. 

 The analytical results above can also be used to \emph{optimally design} the compositional risk measure $\cons$ that is the tightest approximation to a given $\incons = \cvar_\lev$. More precisely, one can characterize the choice of $\cons$ (i.e., levels $\lev_{1,2}^{\textup{LB}}$) that results in the smallest possible factor $\alphaopt{\cons}{\incons}$ among all compositional $\cvar$ that are lower bounds for $\cvar_\lev$, and, similarly, the values $\lev_{1,2}^{\textup{UB}}$ yielding the smallest possible $\alphaopt{\incons}{\cons}$ among all upper-bounding compositional $\cvar$s. The optimal choices satisfy several interesting properties:
 \begin{itemize}
 \item for values of $\lev$ that are common in financial applications, i.e., satisfying $\lev \leq 1/N$ \citep{Jorion_VaR_2006}, the optimal $\alphaopt{\cons}{\incons}$ is obtained by taking $\lev_1^{\textup{LB}} = \lev_2^{\textup{LB}} = \sqrt{\lev}$, corresponding to an iterated $\cvar$ measure. 
 \item the optimal $\alphaopt{\incons}{\cons}$ requires choosing $\lev_1^{\textup{UB}} = \lev$ and $\lev_2^{\textup{UB}} = \max( 1/N, \, N \lev - N + 1)$. Typical values of $\lev$ used in practice would entail $\lev_2^{\textup{UB}} = 1/N$, i.e., the worst-case scenario in the second stage. 
 \item the optimally designed $\alphaopt{\cons}{\incons}$ is \emph{always} smaller than $\alphaopt{\incons}{\cons}$, i.e., for every $\lev$ and $N$, which suggests that starting with an under-estimating $\cvar_\lev$ results in tighter dynamically consistent approximations for $\cvar$.
 \end{itemize}

The results discussed in Theorem~\ref{thm:alpha_star_bound_thm2} for $T=2$ can also be (partially) extended to a case of an arbitrary $T$, which is summarized in the following claim.
\begin{theorem}
  \label{thm:alphaopt_multiperior_cvar}
  Consider an arbitrary $T$, and the pair of risk measures in~\eqref{eq:inconsistent_avar} and~\eqref{eq:consistent_avar}. Then, 
  \begin{enumerate}
  \item There is an algorithm that computes $\alphaopt{\cons}{\incons}$ in time $\mathcal{O}(N^{T^2})$.
  \item $\alphaopt{\incons}{\cons} = \max\Bigl\{ 1, \frac{\lev}{\prod_{t=1}^T \lev_t} \Bigl\}$, and the expression remains valid for an arbitrary scenario tree and reference measure.
  \end{enumerate}
\end{theorem}

It is interesting to note that computing $\alphaopt{\incons}{\cons}$, and hence also testing $\cons \leq \incons$, remains as easy for general $T$ as for $T=2$: an analytical expression is available, which actually holds in considerably more general settings (arbitrary tree and reference measure). By contrast, computing $\alphaopt{\cons}{\incons}$ and testing $\incons \leq \cons$ now requires an algorithm that is polynomial only for a fixed $T$. In light of our earlier result, this suggests that, although starting with lower-bounds for $\incons$ may lead to a tighter approximating $\cons$, the gain does not come for free, as the computation of the resulting $\alphaopt{\cons}{\incons}$ is typically harder than that for $\alphaopt{\incons}{\cons}$.

In a multiperiod setting, the question of designing the tightest possible lower-bounding approximation $\cons$ to a given $\incons$ becomes harder -- even computing one scaling factor $\alphaopt{\cons}{\incons}$ requires a polynomial-time algorithm. By contrast, a complete characterization of the tightest upper-bound $\cons$ is available! Quite surprisingly, it turns out that this choice exactly corresponds to the risk measure $\hat{\mu}_I$ introduced by the construction in Proposition~\ref{prop:varrho_construction}, by expanding the set of measures of $\incons$. This is summarized in the following result (for a proof, see~\citep{Huang_I_Petrik_Subraman_cvar_2012}).

\begin{theorem}
  \label{sec:varrho_for_cvar}
  Consider the risk measure $\incons = \cvar_\lev$, under an arbitrary reference measure $\Pr$, and the construction for the risk measure $\hat{\mu}_I$ characterized in Proposition~\ref{prop:varrho_construction}. Then, $\hat{\mu}_I = \cvar_1 \circ \cvar_2 \circ \dots \circ \cvar_T$, where $\cvar_t = (\hat{\mu}_i)_{i \in \nodes{t-1}}$, and 
  \begin{align*}
    \forall \, i \in \nodes{t-1}, ~~ \hat{\mu}_i =
    \begin{cases}
      \max, & \textup{if}~ \Pr(\desc{i}) \leq 1 - \lev \\
      \cvar_{\gamma_i}, & \textup{otherwise}.
    \end{cases}
  \end{align*}
  Here, $\gamma_i = \frac{\Pr(\desc{i}) - 1 + \lev}{\Pr(\desc{i})}$, and $\cvar_{\gamma_i}$ is computed under the conditional probability induced by $\Pr$, i.e., $\bigl(\frac{\Pr(\desc{j})}{\Pr(\desc{i})}\bigr)_{j \in \child{i}}$.
\end{theorem}

This result, which holds under any reference measure $\Pr$, suggests that starting with $\incons = \cvar_\lev$ and expanding its set of representing probability measures until it becomes rectangular exactly results in a risk measure $\hat{\mu}_I$ that is a composition of one-step $\cvar$s. These one-step $\cvar$s are computed under levels $\gamma_i$ that can be different at each node $i$ in the tree, and under the natural conditional probability induced by the reference measure $\Pr$. 

There are several immediate implications. First, since $\hat{\mu}_I$ is the tightest possible coherent upper-bound for any given coherent $\incons$ (see Lemma~\ref{lem:varrho_best_upper_bound}), this implies that the tightest possible choice for a compositional $\cvar$ that upper bounds a given $\cvar_\lev$ is exactly $\hat{\cvar}_\lev$. In a different sense, this also provides an instance when starting with a comonotonic (in fact, distortion) risk measure $\incons$ results in a comonotonic (distortion) risk measure $\hat{\mu}_I$, which furthermore belongs to the same class as $\incons$. 

Lastly, the theorem confirms that the best possible compositional $\cvar$ that upper bounds $\cvar_\lev$ does involve compositions with the worst-case operator, in any node $i$ that has probability at most $1 - \lev$. Furthermore, it suggests, in a precise sense, that the compositional $\cvar$ gets increasingly conservative as the risk measurement process proceeds in time: note that $\gamma_i \geq \gamma_j, \, \forall \, j \in \child{i}$, and once node $i$ requires a worst-case operator, so will any descendant of $i$, since $\Pr(\desc{i}) \geq \Pr(\desc{j}), \, \forall \, j \in \child{i}$. In particular, \emph{all} future stages are more conservative than the measurement at time $t=0$ (i.e., the root node), which exactly corresponds to the inconsistent evaluation $\incons = \cvar_\lev$.

This last point may be of particular relevance when designing risk measures for use in dynamic financial settings: it suggests that regulators looking for safe counterparts (i.e., upper-bounds) for a static $\cvar_\lev$ should use risk measurement processes that are compositions of increasingly conservative $\cvar_\lev$ measurements.

\section{Conclusions.}
\label{sec:conclusions}
In this paper, we examined two different paradigms for measuring risk in dynamic settings: a time-consistent formulation, whereby the risk assessments are designed so as to avoid na\"{i}ve reversals of preferences in the measurement process, and a time-inconsistent one, which is easier to specify and calibrate from preference data. We discussed necessary and sufficient conditions under which one measurement uniformly bounds the other from above or below, and provided a notion of the multiplicative tightness with which one measure can be approximated by the other. We also showed that it is generally hard to compute the scaling factors even for distortion risk measures, but provided concrete examples when polynomial-time algorithms are possible.

\section{Appendix}

\subsection{Submissives, Downward Monotone Closures and Anti-blocking Polyhedra.}
\label{sec:down_monot_closures}
In the current section, we discuss the important notion of the \emph{down monotone closure} of a polytope, also known as its \emph{anti-blocking polyhedron} or its \emph{submissive}. Our exposition mostly follows Chapter 9 in~\citet{Schriver_book_LP_IP}, to which we direct the interested reader for a more comprehensive treatment and references to related literature.

A polyhedron $Q$ in $\reals^n$ is said to be \emph{down-monotone} or of \emph{anti-blocking type} if
\begin{align*}
  Q \neq \emptyset, \, Q \subseteq \reals^n_{+}, \, \textup{and} \, 0 \leq \mb{y} \leq \mb{x} ~\textup{and}~ \mb{x} \in Q ~\textup{imply}~ \mb{y} \in Q.
\end{align*}
The following proposition summarizes a useful representation for down-monotone polyhedra.
\begin{proposition}
  \label{prop:representations_down_monotone_polyhedra}
  A polyhedron $Q$ in $\reals^n$ is down-monotone if and only if there is a finite set $\mathcal{I}$ of vectors $\{\mb{a}_i\}_{i \in \mathcal{I}}$ and coefficients $\{b_i\}_{i \in \mathcal{I}}$ such that $\mb{a}_i \geq 0, \, \mb{a}_i \neq 0, \, b_i \geq 0, \, \forall \, i \in \mathcal{I}$, and
    \begin{align*}
      Q = \bigl\{ \, \mb{x} \in \reals^n \suchthat \mb{a}_i \tr \mb{x} \leq b_i, \, \forall \, i \in \mathcal{I} \, \bigr\}.
      %\label{eq:down_montone_inequal_representation}
    \end{align*}
    % \end{itemize}
\end{proposition}
\begin{proof}{Proof.}
  The proof follows closely from the definitions. We omit it here, and direct the interested reader to~\citep{Schriver_book_LP_IP}. 
  \qed
\end{proof}

We remark that, whenever $Q$ is full-dimensional, the right-hand sides $b_i$ in the representation above can be taken to be strictly positive.

For any polyhedron $Q \subseteq \reals^n$, we can define its \emph{down-monotone closure}, also known as its \emph{submissive}, by
\begin{align}
  \sub{Q} \bydef \bigl\{ \, \mb{y} \in \reals^n_+ \suchthat \exists \, \mb{x} \in Q, \, \mb{x} \geq \mb{y} \,\bigr\}.
  \label{eq:submissive_definition}
\end{align}
It can be easily checked that $\sub{Q} = (Q + \reals^n_{-}) \cap \reals^n_{+}$, and that $\sub{Q}$ is full-dimensional if and only if $Q \mysetminus \{ \mb{x} \in \reals^n \suchthat x_j = 0 \} \neq \emptyset$, for all $j \in [1,n]$ (see \citet{Balas_Fischetti_97}). 
%A relevant characterization of $\sub{Q}$ can be derived from $Q$, by means of the following result.
% \begin{theorem}[Theorem~5.3 in \citet{Cunningham_Green_Krotki}]
%   Let $P = \bigl\{ \mb{x} \in \reals^n \suchthat A \mb{x} \leq \mb{b} \bigr\}$. Then, 
%   \begin{align*}
%     \sub{P} = \bigl\{ \mb{x} \in \reals^n_+ \,:\, \mb{w} \tr A \mb{x} \leq \scprod{w}{b}, ~\textup{for every}~ \mb{w}~ \textup{satisfying}~ \mb{w} \geq \mb{0}, \, \mb{w}\tr A \geq \mb{0} \bigr\}.
%   \end{align*}
% \end{theorem}
A very interesting characterization of the down-monotone closure of a polyhedron is also possible in terms of the polar of the polyhedron $P$. However, since these results are not directly needed in our treatment here, we direct the interested reader to \citep{Balas_Fischetti_97,Balas_Bockmayr_Pisaruk_Wolsey_2004} or Chapter 9 in \citep{Schriver_book_LP_IP} for more details.

% Under appropriate conditions, optimizing a linear function over a polyhedron $Q$ is equivalent to optimizing a related linear function over the downward monotone closure of $Q$. The following result, which follows this paradigm, is very useful in our analysis.
% \begin{theorem}
%   \label{thm:worst_case_same_on_down_closure}
%   For any polyhedron $Q \subseteq \reals^n_+$ and for any $\mb{w} \geq 0$, we have
%   \begin{align*}
%     \max_{\mb{q} \in Q} \scprod{w}{q} = \max_{\mb{q} \in \sub{Q}} \scprod{w}{q}.
%   \end{align*}
% \end{theorem}
% \begin{proof}{Proof.}
%   Clearly, since $Q \subseteq \sub{Q}$, the left side is always at most equal to the right side. To argue the reverse, note that for any $\mb{x} \in \sub{Q}$, there exists $\mb{q} \in Q$ satisfying $\mb{q} \geq \mb{x} \geq 0$, which implies $\scprod{w}{x} \leq \scprod{w}{q}, \, \forall \, \mb{w} \geq 0$. Since this is true for arbitrary $\mb{x}$, the reverse inequality must also hold.  \qed
% \end{proof}

Down-monotone polyhedra have been used for studying the strength of relaxations in integer programming and combinatorial optimization -- see \citet{Goemans_Hall_1996} are references therein. The following result is relevant for our purposes.
\begin{theorem}
 \label{thm:goemans_hall_strength_valid_ineq}
  Let $P$ and $Q$ be two nonempty, downward monotone polytopes in $\reals^n_{+}$. Then
  \begin{enumerate}
  \item $P \subseteq \alpha \, Q$ if and only if, for any nonnegative vector $\mb{w} \in \reals^n$,
    \begin{align*}
      \max \, \{ \scprod{w}{x} \suchthat \mb{x} \in Q \} \geq \frac{1}{\alpha} \max \, \{ \scprod{w}{x} \suchthat \mb{x} \in P \}.
    \end{align*}
  % \item Letting $\alpha^{\opt}$ denote the minimum value of $\alpha$ such that $P \subseteq \alpha \, Q$, we have
  %   \begin{align*}
  %     \alpha^{\opt} = \sup_{\mb{w} \geq 0} \frac{\max \, \{\scprod{w}{x} \,:\, \mb{x} \in P\}}{\max \, \{\scprod{w}{x} \,:\, \mb{x} \in Q\}},
  %   \end{align*}
  %   where, by convention, $\frac{0}{0} = 0$.
  \item If $Q = \bigl\{ \mb{x} \in \reals^n_{+} \,:\, \mb{a}_i \tr \mb{x} \leq b_i, \, \forall \, i \in \mathcal{I} \bigr\}$, where $\mb{a}_i, \, b_i \geq 0$, then
    \begin{align*}
      \alpha^{\opt} = \max_{i \in \mathcal{I}} \, \frac{d_i}{b_i}, ~\textup{where}~ d_i \bydef \max_{\mb{x} \in P} \, \mb{a}_i \tr \mb{x}.
    \end{align*}
  \end{enumerate}
\end{theorem}
\begin{proof}{Proof.}
  Part (1) is essentially Lemma~1 in~\citep{Goemans_Hall_1996}. Since the latter reference omits a proof, we include one below, for completeness. ``$\Rightarrow$'' follows trivially. ``$\Leftarrow$'' Note first that $\alpha > 0$. Assume (by contradiction) that $\exists \, \bar{\mb{x}} \in  P  \mysetminus \alpha \, Q$. Since $Q$ is down-monotone, by Proposition~\ref{prop:representations_down_monotone_polyhedra}, it can be written as $Q = \{ \, \mb{x} \in \reals^n_{+} \,:\, \mb{a}_i \tr \mb{x} \leq b_i, \, \forall \, i \in \mathcal{I} \, \}$, where $\mb{a}_i, \, b_i \geq 0, \, \forall \, i \in \mathcal{I}$. Since $\bar{\mb{x}} \notin \alpha \, Q$, there exists $j \in \mathcal{I}$ such that $\mb{a}_j \tr \bar{\mb{x}} > \alpha \, b_j$. Since $\bar{\mb{x}} \in P$, we obtain the desired contradiction, $\frac{1}{\alpha} \max \, \{ \mb{a}_j \tr \mb{x} \suchthat \mb{x} \in P \} \geq \frac{1}{\alpha} \mb{a}_j \tr \bar{\mb{x}} > b_j \geq \max \, \{ \mb{a}_j \tr \mb{x} \suchthat \mb{x} \in Q \}$.

 % Part (2) follows as an immediate corollary of Part (1).

  Part (2) is exactly Theorem 2 in~\citep{Goemans_Hall_1996}, to which we direct the reader for a complete proof.
  \qed
\end{proof}

The above result shows that $\alpha^{\opt}$ can be $+\infty$, which is the case if $Q$ has a strictly smaller dimension than $P$ (in this case, some $b_i$ are $0$, while the corresponding $d_i$ are strictly positive \citep{Schriver_book_LP_IP}). However, if $Q$ is full-dimensional, $\alpha^{\opt}$ is always finite.

%
%  Submodular functions
%
\subsection{Submodular Functions and Polymatroids}
\label{sec:subm-polyh-polym}
In this section of the Appendix, we discuss the basic properties of Choquet capacities in light of their connection with rank functions of polymatroids. The exposition is mainly based on volume B of \citep{Schrijver_Alex_book_2003} (Chapter 44) and Chapter 2 of \citep{Fujishige_book_05} (Section 3.3), to which we direct the interested reader for more information.

Consider a ground set $\Omega$ with $|\Omega| = n$, and let $c$ be a set function on $\Omega$, that is, $c : \filt \mapsto \reals$, where $\filt = 2^{\Omega}$ is the set of all subsets of $\Omega$. The function $c$ is called \emph{submodular} if
\begin{align*}
  c(T) + c(U) \geq c(T \cap U) + c(T \cup U), \, \forall \, T, U \in \filt.
\end{align*}
The function $c$ is called \emph{nondecreasing} if $c(T) \leq c(U)$ whenever $T \subseteq U \subseteq \Omega$. For a given set function $c$ on $\Omega$, we define the following two polyhedra
\begin{equation}
  \label{eq:definition_polymatroid_extended_polymatroid}
  \begin{aligned}
    \polymat{c} &\bydef \bigl\{\, \mb{x} \in \reals^{|\Omega|} \suchthat \mb{x} \geq 0, \, \mb{x}(S) \leq c(S), \, \forall \, S \subseteq \Omega \,\bigr\} \\
    \extpolymat{c} &\bydef \bigl\{\, \mb{x} \in \reals^{|\Omega|} \suchthat \mb{x}(S) \leq c(S), \, \forall \, S \subseteq \Omega \,\bigr\}.
  \end{aligned}
\end{equation}
Note that $\polymat{c}$ is nonempty if and only if $c \geq 0$, and that $EP_c$ is nonempty if and only if $c(\emptyset) \geq 0$. These conditions are trivially satisfied in our exposition, since all set functions $c$ of interest are Choquet capacities, i.e., by Definition~\ref{def:choquet_capacities}, they are are nondecreasing and normalized, $c(\emptyset) = 0, \, c(\Omega) = 1$. %Therefore, throughout the remainder of the discussion here, we always assume that

If $c$ is a submodular function, then $\polymat{c}$ is called the \emph{polymatroid associated with $c$}, and $\extpolymat{c}$ the \emph{extended polymatroid associated with $c$}. Note that a nonempty extended polymatroid is always unbounded, while a polymatroid is always a polytope, since $0 \leq x_i \leq c(\{i\}), \, \forall \, i \in \Omega$. The next theorem provides a very useful result concerning the set of tight constraints in the representation of $\extpolymat{c}$.
\begin{theorem}[Theorem 44.2 in \citep{Schrijver_Alex_book_2003}.]
  \label{thm:closed_union_intersection}
  Let $c$ be a submodular set function on $\Omega$ and let $\mb{x} \in \extpolymat{c}$. Then the collection of sets $U \subseteq \Omega$ satisfying $\mb{x}(U) = c(U)$ is closed under taking unions and intersections.
\end{theorem}
\begin{proof}{Proof.}
  Suppose $\mb{x}(T) = c(T)$ and $\mb{x}(U) = c(U)$. Then
  \begin{align*}
    c(T) + c(U) \geq c(T \cap U) + c(T \cup U) \geq  \mb{x}(T \cap U) + \mb{x}(T \cup U) = \mb{x}(T) + \mb{x}(U) = c(T) + c(U),
  \end{align*}
  hence equality most hold throughout, and $\mb{x}(T \cap U) = c(T \cap U)$ and $\mb{x}(T \cup U) = c(T \cup U)$.
  \qed
\end{proof}

A vector $\mb{x} \in \extpolymat{c}$ (or in $\polymat{c}$) is called a \emph{base vector} of $\extpolymat{c}$ (or of $\polymat{c}$) if $\mb{x}(\Omega) = c(\Omega)$. The set of all base vectors is called the \emph{base polytope} of $c$ and is denoted by $\basepoly{c}$,
\begin{align*}
  \basepoly{c} \bydef \bigl\{\, \mb{x} \in \reals^{|\Omega|} \suchthat \mb{x}(S) \leq c(S), \, \forall \, S \subseteq \Omega, \, \mb{x}(\Omega) = c(\Omega) \,\bigr\}.
\end{align*}
The following theorem summarizes several simple properties of $\basepoly{c}$, and its relation to $\extpolymat{c}$ and $\polymat{c}$.
\begin{theorem}
  \label{thm:properties_base_poly}
  For any submodular function $c$ satisfying $c(\emptyset) = 0$,
  \begin{itemize}
  \item[(i)] $\basepoly{c}$ is a face of $\extpolymat{c}$, and is always a polytope.
  \item[(ii)] $\extpolymat{c} = \basepoly{c} + \reals^n_{-}$, so that $\extpolymat{c}$ and $\basepoly{c}$ have the same extreme points.
  \item[(iii)] $\polymat{c} = \sub{\basepoly{c}}$.
  \item[(iv)] For any $\lambda \geq 0$, $\basepoly{\lambda c} = \lambda \cdot \basepoly{c}$, $\extpolymat{\lambda c} = \lambda \cdot \extpolymat{c}$, and $\polymat{\lambda c} = \lambda \cdot \polymat{c}$.
  \end{itemize}
\end{theorem}
\begin{proof}{Proof.}
  (i) The fact that $\basepoly{c}$ is a face of $\extpolymat{c}$ follows directly from the definitions. To see that $\basepoly{c}$ is a polytope, note that, for any $i \in \Omega$, $x_i \leq c(\{i\})$, and $x_i = \mb{x}(\Omega) - \mb{x}(\Omega \mysetminus \{i\}) \geq c(\Omega) - c(\Omega \mysetminus \{i\})$.

  (ii) ``$\supseteq$'' Follows trivially. ``$\subseteq$'' Consider any $\mb{y} \in \extpolymat{c}$. Without loss of generality\footnote{Such a $\mb{y}$ can always be obtained by adding a certain $\mb{\xi} \geq 0$, and if the resulting $\mb{y} + \mb{\xi} \in \basepoly{c} + \reals^n_{-}$, then also $\mb{y} \in \basepoly{c} + \reals^n_{-}$.}, assume $\mb{y}$ does \emph{not} lie in the strict interior of $\extpolymat{c}$, and let $\mathcal{I}_{\mb{y}} \bydef \bigl\{ S \in \filt \suchthat \mb{y}(S) = c(S) \bigr\}$ denote the collection of sets corresponding to tight constraints at $\mb{y}$. If $\Omega \in \mathcal{I}_{\mb{y}}$, then $\mb{y} \in \basepoly{c}$, and the proof would be complete. Therefore, let us assume $\Omega \notin \mathcal{I}_{\mb{y}}$.

  We claim that there exists $s \in \Omega$ such that $s \notin S, \, \forall \, S \in \mathcal{I}_{\mb{y}}$. To see this, note that, if any $s \in \Omega$ were contained in some $S \in \mathcal{I}_{\mb{y}}$, then $\Omega \in \mathcal{I}_{\mb{y}}$, since the set of tight constraints is closed under union and intersection, by Theorem~\ref{thm:closed_union_intersection}. We can then consider the vector $\mb{y}_{\lambda} = \mb{y} + \lambda \, \oneN{s}$ for $\lambda \geq 0$. It is easy to test that, for small enough $\lambda$, $\mb{y}_{\lambda} \in \extpolymat{c}$. By making $\lambda$ sufficiently large, at least one constraint a set $S$ containing $s$ becomes tight, hence enlarging the set $\mathcal{I}_{\mb{y}}$. Repeating the argument for the point $\mb{y}_{\lambda}$ recursively, we eventually recover a vector $\tilde{\mb{y}}$ that belongs to $\basepoly{c}$. Since $\tilde{\mb{y}} = \mb{y} + \mb{\xi}$ for some $\mb{\xi} \geq 0$, we have that $\mb{y} \in \basepoly{c} + \reals^{n}_{-}$, which completes the proof of the first part of (ii). Since $\reals^n_{-}$ is a cone, and $\basepoly{c}$ is a polytope, the representation exactly corresponds to the Motzkin decomposition of an arbitrary polyhedron, so that $\ext(\extpolymat{c}) = \ext(\basepoly{c})$.

  (iii) Follows immediately from (ii), since $\polymat{c} = \extpolymat{c} \cap \reals^n_{+} = \bigl( \basepoly{c} + \reals^n_{-} \bigr) \cap \reals^n_{+} \bydef \sub{\basepoly{c}}$.

  (iv) Since $\lambda \, c$ is also submodular, the results immediately follow from the definitions.
  \qed
\end{proof}

A central result in the theory of submodularity, due to Edmonds, is that a linear function $\scprod{w}{x}$ can be optimized over an (extended) polymatroid by an extension of the greedy algorithm. The following theorem summarizes the finding.
\begin{theorem}[Theorem~44.3, Corollaries~44.3(a,b) in~\citep{Schrijver_Alex_book_2003}.]
  \label{thm:greedy_evaluation_comonotonic_case}
  Let $c : 2^{\Omega} \rightarrow \reals$ be a submodular set function with $c(\emptyset) = 0$, and let $\mb{w} \in \reals^{|\Omega|}_{+}$. Then the optimum solution of $\max_{\mb{x} \in \extpolymat{c}} \, \scprod{w}{x}$ is
  \begin{align*}
    \mb{x}(s_i) \bydef c\bigl( \{s_1,\dots,s_i\} \bigr) - c\bigl( \{s_1,\dots,s_{i-1}\} \bigr), \, i \in [1,n],
    % \label{eq:optimal_greedy_solution}
  \end{align*}
  where $(s_1,\dots,s_n)$ is a permutation of the elements of $\Omega$ such that $\mb{w}(s_1) \geq \mb{w}(s_2) \geq \dots \mb{w}(s_n)$. If $c$ is also nondecreasing, then the above $\mb{x}$ is also an optimal solution to the problem $\max_{\mb{x} \in \polymat{c}} \, \scprod{w}{x}$.
\end{theorem}
\begin{proof}{Proof}
  The proof follows by duality arguments. We omit it here, and direct the interested reader to \citep{Schrijver_Alex_book_2003}. \qed
\end{proof}
In view of this result, the following characterization for the extreme points of $\basepoly{c}, \, \extpolymat{c}$ and $\polymat{c}$ is immediate.
\begin{theorem}
  \label{thm:extreme_points_base_polytope}
  For a submodular set function $c$ satisfying $c(\emptyset) = 0$, the extreme points of $\basepoly{c}$ and $\extpolymat{c}$ are given by
  \begin{align*}
    x_{\sigma(i)} = c\bigl( \{ \sigma(1),\dots,\sigma(i) \} \bigr) - c\bigl( \{ \sigma(1),\dots,\sigma(i-1) \} \bigr), \, i \in [1,n],
  \end{align*}
  where $\sigma \in \Pi(\Omega)$ is any permutation of the elements of $\Omega$. When $c$ is also nondecreasing, the extreme points of $\polymat{c}$ are given by
  \begin{align*}
    x_{\sigma(i)} =
    \begin{cases}
      ~ c\bigl( \{ \sigma(1),\dots,\sigma(i) \} \bigr) - c\bigl( \{ \sigma(1),\dots,\sigma(i-1) \} \bigr) & \textup{if}~ i \leq k, \\
      ~ 0 & \textup{if}~ i > k,
    \end{cases}
  \end{align*}
  where $\sigma \in \Pi(\Omega)$ is any permutation of the elements of $\Omega$, and $k$ ranges over $[0,n]$.
\end{theorem}
\begin{proof}{Proof.}
  For a complete proof, we direct the reader to Theorem 3.22 in \citep{Fujishige_book_05} and Section 44.6c in \citep{Schrijver_Alex_book_2003}.
\end{proof}
The previous result shows that there is a one-to-one correspondence between vertices of $\basepoly{c}$ and permutations of $[1,n]$, and also that every inequality constraint in the characterization of $\basepoly{c}$ is tight at some $\mb{x} \in \basepoly{c}$. The following corollary also immediately follows from the above result.

\begin{corollary}
  \label{corol:nonnegative_base_poly}
  For any submodular $c$ such that $c(\emptyset) = 0$, $\basepoly{c} \subset \reals^n_{+}$ if and only if $c$ is nondecreasing.
\end{corollary}
\begin{proof}{Proof.}
  ``$\Leftarrow$'' Immediate, since $\basepoly{c}$ is the convex hull of its extreme points, which (by Theorem~\ref{thm:extreme_points_base_polytope}) are nonnegative. ``$\Rightarrow$'' Consider any two sets $T \subset U \subseteq \Omega$, and take a chain of sets $S_1 \subset S_2 \subset \dots \subset S_{|U \setminus T|}$ such that $S_1 = T$ and $S_{|U \setminus T|} = U$. By Theorem~\ref{thm:extreme_points_base_polytope}, there exists an extreme point $\mb{x}$ of $\basepoly{c}$ having elements $c(S_{i+1}) - c(S_i), \, i \in [1,\, |U \setminus T| - 1]$ among some of its coordinates. Since $\mb{x} \geq 0$, we immediately obtain that $c(U) - c(T) \geq 0$.
\end{proof}

\subsection{Technical Proofs.}
\label{sec:technical-proofs}
This section contains several technical results from our analysis.

%
%  Representation theorem for consistent measure; characterization of \qcons and its \sub{}
%
\begin{theorem}[Proposition~\ref{prop:rep_comp_measure}.]
   Consider a (two-period) consistent, comonotonic risk measure $\cons(Y) = \mu_1 \circ \mu_2$, where $\mu_t : \rvspace_{t} \rightarrow \rvspace_{t-1}$. Then,
  \begin{enumerate}
  \item There exists $\qcons \subseteq \Delta^{|\nodes{2}|}$ such that $\cons(Y) \bydef \max_{\mb{q} \in \qcons} \, \scprod{q}{Y}, \, \forall \, Y \in \rvspace_2$.
  \item  The set of measures $\qcons$ is given by
    \begin{align*}
      \qcons &\bydef \biggl\{ \mb{q} \in \Delta^{|\nodes{2}|} \suchthat \exists \, \mb{p} \in \Delta^{|\nodes{1}|}, \quad
      \begin{aligned}
        \mb{p}(S) &\leq c_1(S), \, \forall \, S \subseteq \nodes{1} \\
        \mb{q}(U) & \leq p_i \cdot c_{2|i}(U), \, \forall \, U \subseteq \child{i}, \, \forall \, i \in \nodes{1}
      \end{aligned}
      \biggr\}  \\
      & \equiv \Bigl\{ \, \mb{q} \in \Delta^{|\nodes{2}|} \suchthat \exists \, \mb{p} \in \basepoly{c_1} \suchthat \mb{q}\vert_{\child{i}} \in \basepoly{p_i \cdot c_{2|i}}, \, \forall \, i \in \nodes{1} \, \Bigr\},
    \end{align*}
    where $c_1 : 2^{|\nodes{1}|} \rightarrow \reals$ and $c_{2|i} : 2^{|\child{i}|} \rightarrow \reals, \, \forall \, i \in \nodes{1}$ are Choquet capacities, and $\basepoly{c_1}, \basepoly{c_{2|i}}$ are the base polytopes corresponding to $c_1$ and $c_{2|i}$, respectively.
  \item The downward monotone closure of $\qcons$ is given by
    \begin{align*}
      \sub{\qcons} & \bydef \biggl\{ \mb{q} \in \spacenodes{2}_+ \,:\, \exists \, \mb{p} \in \spacenodes{1}_+, \quad
      \begin{aligned}
        \mb{p}(S) &\leq c_1(S), \, \forall \, S \subseteq \nodes{1}, \\
        \mb{q}(U) & \leq p_i \cdot c_{2|i}(U), \, \forall \, U \subseteq \child{i}, \, \forall \, i \in \nodes{1}
      \end{aligned}
      \biggr\} \\
      &= \Bigl\{ \mb{q} \in \spacenodes{2}_+ \,:\, \exists \, \mb{p} \in \polymat{c_1} \suchthat \mb{q}\vert_{\child{i}} \in \polymat{p_i \cdot c_{2|i}}, \, \forall \, i \in \nodes{1} \Bigr\}
    \end{align*}
    where $\polymat{c_1}$ and $\polymat{p_i c_{2|i}}$ are the polymatroids associated with $c_1$ and $p_i c_{2|i}$, respectively.
  \end{enumerate}
\end{theorem}
%\end{theorem}
\begin{proof}{Proof.}
  The first claim is a standard result in the literature \citep{Epstein_Schneid_03,delbaen_multiperiod_2007,Roorda_Schum_Engw_2005_coherent}, but we rederive it here together with the second claim, to keep the paper self-contained.
  To this end, recall that Definition~\ref{def:dynamic_consist_distortion_risk_measures} implies any $\dc$ comonotonic risk measure $\cons$ can be written as $\mu_1 \circ \mu_2$, where $\mu_1 : \rvspace_{1} \rightarrow \reals$ is a first-period comonotonic risk measure, and $\mu_2 \equiv (\mu^i)_{i \in \nodes{1}}$, where $\mu^i : \spacechild{i} \rightarrow \reals, \, \forall \, i \in \nodes{1}$ are comonotonic risk measures. By the representation in Theorem~\ref{thm:representation_comonotonic_rm}, for any $X_1 \in \rvspace_{1}$ and $X_2 \in \rvspace_{2}$, we have
\begin{subequations}
  \begin{align}
    \mu_1(X_1) & = \max_{\mb{p} \in \qcondit{1}} \, \scprod{p}{X}_1, \quad
    && \qcondit{1} \bydef \bigl\{ \, \mb{p} \in \Delta^{|\nodes{1}|} \suchthat \mb{p}(S) \leq c_1(S), \, \forall \, S \subseteq \nodes{1} \, \bigr\}, \label{eq:mu_first_stage_measure_set} \\
    \mu_2^i(X_2) &= \max_{\mb{q} \in \qcondit{2|i}} \, \scprod{q}{X}_2, \quad
    && \qcondit{2|i} \bydef \bigl\{ \, \mb{q} \in \Delta^{|\nodes{2}|} \suchthat \mb{q}(U) \leq c_{2|i}(U), \, \forall \, U \subseteq \child{i}; ~ \mb{q}\vert_{\nodes{2} \setminus \child{i}} = 0\, \bigr\}. \label{eq:mu_second_stage_measure_set}
  \end{align}
\end{subequations}

 In particular, $\mathcal{Q}_1 \equiv \basepoly{c_1}$, and, similarly, the projection of the polytope $\mathcal{Q}_{2|i}$ on the coordinates $\child{i}$ is exactly given by $\basepoly{c_{2|i}}$, for any $i \in \nodes{1}$.
  From these relations, we have that $\cons(Y) = \max_{\mb{q} \in \tilde{Q}} \, \scprod{q}{Y}$, where $\tilde{Q}$ has the following product form structure \citep{Shapiro_Ruszczynski_Dentcheva_2009_Stochastic_Prog}:
  \begin{align}
    \tilde{Q} = \Bigl\{ \mb{q} \in \Delta^{|\nodes{2}|} \suchthat \exists \, \mb{p} \in \qcondit{1}, \, \exists \, \mb{q}^i \in \qcondit{2|i}, \, \forall \, i \in \nodes{1},\, \textup{such that}\, \mb{q} = \sum_{i \in \nodes{1}} p_i \mb{q}^i \Bigr\}.
    \label{eq:qconsistent_product_form}
  \end{align}
  We now show that $\tilde{Q} = \qcons$, by double inclusion. \\
  \noindent  ``$\subseteq$'' Consider any $\mb{q} \in \tilde{Q}$, and let $\mb{p} \in \qcondit{1}$ and $\mb{q}^i \in \qcondit{2|i}$ denote the corresponding vectors in representation~\eqref{eq:qconsistent_product_form}. Since $\mb{q}_{\child{i}} = p_i \cdot \mb{q}^i, \, \forall \, i \in \nodes{1}$, and $\mb{q}^i \in \qcondit{2|i}, \, \forall \, i \in \nodes{1}$, we trivially have that $\mb{p}$ and $\mb{q}$ satisfy the equations defining $\qcons$. \\
  \noindent ``$\supseteq$'' Consider any $\mb{q} \in \qcons$, and let $\mb{p}$ be a corresponding measure satisfying the constraints for $\qcons$. It can be readily checked that $\mb{q}^i \bydef \frac{\mb{q}_{\child{i}}}{p_i} \in \qcondit{2|i}$ (the only non-obvious constraint is $\mb{1}\tr \mb{q}^i = 1$, which must hold, since, otherwise, we would have $\sum_{i \in \nodes{1}} \mb{q}(\child{i}) = \sum_{i \in \nodes{1}} p_i\, \mb{1}\tr \mb{q}^i < \mb{p}(\nodes{1}) = 1$, contradicting $\mb{q} \in \Delta^{|\nodes{2}|}$). Therefore, $\mb{q} = \sum_{i \in \nodes{1}} p_i \, \mb{q}^i \in \tilde{Q}$. For completeness, we also note that $\mb{q}^i = \frac{\mb{q}_{\child{i}}}{p_i} \in \qcondit{2|i} \Leftrightarrow \frac{\mb{q}\vert_{\child{i}}}{p_i} \in \basepoly{c_{2|i}} \Leftrightarrow\mb{q}\vert_{\child{i}} \in \basepoly{p_i \cdot c_{2|i}}$, for any $i \in \nodes{1}$ (by part (iv) of Theorem~\ref{thm:properties_base_poly} in Section~\ref{sec:subm-polyh-polym} of the Appendix).

  To prove the last claim, note that the two sets on the right being identical is immediate from the definition of the polymatroid associated with a rank function $c$ (see Section \ref{sec:subm-polyh-polym}). As such, denote by $\mathcal{A}$ the set on the right of the equation.\\
   \indent ``$\subseteq$''. Consider an arbitrary $\mb{x} \in \sub{\qcons}$. By definition, $\mb{x} \geq 0$ and $\exists \, \mb{q} \in \qcons$ such that $\mb{q} \geq \mb{x}$. Let $\mb{p}$ correspond to $\mb{q}$ in the representation for $\qcons$. To argue that $\mb{x} \in \mathcal{A}$, we show that the pair $(\mb{p}, \mb{x})$ satisfies all the constraints defining $\mathcal{A}$. To this end, since $\mb{p} \in \basepoly{c_1}$ (and $\basepoly{c_1} \subset \reals^{|\Omega_1|}_{+}$), we immediately have $\mb{p} \in \polymat{c_1}$. Furthermore, $\forall \, i \in \nodes{1}$ and $\forall \, U \subseteq \child{i}$, we have $\mb{x}(U) \leq \mb{q}(U) \leq p_i \cdot c_{2|i}(U)$, which proves that $\mb{x} \in \mathcal{A}$.

   ``$\supseteq$''. Consider an arbitrary $\mb{q} \in \mathcal{A}$, and let $\mb{p}$ be such that the pair $(\mb{p}, \, \mb{q})$ satisfies all the constraints defining $\mathcal{A}$. Since $\mb{p} \in \polymat{c_1} \equiv \sub{ \basepoly{c_1} }$, $\exists \, \bar{\mb{p}} \in \basepoly{c_1}$ such that $\bar{\mb{p}} \geq \mb{p} \geq 0$. %In particular, $\bar{\mb{p}}(\nodes{1}) = c_1(\nodes{1}) = 1$, which implies $\bar{\mb{p}} \in \Delta^{|\nodes{1}|}$.
%The pair $(\bar{\mb{p}}, \mb{q})$ also satisfies all the constraints~\eqref{eq:down_closure_consist_set}. In particular, 
Furthermore, $\mb{q}\vert_{\child{i}} \in \polymat{\bar{p}_i \cdot c_{2|i}} \equiv \sub{\basepoly{\bar{p}_i \cdot c_{2|i}}}$, for any $i \in \nodes{1}$. Therefore, $\exists \, \bar{\mb{q}} \in \spacenodes{2}_+$ such that $\bar{\mb{q}}\vert_{\child{i}} \in \basepoly{\bar{p}_i \cdot c_{2|i}}$ and $\bar{\mb{q}}\vert_{\child{i}} \geq \mb{q} \vert_{\child{i}} \geq 0$, for any $i \in \nodes{1}$. %In particular, by (i), $\bar{\mb{q}}\vert_{\child{i}}(\child{i}) = \bar{p}_i \cdot c_{2|i}(\child{i}) = \bar{p}_i, \, \forall \, i \in \nodes{1}$.
It can be readily checked that, by construction, the pair $(\bar{\mb{p}}, \, \bar{\mb{q}})$ satisfies all the constraints defining $\qcons$. Therefore, with $\bar{\mb{q}} \in \qcons$ and $\bar{\mb{q}} \geq \mb{q} \geq 0$, we must have $\mb{q} \in \sub{\qcons}$.
\qed
\end{proof}

%
%  Equivalent conditions for \qcons \subseteq \qincons (and vice-versa)
% 
\begin{theorem}[Corollary~\ref{cor:lower_upper_bounds_explicit}.]
  For any pair of risk measures $\incons$ and $\cons$ as introduced in Section~\ref{sec:charact_qc_qi_subqc_subqi},
  \begin{enumerate}
  \item The inequality $\cons(Y) \leq \incons(Y), \, \forall \, Y \in \rvspace_2$ holds if and only if
    \begin{align*}
      \sum_{j = 1}^{|\nodes{1}|} \, \Bigl[ c_1\bigl(\cup_{k=1}^j s_k \bigr) - c_1\bigl(\cup_{k=1}^{j-1} s_k \bigr) \Bigr] \cdot c_{2|s_j}(U_{s_j}) \leq c \bigl( \cup_{i \in \nodes{1}} U_i \bigr),
    \end{align*}
    where $(s_1,\dots,s_{|\nodes{1}|})$ denotes any permutation of the elements of $\nodes{1}$, and $U_i \subseteq \child{i}$ for any $i \in \nodes{1}$.
  \item The inequality $\incons(Y) \leq \cons(Y), \, \forall \, Y\in\rvspace_{2}$ holds if and only if 
    \begin{align*}
      c\bigl( \cup_{i \in S} \child{i} \bigr) &\leq c_1(S), ~~ \forall \, S \subseteq \nodes{1}, \\
      \frac{c(U)}{c(U) + 1 - c(\nodes{2} \setminus \child{i} \cup U)} &\leq c_{2,i}(U), \, \, \forall \, U \subseteq \child{i}, \, \forall \, i \in \nodes{1}.
    \end{align*}
  \end{enumerate}
\end{theorem}

\begin{proof}{Proof.}
  The main idea proof behind the proof is to rewrite the results in Corollary~\ref{cor:lower_upper_bounds} in terms of the extreme points of $\qcons$ and $\qincons$, and then to suitably simplify the resulting problems.
  
  To prove part (1), by Corollary~\ref{cor:lower_upper_bounds}, we have that $\cons(Y) \leq \incons(Y), \, \forall \, Y \in\rvspace_{2}$ holds if and only if
  \begin{align*}
    \max_{\mb{q} \in \ext(\qcons)} \mb{q}(S) \leq c(S), \, \forall \, S \subseteq \nodes{2}.
  \end{align*}
  To this end, consider any $S \subseteq \nodes{2}$, and partition it as $S = \cup_{\ell \in \nodes{1}} U_\ell$, for some $U_\ell \subseteq \child{\ell}, \, \forall \, \ell \in \nodes{1}$. The expression for $\ext(\qcons)$ is given in Proposition~\ref{prop:extreme_points}, which we paste below for convenience
  \begin{align*}
    q_{\sigma_{\ell}(i)} = \Bigl[ c_1 \bigl(\cup_{k=1}^\ell \pi(k) \bigr) - c_1\bigl(\cup_{k=1}^{\ell-1} \pi(k) \bigr) \Bigr] \cdot \Bigl[ c_{2|\ell} \bigl(\cup_{k=1}^i \sigma_\ell(k) \bigr) - c_{2|\ell}\bigl(\cup_{k=1}^{i-1} \sigma_\ell(k) \bigr) \Bigr] , \, \forall \, i \in [1,|\child{\ell}|], \forall \, \ell \in\nodes{1}.
  \end{align*}
    where $\pi$ is any permutation of $\nodes{1}$, and $\sigma_\ell$ is any permutation of $\child{\ell}$, for each $\ell \in \nodes{1}$. 

    Consider a fixed permutation $\pi \in \nodes{1}$. We claim that the permutation $\sigma_\ell$ yielding a maximal value of $\mb{q}(U_\ell)$ is always of the form $\bigl( \sigma(U_\ell), \, \sigma(\child{\ell} \setminus U_\ell) \bigr)$, i.e., it has the elements of $U_\ell$ in the first $|U_\ell|$ positions. This is because the functions $c_{2|\ell}$ are submodular, so that $c(U_\ell) - c(\emptyset) \geq c(U_\ell \cup A) - c(A)$, for any $A \subseteq \child{\ell} \setminus U_\ell$. With this recognition, the optimal permutations $\sigma_\ell$ always result in $\mb{q}(U_\ell) = \Bigl[ c_1 \bigl(\cup_{k=1}^\ell \pi(k) \bigr) - c_1\bigl(\cup_{k=1}^{\ell-1} \pi(k) \bigr) \Bigr] \, c_{2|\ell} (U_\ell)$. Maximizing over all permutations $\pi \in \Pi(\nodes{1})$ then leads to the first set of desired conditions.

    To prove part (2), one can use the expression from Corollary~\ref{cor:lower_upper_bounds}, and show that it reduces to the desired condition. Instead, we find it more convenient to work with the results of Lemma~\ref{lem:varrho_best_upper_bound} concerning $\widehat{\incons}$, the tightest possible coherent upper bound to $\incons$. To this end, first recall the representation for $\qcons$ in Proposition~\ref{prop:rep_comp_measure}, pasted below for convenience:
    \begin{align*}
      \qcons = \Bigl\{ \, \mb{q} \in \Delta^{|\nodes{2}|} \suchthat \exists \, \mb{p} \in \basepoly{c_1} \suchthat \mb{q}\vert_{\child{i}} \in \basepoly{p_i \cdot c_{2|i}}, \, \forall \, i \in \nodes{1} \, \Bigr\}.
    \end{align*}
    Lemma~\ref{lem:varrho_best_upper_bound} implies that that $\incons(Y) \leq \cons(Y) ~\Leftrightarrow ~ \widehat{\qcal}^i_{\incons} \subseteq \basepoly{c_{t|i}}, \, \forall\, i \in \nodes{t-1}, \, \forall \, t \in [1,2]$. Here, $\widehat{\qcal}^i_{\incons}$ are the one-step conditional risk measures yielding $\widehat{\incons}$, and are given by~\eqref{eq:varrho_sets_of_measures}. This is equivalent to
\begin{align*}
  \max_{\mb{q} \in \qincons} \mb{q}\bigl( \cup_{i \in S} \child{i} \bigr) &\leq c_1(S), \, \forall \, S \subseteq \nodes{1}  && (*)\\
  \max_{\mb{q} \in \qincons \,:\, \mb{q}(\child{i}) \neq 0} \frac{\mb{q}(U)}{\mb{q}(\child{i})} &\leq c_{2|i}(U), \, \forall \, U \subseteq \child{i}, \, \forall \, i \in \nodes{1} && (**)
\end{align*}
We now argue that $(*)$ and $(**)$ are equivalent to the conditions in part (2). Recalling the description of $\qincons$ in Proposition~\ref{prop:incon_rep_thm}, and the fact that any inequality $\mb{q}(S) \leq c(S)$ is tight at some set $S$ (also see Theorem~\ref{thm:extreme_points_base_polytope}), it can be seen that the maximum value of $\mb{q}(\cup_{i \in S} \child{i})$ in $(*)$ is exactly $c(\cup_{i \in S} \child{i})$, which yields the first desired condition. The proof that $(**)$ are equivalent to the second condition is the subject of Proposition~\ref{prop:properties_maximiz_ratio} below, which completes our proof. \qed
\end{proof}

\begin{proposition}
  \label{prop:properties_maximiz_ratio}
  Consider any $i \in \nodes{t-1}$ for some $t \in [1,T]$. Then, for any $U \subseteq \child{i}$, we have
  \begin{align}
    \max_{\mb{q} \in \qincons \,:\, \mb{q}(\desc{i}) \neq 0} \frac{\mb{q}(\desc{U})}{\mb{q}(\desc{i})} = \frac{c(\desc{U})}{c(\desc{U}) + 1 - c(\Omega_T \setminus \desc{i} \cup \desc{U} )}.
    \label{eq:max_ratio_qU_qchildi}
  \end{align}
\end{proposition}

\begin{proof}{Proof.}
  Since the problem on the left is a fractional linear program, the maximum is reached at an extreme point of $\qincons$ \citep{Boyd_Vanden_book_2004}. Note also that the objective is increasing in any $q_j, \, j \in \desc{U}$, and decreasing in any $q_s, \, s \in \desc{i} \setminus \desc{U}$.

  Recalling the expression for $\ext(\qincons)$ in Proposition~\ref{prop:extreme_points}, 
  \begin{align*}
    q_{\sigma(i)} = c\bigl( \cup_{k=1}^{i} \sigma(k) \bigr) - c\bigl( \cup_{k=1}^{i-1} \sigma(k) \bigr), \, \forall \, i \in [1,|\nodes{T}|],
  \end{align*}
  let $\mb{v}^{\sigma} \in \ext(\qincons)$ be the extreme point corresponding to $\sigma \in \Pi(\nodes{T})$. We claim that there exists an optimal solution in~\eqref{eq:max_ratio_qU_qchildi} such that the permutation $\sigma$ is of the form
 \begin{equation}
    \label{eq:optimal_permut_max_qUi}
    \begin{aligned}
      \desc{U} &= \{ \sigma(1), \dots, \sigma(|\desc{U}|) \} \\
      \desc{i} \setminus \desc{U} &= \{ \sigma(|\nodes{T} \setminus \desc{i} \cup \desc{U}|+1), \dots, \sigma(|\nodes{T}|) \}, 
    \end{aligned}
  \end{equation}
  i.e., the elements of $\desc{U}$ appear in the first $|\desc{U}|$ positions of $\sigma$, and the elements of $\desc{i} \setminus \desc{U}$ appear in the last positions of $\sigma$.

  The proof involves a repeated interchange argument. We first argue that there exists an optimal permutation $\sigma$ such that the elements of $\desc{U}$ appear \emph{before} those of $\desc{i} \setminus \desc{U}$.

  To see this, consider any permutation $\sigma$ such that $\mb{v}^{\sigma}$ is optimal in~\eqref{eq:max_ratio_qU_qchildi}, yet there exist $j \in \desc{U}$ and $s \in \desc{i} \setminus \desc{U}$ such that $j = \sigma(k)$, $s = \sigma(\bar{k})$, and $\bar{k} < k$. In fact, let $k$ be the smallest, and $\bar{k}$ the largest such index among all indices satisfying the property (this ensures that there are no indices from $\desc{i}$ appearing in $\sigma$ between $\bar{k}$ and $k$). Consider a new permutation $\pi$ where the positions $k$ and $\bar{k}$ are interchanged, and let $\mb{v}^{\pi}$ denote the corresponding vertex of $\qincons$. By submodularity of $c$,
  \begin{align*}
    v^{\pi}_j \bydef c\bigl( \cup_{\ell=1}^{\bar{k}} \sigma(\ell) \bigr) -
    c\bigl( \cup_{\ell=1}^{\bar{k}-1} \sigma(\ell) \bigr) \geq
    c\bigl( \cup_{\ell=1}^{k} \sigma(\ell) \bigr) -
    c\bigl( \cup_{\ell=1}^{k-1} \sigma(\ell) \bigr) \bydef v^{\sigma}_j.
  \end{align*}
  By a similar argument, $v^{\pi}_s \leq v^{\sigma}_s$. Furthermore, by construction, $v^{\pi}_{r} = v^{\sigma}_r, \, \forall \, r \in \desc{i} \setminus \{j, s\}$, since no indices from $\desc{i}$ appear between $\bar{k}$ and $k$. Therefore, we have $\mb{v}^{\pi}(\desc{U}) \geq \mb{v}^{\sigma}(\desc{U})$, and $\mb{v}^{\pi}(\desc{i} \setminus \desc{U}) \leq \mb{v}^{\sigma}(\desc{i} \setminus \desc{U})$, so that the objective at $\mb{v}^{\pi}$ is at least as large as at $\mb{v}^{\sigma}$. Repeating the argument as often as needed, we obtain an optimal permutation satisfying the desired property.

  Having argued that (w.l.o.g.) $\sigma$ contains the elements of $\desc{U}$ before those of $\desc{i} \setminus \desc{U}$, a similar interchange argument can be done with respect to $\nodes{T} \setminus \desc{i}$, to reach the conclusion~\eqref{eq:optimal_permut_max_qUi}. The final result of the lemma exactly denotes the value corresponding to such a configuration (it follows immediately by recognizing the telescoping sums appearing in the expressions).
  \qed
\end{proof}

%
%  Checking that mu(E) and E(mu) are consistent lower bounds
%
\begin{theorem}[Proposition~\ref{prop:lower_bounds_by_cond_expectation}.]
  Consider any \emph{distortion} risk measure $\incons : \rvspace_2 \rightarrow \reals$, and the time-consistent, comonotonic measures $\incons \circ \E$ and $\E \circ \incons$, where $\E$ denotes the conditional expectation operator. Then, for any cost $Y \in \rvspace_{2}$,
  \begin{align*}
    (\incons \circ \E)(Y) \leq \incons(Y) \quad \textup{and} \quad (\E \circ \incons)(Y) \leq \incons(Y).    
  \end{align*}
\end{theorem}
\begin{proof}{Proof.}
  First note that both measures are readily time-consistent and comonotonic, by Definition~\ref{def:dynamic_consist_distortion_risk_measures}. The proof entails arguing that these choices correspond to Choquet capacities that verify the first set of conditions in Corollary~\ref{cor:lower_upper_bounds_explicit}.

  To this end, let $\Psi$ denote the distortion function corresponding to the (distortion) measure $\incons$, i.e., the Choquet capacity is given by $c(S) = \Psi(\Pr(S)), \, \forall \, S \subseteq \Omega_2$, where $\Psi : [0,1] \rightarrow [0,1]$ is concave, nondecreasing, with $\Psi(0) = 0, \, \Psi(1) = 1$. Recall from Section~\ref{sec:examples} that the (compositional) risk measure $\incons \circ \E$ (or, more correctly, $\incons^1 \circ \E$) exactly corresponds to the following choice of Choquet capacities for the first and second stage, respectively:
  \begin{align*}
    c_1 : 2^{\nodes{1}} \rightarrow \reals, \, c_1(S) &= \Psi\Bigl(\, \sum_{i \in S} \Pr(\child{i}) \Bigr), \, \forall \, S \subseteq \nodes{1} \\
    c_{2|i} : 2^{\child{i}} \rightarrow \reals, \, c_{2|i}(U_i) &= \frac{\Pr(U_i)}{\Pr(\child{i})}, \, \forall \, U_i \subseteq \child{i}, \, \forall \, i \in \nodes{1}.
  \end{align*}
  Note that the same distortion function $\Psi$ (yielding the risk measure $\incons$) is applied in the first stage, but to the appropriate conditional probability measure. The second stage is simply a standard conditional expectation.

  With $p_i \bydef \Pr(\child{i})$ and $u_i \bydef \Pr(U_i)$, the desired condition in Corollary~\ref{cor:lower_upper_bounds_explicit} becomes:
\begin{align*}
  \sum_{i=1}^{|\nodes{1}|} \frac{\Psi\bigl( \sum_{j=1}^i p_{\sigma(j)} \bigr) - \Psi\bigl( \sum_{j=1}^{i-1} p_{\sigma(j)} \bigr)}{p_{\sigma(i)}} \, u_{\sigma(i)}
  & \leq \Psi\Bigl( \sum_{i=1}^{|\nodes{1}|} u_{\sigma(i)} \Bigr), \, \forall \, \sigma \in \Pi(\Omega_1), \, \forall \, u_i \in [0,p_i], \, \forall \, i \in \nodes{1}. ~(*)
\end{align*}
To see this, one can use the decreasing marginal returns property of $\Psi$, i.e.,
\begin{align*}
  \frac{\Psi(y_2)-\Psi(y_1)}{y_2-y_1} \leq \frac{\Psi(x_2)-\Psi(x_1)}{x_2-x_1}, \, \forall \, x_1 < x_2, \, x_1 \leq y_1,\, \, x_2 \leq y_2, \, y_1 < y_2,
\end{align*}
to argue that 
% \begin{align*}
$  \frac{\Psi\bigl( \sum_{j=1}^i p_{\sigma(j)} \bigr) - \Psi\bigl( \sum_{j=1}^{i-1} p_{\sigma(j)} \bigr)}{p_{\sigma(i)}} \leq \frac{\Psi\bigl( \sum_{j=1}^i u_{\sigma(j)} \bigr) - \Psi\bigl( \sum_{j=1}^{i-1} u_{\sigma(j)} \bigr)}{u_{\sigma(i)}}$.
% \end{align*}
Replacing this in the left-hand side of $(*)$ and telescoping the sum directly yields the desired result.

In a similar fashion, the risk measure $\E \circ \incons$ corresponds to a choice of capacities
\begin{align*}
  c_1 : 2^{\nodes{1}} \rightarrow \reals, \, c_1(S) &= \sum_{i \in S} \Pr( \child{i} ), \, \forall \, S \subseteq \nodes{1} \\
  c_{2|i} : 2^{\child{i}} \rightarrow \reals, \, c_{2|i}(U_i) &= \Psi\Bigl(\frac{\Pr(U_i)}{\Pr(\child{i})}\Bigr), \, \forall \, U_i \subseteq \child{i}, \, \forall \, i \in \nodes{1}.
\end{align*}
With the same notation $p_i \bydef \Pr(\child{i})$ and $u_i \bydef \Pr(U_i)$, the conditions to test become:
\begin{align*}
  \sum_{i=1}^{|\nodes{1}|} p_{\sigma(i)} \Psi \Bigl( \frac{u_{\sigma(i)}}{p_{\sigma(i)}} \Bigr) &\leq \Psi\biggl( \sum_{i=1}^{|\nodes{1}|} u_i \biggr), \, \forall \, \sigma \in \Pi(\Omega_1), \, \forall \, u_i \in [0,p_i], \, \forall \, i \in \nodes{1}.
\end{align*}
These are readily true, since $\{p_i\}_{i \in \nodes{1}}$ are convex combination coefficients, and $\Psi$ is concave. 
\qed
\end{proof}

%
%  Proof of alpha*(E(rho)) = alpha*(rho(E)) for uniform tree and reference measure
%
\begin{theorem}[Theorem~\ref{thm:alpha_rhoE_equal_Erho_uniform_rm}.]
  Consider a uniform scenario tree, i.e., $|\nodes{1}| = N, \, |\child{i}| = N, \, \forall \, i \in \nodes{1}$, under a uniform reference measure. Then, for any distortion risk measure $\incons$, we have 
  \begin{equation*}
    \alphaopt{\incons \circ \E}{\incons} = \alphaopt{\E \circ \incons}{\incons} = N \cdot \max \Bigl\{ \frac{\Psi( 1/N^2)}{\Psi(1/N)}, \, \frac{\Psi( 2/N^2)}{\Psi(2/N)}, \dots, \Psi( 1/N) \Bigr\}.
  \end{equation*}
\end{theorem}
Before presenting the proof, we introduce two lemmas that outline several relevant properties for the two expressions that need to be compared. To fix ideas, assume the distortion risk measure $\incons$ is given by a concave distortion function $\Psi : [0,1] \rightarrow [0,1]$. To this end, by applying the result in Theorem~\ref{thm:optimal_alpha}, our goal is to argue that 
\begin{align}
  \alphaopt{\E \circ \incons}{\incons} \bydef \max_{\mb{q} \in \ext(\qincons)} \max_{S \subseteq \nodes{1}} \frac{\sum_{i \in S} \max_{U_i \subseteq \child{i}} \frac{\mb{q}(U_i)}{\Psi\bigl( \Pr(U_i)/\Pr(\child{i}) \bigr)} }{ \sum_{i \in S} \Pr(\child{i}) } = \max_{\mb{q} \in \ext(\qincons)} \max_{S \subseteq \nodes{1}} \frac{\sum_{i \in S} \max_{U_i \subseteq \child{i}} \frac{\mb{q}(U_i)}{\Pr(U_i)/\Pr(\child{i})} }{ \Psi \bigl( \sum_{i \in S} \Pr(\child{i}) \bigr) } \bydef \alphaopt{\incons \circ \E}{\incons}.
  \label{eq:comparison_alpha_rhoE_Erho}
\end{align}

The following lemma discusses the factor $\alphaopt{\E \circ \incons}{\incons}$ in the expression above.
\begin{lemma}
  \label{lem:properties_maximiz_E_of_rho}
  Consider the maximization problems yielding $\alphaopt{\E \circ \incons}{\incons}$ in~\eqref{eq:comparison_alpha_rhoE_Erho}. We claim that:
  \begin{enumerate}
  \item For any given $\mb{q} \in \qincons$, the inner maximization over $S \subseteq \nodes{1}$ is reached at a singleton set $S = \{i\}$ for some $i \in \nodes{1}$.
  \item The optimal $\mb{q} \in \qincons$ in the outer maximization always corresponds to a permutation $\sigma \in \Pi(\nodes{2})$ satisfying the property
    \begin{align}
      \{ \sigma(1), \dots, \sigma(N) \} = \child{i},
      \label{eq:childi_first_N}
    \end{align}
    for some $i \in \nodes{1}$. That is, the first $N$ elements in the permutation belong to the same subtree $\child{i}$.
  \item For any fixed $i \in \nodes{1}$,
    \begin{align*}
      \max_{\mb{q} \in \ext(\qincons)} \max_{U_i \subseteq \child{i}} \frac{\mb{q}(U_i)}{\Psi\bigl( \frac{\Pr(U_i)}{\Pr(\child{i})} \bigr)} = \max_{U_i \subseteq \child{i}} \frac{\Psi\bigl( \Pr(U_i) \bigr)}{\Psi\bigl( \frac{\Pr(U_i)}{\Pr(\child{i})} \bigr)}.
    \end{align*}
  \item $\alphaopt{\E \circ \incons}{\incons} = \max_{i \in \nodes{1}} \max_{U_i \subseteq \child{i}} \frac{\Psi( \Pr(U_i) )}{ \Pr(\child{i})  \Psi\bigl( \frac{\Pr(U_i)}{\Pr(\child{i})} \bigr)}$.
  \end{enumerate}
\end{lemma}
\begin{proof}{Proof of Lemma~\ref{lem:properties_maximiz_E_of_rho}.}
  Claim (1) follows from the mediant inequality. To see this, for a fixed $\mb{q}$, let $v_i \bydef \max_{U_i \subseteq \child{i}} \frac{\mb{q}(U_i)}{\Psi\bigl( \Pr(U_i)/\Pr(\child{i}) \bigr)}, \, \forall\, i \in \nodes{1}$, and note that the maximum over $S \subseteq \nodes{1}$ is achieved at any singleton $\{i\} \subseteq \argmax \{ v_\ell / \Pr(\child{\ell}) \,:\, \ell \in \nodes{1} \}$.

  To see Claim (2), first recall that the set $\ext(\qincons)$ corresponds to all possible permutations of $\child{2}$ (Proposition~\ref{prop:extreme_points}). By Claim (1), since the inner maximum always occurs at a singleton $i^{\opt}(\mb{q})$, the optimal $\mb{q}^{\opt}$ must be such that components in $\child{i^{\opt}(\mb{q}^{\opt})}$ are ``as large as possible''. Due to the concavity of $\Psi$, this occurs when they appear in the first $N$ positions in the permutation $\sigma$ (also see the proof of Corollary~\ref{cor:lower_upper_bounds_explicit}).

  Claim (3) follows directly from Claim (2), by switching the order of the two maximizations, and using the expression for the extreme points of $\qincons$ from Proposition~\ref{prop:extreme_points}.

  Claim (4) follows from Claims (1) and (3), after switching the order of the maximizations over $S$ and $\mb{q}$.
  \qed
\end{proof}

%
%  Lemma about mu(E)
%
The following lemma similarly summarizes properties of the second quantity of interest, $\alphaopt{\incons \circ \E}{\incons}$.
\begin{lemma}
  \label{lem:properties_maximiz_rho_of_E}
  Consider the maximization problems yielding $\alphaopt{\incons \circ \E}{\incons}$ in~\eqref{eq:comparison_alpha_rhoE_Erho}. We claim that:
  \begin{enumerate}
  \item For any given $\mb{q} \in \qincons$, and any $i \in \nodes{1}$,
the inner maximization over $U_i \subseteq \child{i}$ is reached at a singleton set $U_i = \{j\}$ for some $j \in \child{i}$.
  \item Fix $S \subseteq \nodes{1}$. The optimal $\mb{q}^{\opt}(S) \in \qincons$ corresponds to a permutation $\sigma^S \in \Pi(\nodes{2})$ such that
    \begin{align*}
      \nexists \, j_{1,2} \in \{1,\dots,|S|\} ~\textup{such that}~ \sigma(j_1), \, \sigma(j_2) \in \child{i}, ~ \textup{for some}~ i \in \nodes{1}.
     %\label{eq:first|S|_from_different_childi}
    \end{align*}
    In other words, the first $|S|$ elements in the permutation $\sigma$ belong to distinct subtrees $\child{i}$.
  \item Under the same setup as (2), the first $|S|$ elements in $\sigma^S \in \Pi(\nodes{2})$ correspond to the minimum-probability in their respective subtree, i.e.,
    \begin{align*}
      \forall \, k \in [1,|S|], ~~  \sigma^S(k) \in \argmin_{j \in \child{i}} \Pr_j, ~\textup{where}~ i ~\textup{is such that}~ \sigma(k) \in \child{i}.
    \end{align*}
  \item Let $m(i) \bydef \argmin_{j \in \child{i}} \Pr_j$. Then
    \begin{align*}
      \alphaopt{\incons \circ \E}{\incons} = \max_{S \subseteq \nodes{1}} \max_{\sigma \in \Pi(S)} \frac{\sum_{i=1}^{|S|} \Pr(\child{m(\sigma(i))}) \, \frac{\Psi\bigl( \sum_{k=1}^{i} \Pr_{m(\sigma(k))} \bigr) - \Psi\bigl(\sum_{k=1}^{i-1} \Pr_{m(\sigma(k))}\bigr)}{\Pr_{m(\sigma(i))}} }{ \Psi \bigl(\sum_{i=1}^{|S|} \Pr(\child{m(\sigma(i))}) \bigr) }.
    \end{align*}
  \end{enumerate}
\end{lemma}
\begin{proof}{Proof.}
  Claim (1) follows, again, by the mediant inequality. The logic is the same as in Claim (1) of Lemma~\ref{lem:properties_maximiz_E_of_rho}, and is omitted.

  Claim (2) follows from Claim (1), and by recognizing again that $\mb{q}$ should have components ``as large as possible'' in the singletons $j$ that yield the maximums.

  To see Claim (3), first note that Claim (2) allows restricting attention to permutations $\sigma^S$ that have elements from distinct subtrees in the first $|S|$ components. For any such $\sigma(j)$, with $j \in \{1,\dots,|S|\}$,
 \begin{align*}
   q_{\sigma(j)} = \frac{\Psi( \Pr_{\sigma(j)} + \sum_{k=1}^{j-1} \Pr_{\sigma(k)} ) - \Psi( \sum_{k=1}^{j-1} \Pr_{\sigma(k)} )} { \Pr_{\sigma(j)} / \Pr(\child{i}) },
 \end{align*}
 where $\sigma(j) \in \child{i}$. By the concavity of $\Psi$, the above expression is decreasing in $\Pr_{\sigma(j)}$, which implies that $\sigma(j)$ always corresponds to the element in $\child{i}$ with smallest probability.

 Claim (4) follows from the previous three. \qed
\end{proof}

%
%  Completing the proof of alpha*(E(rho)) = alpha*(rho(E))
%
With the previous results, we are now ready to provide a complete proof for our desired result, namely that under a uniform reference measure, $\alphaopt{\E \circ \incons}{\incons} = \alphaopt{\incons \circ \E}{\incons}$.
\begin{proof}{Proof of Theorem~\ref{thm:alpha_rhoE_equal_Erho_uniform_rm}.}
  By Lemma~\ref{lem:properties_maximiz_E_of_rho}, $ \alphaopt{\E \circ \incons}{\incons} = \max_{i \in \nodes{1}} \max_{U_i \subseteq \child{i}} \frac{\Psi\bigl( \Pr(U_i) \bigr)}{ \Pr(\child{i})  \Psi\bigl( \frac{\Pr(U_i)}{\Pr(\child{i})} \bigr)}$. For a uniform reference measure, due to the symmetry, this expression becomes
  \begin{align*}
    \alphaopt{\E \circ \incons}{\incons} = N \cdot \max \Bigl\{ \frac{\Psi( 1/N^2)}{\Psi(1/N)}, \, \frac{\Psi( 2/N^2)}{\Psi(2/N)}, \dots, \Psi( 1/N) \Bigr\}.
  \end{align*}
  Similarly, by Lemma~\ref{lem:properties_maximiz_rho_of_E}, $\alphaopt{\incons \circ \E}{\incons} = \max_{S \subseteq \nodes{1}} \max_{\sigma \in \Pi(S)} \frac{\sum_{i=1}^{|S|} \Pr(\child{m(\sigma(i))}) \, \frac{\Psi\bigl( \sum_{k=1}^{i} \Pr_{m(\sigma(k))} \bigr) - \Psi\bigl(\sum_{k=1}^{i-1} \Pr_{m(\sigma(k))}\bigr)}{\Pr_{m(\sigma(i))}} }{ \Psi \bigl(\sum_{i=1}^{|S|} \Pr(\child{m(\sigma(i))}) \bigr) }$, which becomes, under uniform reference measure,
  \begin{align*}
    \alphaopt{\incons \circ \E}{\incons} = N \cdot \max \Bigl\{ \frac{\Psi( 1/N^2)}{\Psi(1/N)}, \, \frac{\Psi( 2/N^2)}{\Psi(2/N)}, \dots, \Psi( 1/N) \Bigr\}.
  \end{align*}
  Comparing the two expressions above immediately yields the desired equality.
  \qed
\end{proof}

\begin{theorem}[Proposition~\ref{prop:upper_bounds_by_max}.]
  Consider any distortion risk measure $\incons$, and the time-consistent, comonotonic measures $\incons \circ \max$ and $\max \circ \incons$, where $\max$ denotes the conditional worst-case operator. Then:
  \begin{enumerate}
  \item[(i)] For any cost $Y \in \rvspace_{2}$, $\incons(Y) \leq (\incons \circ \max)(Y)$.
  \item[(ii)] There exists a choice of $\incons$ and of random costs $Y_{1,2} \in \rvspace_2$ such that
  $(\max \circ \incons)(Y_1) < \incons(Y_1)$ and $(\max \circ \incons)(Y_2) > \incons(Y_2)$.
  \end{enumerate}
\end{theorem}
\begin{proof}{Proof.}
Let the Choquet capacity yielding the distortion measure $\incons$ be of the form $c(S) = \Psi(\Pr(S)), \, \forall \, S \subseteq \Omega_2$. We show Part (i) of the corollary by checking the conditions of Corollary~\ref{cor:lower_upper_bounds_explicit}. Recall that the risk measure $\incons^1 \circ \max$ corresponds to a choice of capacities
\begin{align*}
  c_1 : 2^{\nodes{1}} \rightarrow \reals, \, c_1(S) &= \Psi\Bigl(\, \sum_{i \in S} \Pr(\child{i}) \Bigr), \, \forall \, S \subseteq \nodes{1} \\
  c_{2|i} : 2^{\child{i}} \rightarrow \reals, \, c_{2|i}(U_i) &= 1, \, \forall \, U_i \neq \emptyset \subseteq \child{i}, \, \forall \, i \in \nodes{1}.
\end{align*}
The conditions to check from Corollary~\ref{cor:lower_upper_bounds_explicit} are
\begin{align*}
    \Psi\bigl(\Pr\bigl( \cup_{i \in S} \child{i} \bigr) \bigr) &\leq \Psi\Bigl(\, \sum_{i \in S} \Pr(\child{i}) \Bigr), ~~ \forall \, S \subseteq \nodes{1}, \\
    \frac{\Psi(\Pr(U))}{\Psi(\Pr(U)) + 1 - \Psi(\Pr(\nodes{2} \setminus \child{i} \cup U))} &\leq 1, \, \, \forall \, U \subseteq \child{i}, \, \forall \, i \in \nodes{1}.
\end{align*}
The first inequality holds since $\Pr\bigl( \cup_{i \in S} \child{i} \bigr) = \sum_{i \in S} \Pr(\child{i})$. The second inequality readily follows since $\Psi$ is upper bounded by 1. \qed
\end{proof}

%
%  beta* for mu(max)
% 
\begin{theorem}[Proposition~\ref{prop:beta_star_mu_of_max}.]
  Consider a uniform scenario tree, i.e., $|\nodes{1}| = N, \, |\child{i}| = N, \, \forall \, i \in \nodes{1}$, under a uniform reference measure. Then, for any distortion risk measure $\incons$, we have 
  \begin{equation*}
    \alphaopt{\incons}{\incons \circ \max} = \max \Bigl\{ \frac {\Psi(1/N)}{\Psi( 1/N^2)}, \, \frac {\Psi(2/N)}{\Psi( 2/N^2)}, \dots, \frac{1}{\Psi( 1/N)} \Bigr\}.
  \end{equation*}
\end{theorem}
\begin{proof}{Proof.}
  Recall that the risk measure $\cons \equiv \incons \circ \max$ (or, more correctly, $\incons^1 \circ \max$) corresponds to a choice of capacities
  \begin{align*}
    c_1 : 2^{\nodes{1}} \rightarrow \reals, \, c_1(S) &= \Psi\Bigl(\, \sum_{i \in S} \Pr(\child{i}) \Bigr) \equiv \Psi \Bigl( \frac{|S|}{N}\Bigr), \, \forall \, S \subseteq \nodes{1} \\
    c_{2|i} : 2^{\child{i}} \rightarrow \reals, \, c_{2|i}(U_i) &= 1, \, \forall \, U_i \neq \emptyset \subseteq \child{i}, \, \forall \, i \in \nodes{1}.
  \end{align*}
  By Theorem~\ref{thm:optimal_alpha}, the optimal scaling factor is given by $ \alphaopt{\incons}{\cons} = \max_{\mb{q} \in \qcons} \max_{S \subseteq \nodes{2}} \mb{q}(S) / \Psi(\frac{|S|}{N^2})$. Let us switch the order of the maximizations, and fix an arbitrary $S = \cup_{i \in \nodes{1}} U_i \subseteq \nodes{2}$. Using the representation of $\qcons$ provided by Proposition~\ref{prop:rep_comp_measure}, it can be readily seen that $\mb{q}(U_i) = 0$ if $U_i = \emptyset$, and $\mb{q}(U_i) \leq p_i$, otherwise, where $\mb{p}(S) \leq c_1(S), \, \forall S \subseteq \nodes{1}$. Therefore,
  \begin{align*}
    \max_{\mb{q} \in \qcons} \mb{q}(S) = c_1(S) = \Psi\Bigl( \frac{|S|}{N} \Bigr),
  \end{align*}
  which, when used in the expression for $\alphaopt{\incons}{\cons}$, immediately leads to the desired result.
\end{proof}

\section{Acknowledgement}
The authors are grateful to Viswanath Nagarajan and Nikhil Bansal for their help with the hardness proof in Section~\ref{sec:computational_complexity}.

\small
\bibliographystyle{plainnat} % outcomment this and next line in Case 1
%\bibliography{/Users/daniancu/Dropbox/Academic/biblio}
%\bibliography{risk_measures.bib}

\end{document}